\documentclass[pre,reprint,amsmath,amssymb,floats,superscriptaddress,nofootinbib,10pt,floatfix]{revtex4-2}

\pdfoutput=1

\usepackage[utf8]{inputenc} 
\usepackage[T1]{fontenc} 
\usepackage{microtype}

\usepackage{graphicx}
\usepackage{dcolumn}
\usepackage{bm}
\usepackage{revsymb}
\usepackage[usenames]{color}
\usepackage{subfigure}
\usepackage{color}
\usepackage{physics}
\usepackage{amsmath}
\usepackage{soul}

\usepackage[usenames]{color}
\usepackage{amsfonts}
\usepackage{graphicx}
\usepackage{dcolumn}
\usepackage{bm}
\usepackage{revsymb}
\usepackage[usenames]{color}
\usepackage{subfigure}
\usepackage{color}
\usepackage{physics}
\usepackage{amsmath}
\usepackage{amssymb,bbm}

\usepackage{xcolor}

\newcommand{\bea}{\begin{eqnarray}}
\newcommand{\eea}{\end{eqnarray}}

\def\dow{\partial}
\def\nn{\nonumber}
\def\lb{\left(}
\def\rb{\right)}

\usepackage{mathrsfs}

\definecolor{nblue}{RGB}{28,130,185}

\definecolor{cgreen}{RGB}{76,153,0}

\definecolor{myorange}{RGB}{245,156,74}

\usepackage{hyperref}
\hypersetup{
  colorlinks=true,
  citecolor=magenta,
  urlcolor=-myorange
}

\usepackage{xcolor}
\definecolor{ogreen} {RGB}{71,191,145}
\definecolor{edit} {RGB}{123,150,145}
\definecolor{purple} {RGB}{148,0,211}

\usepackage{mathbbol}

\newcommand\df{\mathrm{d}}

\newcommand\bbh{\mathbb{h}}
\newcommand\bbr{\mathbb{r}}
\newcommand\ext{\text{ext}}
\newcommand\half{\frac12}
\DeclareMathOperator\Diff{Diff}
\DeclareMathOperator\SO{SO}

\newcommand\sfT{\mathsf T}
\newcommand\sfTT{\mathsf{TT}}
\newcommand\dloc{\text{dloc}}

\begin{document}

\title{Hydrodynamics of plastic deformations in electronic crystals}

\author{Jay Armas}
\email{j.armas@uva.nl}
\affiliation{Institute for Theoretical Physics, University of Amsterdam, 1090 GL Amsterdam, The Netherlands}
\affiliation{Dutch Institute for Emergent Phenomena (DIEP), University of Amsterdam, 1090 GL Amsterdam, The Netherlands}

\author{Erik van Heumen}
\email{e.vanheumen@uva.nl}
\affiliation{Van der Waals-Zeeman Institute, Institute of Physics, University of Amsterdam, 1090 GL Amsterdam, The Netherlands}
\affiliation{Dutch Institute for Emergent Phenomena (DIEP), University of Amsterdam, 1090 GL Amsterdam, The Netherlands}

\author{Akash Jain}
\email{a.jain2@uva.nl}
\affiliation{Institute for Theoretical Physics, University of Amsterdam, 1090 GL Amsterdam, The Netherlands}
\affiliation{Dutch Institute for Emergent Phenomena (DIEP), University of Amsterdam, 1090 GL Amsterdam, The Netherlands}

\author{Ruben Lier}
\email{rubenl@pks.mpg.de}
\affiliation{Max Planck Institute for the Physics of Complex Systems, 01187 Dresden, Germany}
\affiliation{W\"{u}rzburg-Dresden Cluster of Excellence ct.qmat, Germany}

\begin{abstract}
We construct a new hydrodynamic framework describing plastic deformations in electronic crystals. The framework accounts for pinning, phase, and momentum relaxation effects due to translational disorder, diffusion due to the presence of interstitials and vacancies, and strain relaxation due to plasticity and dislocations. We obtain the hydrodynamic mode spectrum and correlation functions in various regimes in order to identify the signatures of plasticity in electronic crystal phases. In particular, we show that proliferation of dislocations de-pins the spatially resolved conductivity until the crystal melts, after which point a new phase of a pinned electronic liquid emerges. In addition, the mode spectrum exhibits a competition between pinning and plasticity effects, with the damping rate of some modes being controlled by pinning-induced phase relaxation and some by plasticity-induced strain relaxation. We find that the recently discovered damping-attenuation relation continues to hold for pinned-induced phase relaxation even in the presence of plasticity and dislocations. We also comment on various experimental setups that could probe the effects of plasticity. The framework developed here is applicable to a large class of physical systems including electronic Wigner crystals, multicomponent charge density waves, and ordinary crystals. 
\end{abstract}

\maketitle


\section{Introduction}

Strong interactions in electronic systems can lead to collective electron states with properties resembling solid, liquid-crystal, or glassy phases of matter. Such states have been observed in metals, semiconductors, as well as superconductors, and appear to be generic in strongly correlated materials. The intricate symmetry breaking patterns that characterise these phases pose both experimental as well as theoretical challenges \cite{2015RvMP...87..457F, 2015Natur.518..179K, 2012AdPhy..61..325M}. Amongst these exotic states of matter, the phases characterised by some form of crystallisation are particularly fascinating. A typical example is the formation of collective charge density wave states, the observational signatures of which are widespread across the phase diagram of various materials~\cite{2015RvMP...87..457F, 2015Natur.518..179K, 2012AdPhy..61..325M, Gruner:1988zz, gor1989charge}. Phenomenologically, charge density wave states are one-dimensional phenomena, akin to uniaxial crystals, and are characterised by an approximate translational order along the direction in which the density wave propagates~\cite{Gruner:1988zz, gor1989charge}.

Multicomponent charge density wave phases, where the translation symmetry is spontaneously broken in all spatial directions, have also been experimentally observed in various contexts; see e.g. \cite{2015RvMP...87..457F, 2015Natur.518..179K, 2019PhRvB..99k5417J}. Of particular importance are ordered phases characterised by rotational symmetry among all the density wave components, such as tridirectional charge density waves \cite{2019PhRvR...1c3114P}. Rotational symmetry combined with spontaneously broken translations makes such phases analogous to Wigner crystals, despite the different microscopic origins. In particular, Wigner crystals typically form when the Coulomb interactions between electrons dominate over their kinetic energy, resulting in electrons spontaneously crystallising and giving rise to an electronic crystal \cite{1934PhRv...46.1002W, TF9383400678, 1978PhRvB..17..494H, Ceperley:1980zz}. 

Isotropic electronic crystal phases, while expected to be ubiquitous in a broad class of materials, continue to pose considerable experimental challenges compared to their one-dimensional charge density wave counterparts. In particular, Wigner crystals are quite fragile in nature and the presence of impurities or topological defects in the crystalline structure can readily destroy the ordered state. Nevertheless, they have been observed in a variety of experimental setups ranging from two-dimensional electron gases \cite{1990PhRvL..65.2189G, 2017NatPh..13..340J}, metal dichalcogenide heterostructures \cite{2021Natur.595...48Z, 2020QuIP...19...15S} and moir\'{e} superlattices \cite{2020Natur.579..359R, 2021NatMa..20..940J}, liquid helium interfaces \cite{GRIMES19801, Grimes:1979zz}, van der Waals heterostructures \cite{2020PhRvB.102t1104P, 2021arXiv210610599L},  to soft materials made of charged colloids \cite{2012NatMa..11..948I}, to mention a few. 

The difficulty in unambiguously observing the electronic crystal phases has been a subject of debate in the past (see e.g. \cite{2001PhRvB..65c5312C}) and motivates the identification of clear signatures of such states in potential experimental realisations. In addition, recent developments in the direct observation of Wigner crystals using scanning tunnelling microscopy \cite{2021arXiv210610599L} and the quantum melting of Wigner crystals \cite{2021Natur.595...48Z}, makes it timely to understand the dynamics of these states in the bulk of the material. A useful theoretical approach to this problem, applicable at low-energies and for long-wavelength fluctuations, is to formulate a hydrodynamic theory for electronic crystals where the role of defects and impurities in various phase transitions can be systematically investigated. 

It is well-understood that defects and impurities can act as indirect probes into the phases of electronic crystals. For instance, the physics of pinned Wigner crystals involves a rich interplay between the translational order due to the underlying lattice structure and relaxation effects (e.g. momentum and phase relaxation) due to possible homogeneities and impurities \cite{Gruner:1988zz, gor1989charge, 2001PhRvB..65c5312C, Delacretaz:2017zxd, 1974SSCom..14..703L, Delacretaz:2016ivq}. This combination manifests itself as broadening and pinning of the Drude-like peak in the optical conductivity
\begin{equation} \label{eq:pinning}
    \sigma(\omega)
    = 
    - \frac{n^2}{\rho}
    \frac{i\omega - \Omega_\phi}
    {(i\omega - \Gamma) (i\omega - \Omega_\phi ) + \omega_0^2}~~,
\end{equation}
defined as the flux response function $\sigma(\omega) = \frac{i}{\omega}G^R_{jj}(\omega)$.
Here $n$ denotes the electron charge density, $\rho$ the momentum susceptibility, $\Omega_\phi$ the rate of density wave phase relaxation, $\Gamma$ the rate of momentum relaxation, $\omega$ the probe frequency, and $\omega_0$ is the pinning frequency. Assuming $\Gamma\approx 0$, the real part of the optical conductivity peaks at a nonzero frequency
\begin{equation}
    \omega^2_{\text{peak}} = \omega_0^2 
    - \half \Omega_\phi^2~~.
    \label{eq:peak-position}
\end{equation}
In a recent paper \cite{Armas:2021vku} (see also \cite{Delacretaz:2017zxd, Delacretaz:2021qqu}), we showed that the optical conductivity \eqref{eq:pinning} arises from a hydrodynamic framework where the presence of point-like impurities leads to pseudo-spontaneous breaking of translation symmetry, i.e. translations are both spontaneously as well as explicitly broken.

Our main interest in this work is the role of topological defects (dislocations) and associated plasticity in isotropic electronic crystal phases. These point defects in two spatial dimensions, or line defects in three spatial dimensions, are known to mediate plastic deformations and their proliferation plays a crucial role in phase transitions, in particular crystal melting. The interplay between topological defects and collective electronic states has been extensively studied in the context of charge density waves (see e.g. \cite{2019NatPh..15...27Z,mesaros}), which are subject to both elastic and plastic deformations. In particular, dislocations can de-pin the density waves and produce coherent signals \cite{1990PhRvL..65.2189G, 2019JETP..129..659B} as well as cause softening of the crystalline structure \cite{2002PhyB..324...82H}. In this paper we will distill some of these signatures of topological defects in electronic crystal phases using a novel hydrodynamic framework.

Frameworks dealing with the near-equilibrium dynamics of topological defects have been formulated in the context of ordinary crystals \cite{1979PhRvB..19.2457N, 1980PhRvB..22.2514Z} as well as for charge density waves and Wigner crystal phases \cite{Delacretaz:2017zxd}. Our approach will be distinct from these earlier works and combines insights from various sources \cite{Azeyanagi:2009zd, Fukuma:2011pr, Armas:2019sbe, Armas:2020bmo, Armas:2021vku, Lier:2021wxd}. Specifically, we introduce a bookkeeping parameter $\ell$ that allows us to control the strength of topological defects or plasticity in a crystal. Furthermore, instead of working with singular Goldstone fields that arise due to spontaneous breaking of translation symmetry in a defected crystal, we work with a dynamical reference metric $\bbh_{IJ}$ that tracks the evolution of the reference configuration of the crystal.\footnote{In standard treatments of dislocations \cite{1979PhRvB..19.2457N, 1980PhRvB..22.2514Z}, one introduces a frame field $e^I_i$ that accounts for the derivatives of both the smooth and singular parts of the translational Goldstone fields. In this work, we neglect the antisymmetric part of $e^I_i$ that characterises the deformation of bond angles in the crystal lattice. Instead, we focus on just the bond distance degrees of freedom contained in the symmetric part of $e^I_i$, equivalently captured by a dynamical reference metric $\bbh_{IJ}$.} 

Using symmetry considerations as our guiding principle, we uncover novel transport coefficients arising from the presence of topological defects. Additionally, the small parameter $\ell$ allows us to probe the dynamics of plastic crystals both for low and high density of topological defects. In particular, we show that topological defects do not contribute to pinning or phase/momentum relaxation in the optical conductivity \eqref{eq:pinning}; these effects are purely induced by point-like impurities.
Instead, topological defects lead to the relaxation of the strain tensor, which can only be probed by the optical conductivity at non-zero wave-vector $k$. In a specific limit where the crystal
viscosity is ignored and is nearly Galilean, the $k$-dependent optical conductivity in the transverse sector reads
\begin{widetext}
\begin{align}
    \sigma_\perp(\omega,k)
    &= 
    - \frac{n^2}{\rho}
    \frac{(i\omega - \Omega_\phi )
    - \frac{i\omega}{i\omega - \Omega_G} D_\phi^\perp k^2}
    {(i\omega - \Gamma) (i\omega - \Omega_\phi )
    + \omega_0^2
    + \frac{i\omega}{i\omega - \Omega_G}  \Big(v_\perp^2
    - (i\omega - \Gamma) D_\phi^\perp \Big)k^2 }~,
    %
    \label{eq:omega-k-conductivity}
\end{align}
\end{widetext}
where $\Omega_G$ is the shear-strain relaxation rate, $v_\perp$ the speed of the crystal sound mode, and $D_\phi^\perp$ is the attenuation rate of Goldstone phases (equivalently, the attenuation rate of interstitial defects). We can see that this expression reduces to its plasticity-free form \eqref{eq:pinning} in the $k\to 0$ limit. However, for nonzero $k$, the optical conductivity receives additional signatures from the plasticity-induced relaxation rate $\Omega_G$. 

\begin{figure*}[!t]
	\center
    {\includegraphics[width=0.4\linewidth]{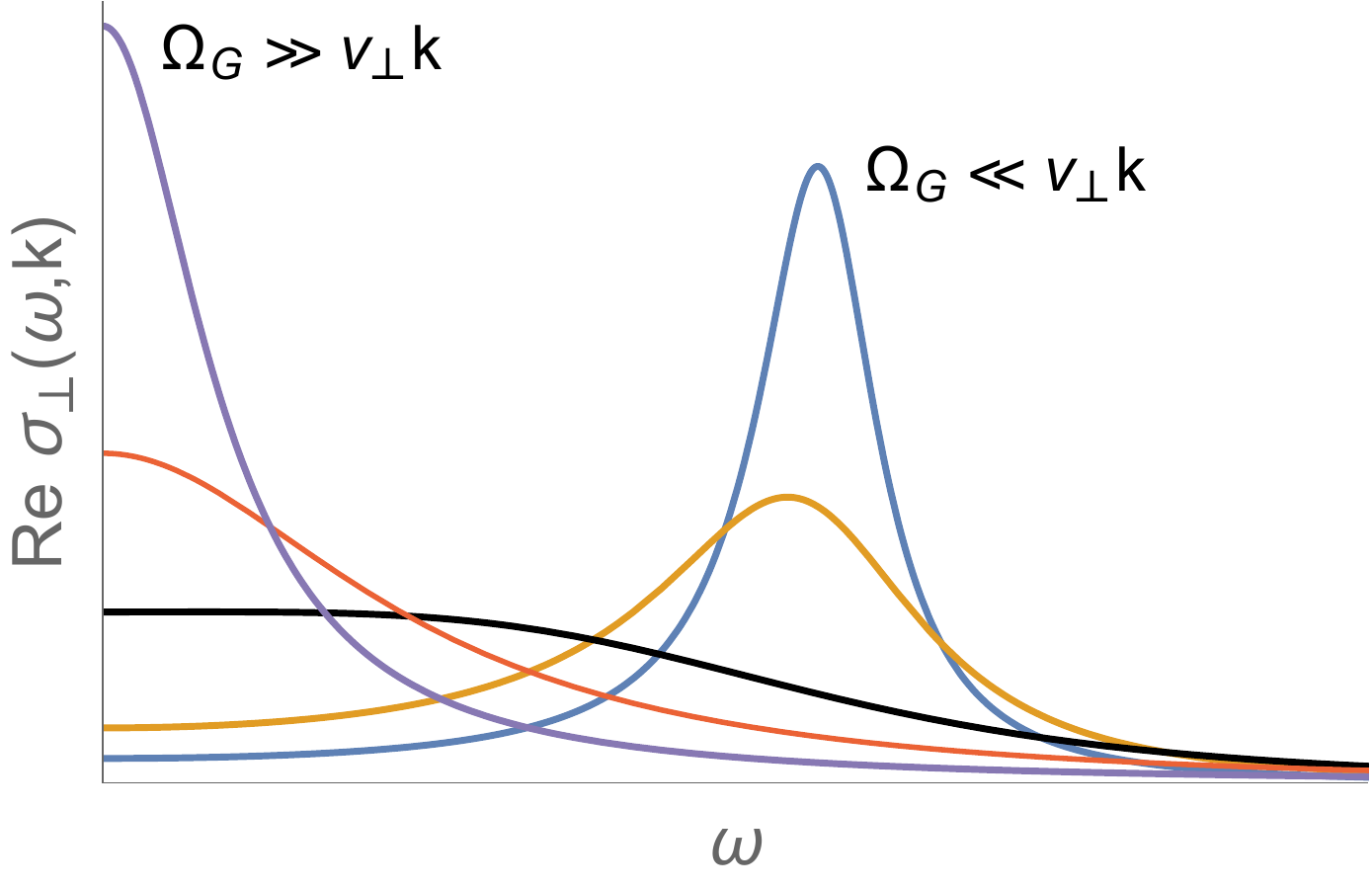}}
    \hspace{5em}
    {\includegraphics[width=0.4\linewidth]{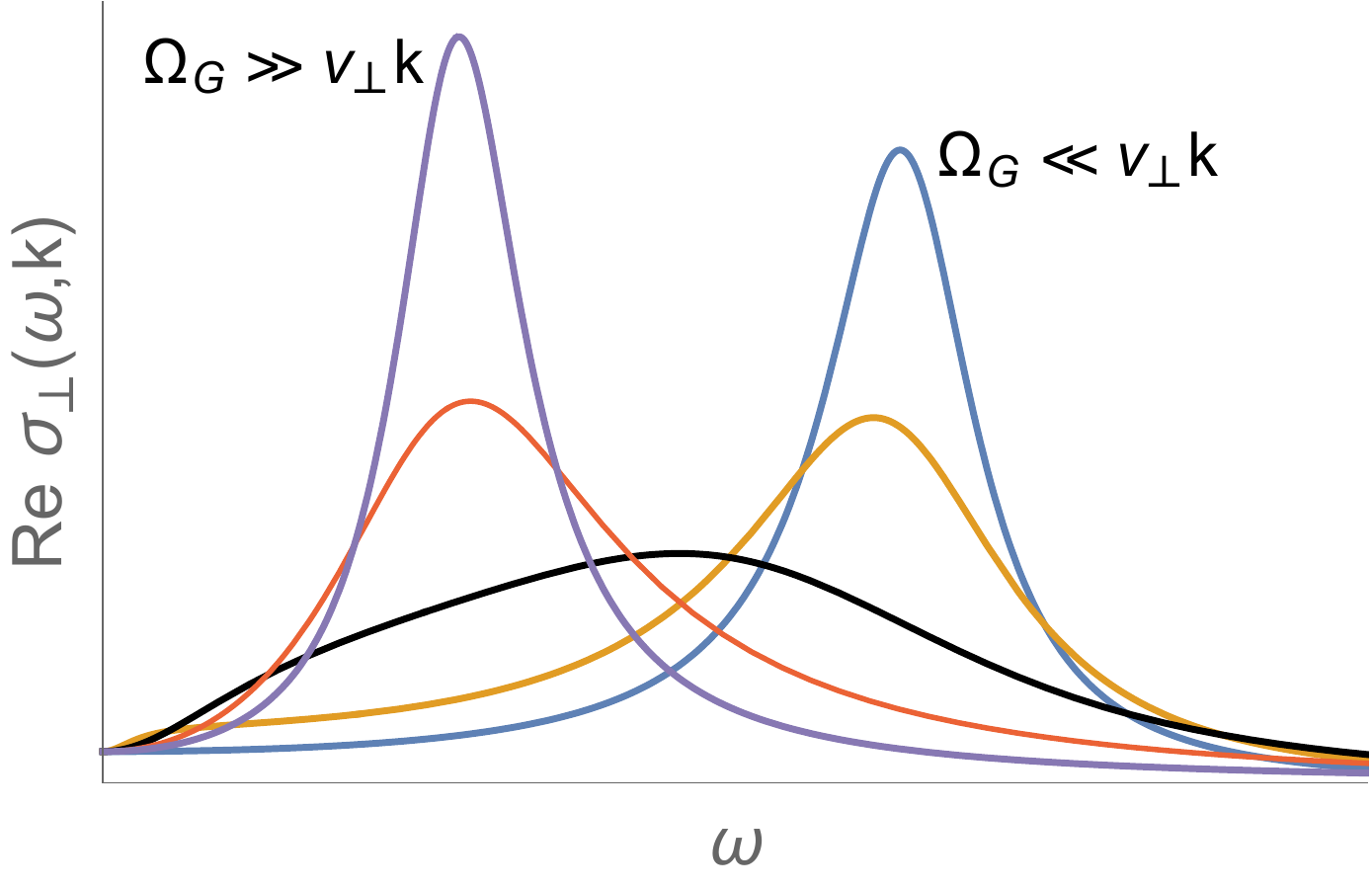}}
    
    \caption{Real part of the transverse optical conductivity at nonzero wavevector for increasing rate of plasticity-induced relaxation $\Omega_G$ in the absence of pinning (left) and in the presence of pinning (right). The peak in the optical conductivity widens and shortens until the solid to liquid phase transition point, sharpening and rising back up again at a lower frequency after the transition. The black curves represent the phase transition point.
    \label{fig:peak-dance}}
\end{figure*}

The result in \eqref{eq:omega-k-conductivity} allows us to draw some interesting conclusions regarding the effects of plasticity on optical conductivity at nonzero wavevector. Firstly, we note that the strength of plasticity (or dislocations) is related to the solid-liquid phase transition of a crystal. If the scale of plasticity-induced relaxation $\Omega_G$ is much smaller than the scale of probe wavevector $v_\perp k$, and all other time-scales induced by pinning, the material essentially behaves like a pinned solid and the position of the $k$-dependent peak is given by 
\begin{equation}
    \omega_{\text{peak}}^2
    = \omega_0^2 + v^2_\perp k^2
    - \frac{
    \Omega_\phi^2(\omega_0^2 + v^2_\perp k^2)^2}{2\omega_0^4}.
\end{equation}
This peak is nothing but the resonance associated with the transverse sound mode in a crystal.
As dislocations start to proliferate, $\Omega_G$ increases and the peak starts to shorten, widen, and shift to the left until we hit the solid to liquid phase transition point; see figure \ref{fig:peak-dance}. At this point, the peak starts to rise-up and sharpen again while still moving to the left, eventually settling back to its zero-wavevector position in \eqref{eq:peak-position} as the melting completes and $\Omega_G\gg v_\perp k$. A qualitatively similar result also holds for the longitudinal optical conductivity and charge susceptibility at nonzero wavevector, and is discussed in detail in section \ref{sec:correlations-pinned}.

To highlight the physical signatures of plasticity and dislocations, for the majority of our discussion we focus on ``pure crystals'' and neglect the presence of point-like impurities or inhomogeneities.\footnote{We also consider electronic crystals with vanishing background magnetic fields for simplicity; see e.g.~\cite{2020PhRvB.102t1104P, 2021arXiv210610599L} for experimental realisations of Wigner crystals in the absence of magnetic fields.} Phenomenologically, this means that we focus on crystals where the translation symmetry is spontaneously, but not explicitly, broken. However, this discussion will incorporate the presence of other point-like defects in crystals, namely interstitials and vacancies. Specifically, we will show how the diffusive nature of interstitials naturally arises from the diffusion of translational Goldstones. In the final section of our discussion, we will generalise this construction to combine the effects of plasticity with point-like impurities and explicitly broken translational invariance. In particular, this will allow us to probe the qualitative differences between strain and Goldstone phase relaxation and how they affect the hydrodynamic equations. We will in particular show that the recently derived damping-attenuation relation $\Omega_\phi =D_\phi^\perp \omega_0^2/ v_\perp^{2}$ \cite{Delacretaz:2021qqu, Armas:2021vku} continues to hold in the presence of dislocations and plasticity, albeit for phase relaxation and not the total strain relaxation. The hydrodynamic framework we construct is applicable to a large class of isotropic physical systems with spontaneously broken (approximate) translational invariance, including (electronic) liquid crystals, metals, and multicomponent charge density wave phases.

This paper is organised as follows. We start our discussion in section~\ref{eq:non-rel-plasticity} with the formulation of a hydrodynamic theory for dissipative plastic crystals. Notably, we work without explicitly imposing any boost symmetry. This allows us to democratically describe both non-relativistic and relativistic crystals at once, while simultaneously enabling us to describe physical situations where boost symmetries might not apply. In section~\ref{sec:linear-viscoplasticity}, we linearise our hydrodynamic theory and work out the rheology equations and stress-strain material diagrams associated with our model. We also report the hydrodynamic predictions for the mode spectrum and response functions in this section, and work out the frequency-dependent viscosities and conductivity. We devote section~\ref{sec:dislocations} to a brief discussion of dislocations in crystals and how they give rise to the dynamical reference metric. In section \ref{sec:pinning}, we combine our results with explicitly broken translations and study the interplay between pinning, phase relaxation, momentum relaxation, and plasticity-induced relaxation. In section \ref{sec:expobs} we discuss experimental setups for probing signatures of plasticity in electronic materials. Finally, we provide an outlook and possibilities for future explorations in section \ref{sec:outlook}. The paper has four appendices. In appendix \ref{app:background}, we revisit our hydrodynamic framework in the presence of a curved spacetime background, enabling us to compute hydrodynamic response functions using the variational approach. In appendix \ref{app:diagramsprocess}, we give details regarding the material diagrams. In appendix \ref{app:Zpp}, we give a detailed comparison of our work to \cite{1980PhRvB..22.2514Z}. The final appendix \ref{app:relativistic} contains a manifestly Lorentz-invariant reconstruction of our hydrodynamic framework specialised to relativistic crystals, which can be useful for approaches to condensed matter systems using holography.

\section{Hydrodynamics of plastic deformations}
\label{eq:non-rel-plasticity}

In this section, we develop the hydrodynamic formalism to describe plastic deformations in a crystal. This is an extension of the earlier work on viscoelasticity with translational Goldstone fields~\cite{Armas:2019sbe, Armas:2020bmo} to include a dynamical reference configuration. Notably, the mentioned references worked exclusively with relativistic crystals. In contrast, with applications to condensed matter systems in mind, we will work without any boost symmetry, relativistic or Galilean. We will comment on the specialisation of our results to Galilean or relativistic crystals as we go. We have also given a separate discussion for relativistic plastic crystals in appendix \ref{app:relativistic}.


\subsection{Elastic vs plastic crystals}
\label{sec:elastic-plastic}

A crystal is a phase of matter where the spatial translational symmetry is spontaneously broken, giving rise to a set of Goldstone fields $\phi^I(\vec x,t)$, which we call the \emph{crystal fields}. The crystal space indices $I,J,\ldots = 1,\ldots, d$ run over the number of spatial dimensions, which we shall distinguish from the physical space indices $i,j,\ldots = 1,\ldots,d$ also running over the number of spatial dimensions. From a phenomenological standpoint, the crystal fields can be understood as a set of Eulerian coordinates  describing the spatial distribution of the lattice cores as a function of time~\cite{Armas:2019sbe, Armas:2020bmo}. In the context of electronic crystals and charge density wave states, the crystal fields, also referred to as \emph{phasons}, are the phases in each spatial direction associated with the spontaneous modulation of electron charge and atomic displacements.

Provided that the crystal is homogeneous, the effective theory describing the crystal must be invariant under constant shifts of the crystal fields $\phi^I\to\phi^I + a^I$. This means that all the dependence on $\phi^I$ in the effective theory must arise via the \emph{crystal frame fields} $e^I_i = \dow_i\phi^I$, which represent a local Cartesian basis carried by each lattice site. We will also make use of the inverse crystal frame fields $e^i_I$. The physical distances between lattice cores throughout the crystal can be measured by the induced metric on the crystal space
\begin{subequations}
\begin{align}
    \df s^2_{\text{crystal}} = h_{IJ} \df\phi^I\df\phi^J,
\end{align}
where $h_{IJ}$ is the inverse of $h^{IJ} = e^{Ii} e^J_i$. We choose to lower/raise the crystal space indices $I,J,\ldots$ using $h_{IJ}$ and $h^{IJ}$. Crystals are also equipped with a reference intrinsic metric 
\begin{align}
    \df s^2_{\text{reference}} = \bbh_{IJ} \df\phi^I\df\phi^J,
\end{align}
for some invertible symmetric matrix $\bbh_{IJ}$. This represents the preferred equilibrium distance between lattice cores that the crystal tries to abide by when no external strains are at play. 
The difference between the two metrics is captured by the \emph{crystal strain tensor}
\begin{align} \label{eq:strain-tensor}
    \kappa_{IJ} = \frac{1}{2} ( h_{IJ} - \bbh_{IJ})  ~~,
\end{align}
\end{subequations}
that serves as a measure for the distortions, shear and expansion, of the crystal. We will often also use its pull-back onto the physical space $\kappa_{ij} = e_i^I e_j^J \kappa_{IJ}$.
The crystal evolves in such a way so as to minimise its strain.

The physical description of a crystal should not depend on the choice of coordinates labelling the lattice sites. Therefore, we must impose a symmetry under local diffeomorphisms on the crystal space 
\begin{subequations}
\begin{equation}
    \Diff(\phi): \df\phi^I \to \Lambda^I_{~J}(\phi)\df\phi^J, \qquad 
    \dow_{[K}\Lambda^I_{~J]} = 0,
\end{equation}
that act on $h^{IJ}$ and $\bbh_{IJ}$ as usual
\begin{align}
    h^{IJ} &\to \Lambda^I_{~K} \Lambda^J_{~L} h^{KL}, \nn\\
    \bbh_{IJ} &\to (\Lambda^{-1})^K_{~I} (\Lambda^{-1})^L_{~J} \bbh_{KL}~.
\end{align}
\label{eq:diffphi}%
\end{subequations}
This means that the crystal fields $\phi^I$ and the reference metric $\bbh_{IJ}$ are not independently physical, while the $\Diff(\phi)$-invariant strain tensor $\kappa_{ij}$ is.
It should be emphasized that $\Diff(\phi)$ is not a local gauge symmetry in the physical space, because these diffeomorphisms can only depend on the crystal fields and not on the spacetime coordinates explicitly. A consequence of this is that the time-derivatives of $\phi^I$ do carry physical information in form of the crystal velocity 
\begin{equation}
    u^i_\phi = -e^i_I \dow_t\phi^I~.
    \label{eq:crystal-velocity}
\end{equation}
It satisfies $(\dow_t + u^i_\phi \dow_i)\phi^I = 0$ and defines the local rest frame of the crystal. In this work we are mainly interested in pure crystals where the translational symmetry in all the $d$ spatial directions is spontaneously broken. We could generalise the above setup to model smectic liquid crystals, where only $k<d$ translations are spontaneously broken, by allowing the crystal space indices to only run over $I,J,\ldots = 1,\ldots,k$. In this case, the derivatives of the crystal fields transverse to the crystalline structure will be $\Diff(\phi)$-invariant. While such situations would be very interesting to study as they would describe generic multicomponent charge density wave states, we leave an explicit analysis for future work. 

For an elastic crystal, the reference metric $\bbh_{IJ}$ is fixed to some known form $\bbh_{IJ}(\phi)$. This means that the reference metric does not evolve in the rest frame of the crystal, specifically\footnote{Throughout this work, we will use ``dot'' to denote the crystal comoving derivative operator $\dow_t + u^i_\phi\dow_i$. For tensors on physical space, ``dot'' will instead denote a Lie derivative along $\dow_t + u^i_\phi\dow_i$.}
\begin{equation}
    \text{Elastic crystals:}\quad 
    \dot\bbh_{IJ}
    \equiv \lb \dow_t + u^i_\phi \dow_i \rb \bbh_{IJ} = 0~.
    \label{eq:elastic-limit}
\end{equation}
In fact, provided that the crystal is homogeneous, we can always choose $\bbh_{IJ}$ to be the Kronecker delta symbol $\delta_{IJ}$, thereby fixing the $\Diff(\phi)$ symmetry down to global rotation in the crystal space $\SO(\phi)$. Following a distortion, an elastic crystal tries to relax back to its original state by aligning the induced metric $h_{IJ}$ with its fixed reference metric $\bbh_{IJ}=\delta_{IJ}$. In contrast, for a plastic crystal, the reference metric $\bbh_{IJ}$ no longer satisfies \eqref{eq:elastic-limit} and evolves independently from the crystal fields. The best we can do in this case is to fix $h_{IJ}$ to $\delta_{IJ}$ at some initial time, say $t=0$. A plastic crystal still tries to align the induced metric $h_{IJ}$ with the reference metric $\bbh_{IJ}$. However, in the time $\Delta t$ that this process takes, the reference metric itself might have evolved from $\bbh_{IJ}(t=0) = \delta_{IJ}$ to some $\bbh_{IJ}(t=\Delta t) \neq \delta_{IJ}$, leading to a ``permanent distortion'' of the crystal; see e.g.~\cite{Fukuma:2011pr, Azeyanagi:2009zd}. 

In the context of plasticity, it is also useful to define a \emph{distortion strain tensor} with respect to the original configuration of the crystal, i.e.
\begin{equation}
    \varepsilon_{IJ} 
    = \frac{1}{2} ( h_{IJ} - \bbh_{IJ}(t=0)) 
    = \frac{1}{2} ( h_{IJ} - \delta_{IJ}) ~~.
    \label{eq:distortion-strain}
\end{equation}
This quantity is more meaningful for experiments as it measures the net distortion of the crystal with respect to some original state. This notion is distinct from the strain tensor $\kappa_{IJ}$ defined in \eqref{eq:strain-tensor}, which is the strain that is felt by the crystal and which it tries to minimise, and hence is more relevant for the effective description. For an elastic crystal, the two definitions coincide. The distortion strain $\varepsilon_{IJ}$ is not directly relevant for the effective field theory because it is not $\Diff(\phi)$-covariant. This makes sense because the absolute notion of ``distortion'' bears no physical meaning without a fixed reference state. However, temporal changes in the distortion are physical and can be captured by the $\Diff(\phi)$-covariant object 
\begin{align}
    \dot\varepsilon_{IJ} 
    &\equiv (\dow_t + u^i_\phi\dow_i) \varepsilon_{IJ}
    = \half (\dow_t + u^i_\phi\dow_i) h_{IJ} \nn\\
    &= e^i_{I}e^j_{J} \dow_{(i} u^\phi_{j)}~.
\end{align}
These are essentially the shear and expansion associated with the crystal velocity $u^i_\phi$.

To get some handle on the problem, we will focus on weakly plastic crystals. To this end, we introduce a small parameter $\ell$ to control the strength of plasticity and decompose the reference metric as
\begin{equation}
    \bbh_{IJ} = \delta_{IJ} + \ell \psi_{IJ}~,
\end{equation}
where $\psi_{IJ}$ parametrises the plastic deformations. In view of the discussion above, we take $\psi_{IJ}(t=0) = 0$. Correspondingly, the intrinsic and distortion strain tensors are related via $\varepsilon_{IJ} = \kappa_{IJ} + \ell/2\,\psi_{IJ}$. We will use the derivative ordering $\ell\sim{\cal O}(\dow)$ to suppress the plastic corrections.

\subsection{Crystals in equilibrium}

The equilibrium configurations of a crystal, elastic or plastic, can be obtained by minimising the grand-canonical free energy 
\begin{align}
    F 
    &= - \int \df^d x \bigg(
    p (T,\mu ,\vec u^2, h^{IJ},\bbh_{IJ}) 
    + T^{ij}_\ext \kappa_{ij}
    \bigg)
    \label{eq:freeEnergy}~~.
\end{align}
Here $p$ is the local thermodynamic pressure of the system, written in terms of the thermodynamic variables: temperature $T$, chemical potential $\mu$, and velocity $u^i$, as well as the induced and reference crystal metrics $h^{IJ}$ and $\bbh_{IJ}$. Note that the crystal velocity $u^i_\phi$ defined earlier is distinct from the thermodynamic velocity $u^i$ introduced here. While the former characterises the local velocity of the lattice sites, the latter characterises the flow of momentum. The explicit relation between the pressure and its arguments $p=p(T,\mu ,\vec u^2, h^{IJ},\bbh_{IJ})$ defines the grand-canonical equation of state of the system. The equation of state must, of course, respect the physical space rotation and translation symmetries, plus the boost symmetry relevant for the problem -- relativistic, Galilean, or none at all. In addition, it must also be invariant under the $\Diff(\phi)$ symmetry given in \eqref{eq:diffphi}.

We have introduced an external stress source $T^{ij}_\ext$ for the strain tensor $\kappa_{ij}$ in \eqref{eq:freeEnergy}. In previous works on elastic crystals~\cite{Armas:2019sbe, Armas:2020bmo, Armas:2021vku}, the authors introduced sources for the crystal fields $\phi^I$ directly. However, for a plastic crystal, it is not consistent to introduce background sources for $\phi^I$ and $\bbh_{IJ}$ independently because of the $\Diff(\phi)$ symmetry mentioned above.

The variation of the thermodynamic pressure can be parameterised by the Gibbs-Duhem relation
\begin{subequations}
\begin{equation}
    \df p  
    = s \df T + n \df\mu + \pi_i \df u^i  
    +  \frac{1}{2} r_{IJ} \df h^{IJ} 
    + \frac{1}{2} {\mathbb r}^{IJ} \df \bbh_{IJ},
    \label{eq:Gibbs-Duhem}
\end{equation}
defining the entropy density $s$, charge/particle density $n$, momentum density $\pi^i$, and the crystal stress tensor $r_{IJ}$. The quantity ${\mathbb r}^{IJ}$ is entirely fixed in terms of $r_{IJ}$ due to $\Diff(\phi)$ symmetry as ${\mathbb r}^{JK}\bbh_{KJ} = r_{IK}h^{KJ}$. Up to leading order in strain, both of these objects are the same. We can also define the energy density $\epsilon$ via the Euler relation
\begin{align}
    \epsilon = - p + Ts + \mu n + u^i\pi_i.
\end{align}
Combining this with \eqref{eq:Gibbs-Duhem} above, we find the first law of thermodynamics
\begin{equation}
    \df\epsilon 
    = T \df s + \mu \df n + u^i \df \pi_i
    - \frac{1}{2} r_{IJ} \df h^{IJ} 
    - \frac{1}{2} {\mathbb r}^{IJ} \df \bbh_{IJ}~.
\end{equation}
\label{eq:thermodynamics}%
\end{subequations}
Due to rotational invariance on physical space, momentum density and fluid velocity must be aligned, i.e. $\pi^i = \rho u^i$.\footnote{Technically, this relation can admit derivative corrections allowed by rotational symmetry, but we can always ``choose'' the thermodynamic velocity to be aligned exactly with momentum.} The quantity $\rho$ is the momentum susceptibility; it is just $n$ for a Galilean system (multiplied with appropriate units of mass per unit particle), while for a relativistic system it is $(\epsilon+p)/c^2$, where $c$ is the speed of light. See~\cite{deBoer:2020xlc,Novak:2019wqg,Armas:2020mpr} for further details on this point.

By varying the free energy with respect to the crystal fields $\phi^I$ and $\psi_{IJ}$, the latter being the plastic part of the reference metric, we can obtain the respective configuration equations
\begin{subequations}
\begin{align}
   - \dow_i\!\lb r_{IJ} e^{Ji} \rb + \frac{\ell}{2} \bbr^{JK} e^i_I \dow_i\psi_{JK} + K_I^\ext &= 0~, \label{eq:config-phi} \\
   \ell\bbr^{IJ} + U^{IJ}_\ext &= 0~, \label{eq:config-bbh}
\end{align}
\label{eq:config-equations}%
\end{subequations}
where we have defined convenient combinations of background fields
\begin{align}
    U^{IJ}_\ext &= - \ell T^{ij}_\ext e^I_i e^J_j, \nn\\
    K_I^\ext &= \dow_i\lb T^{ij}_\ext e_j^J \rb \bbh_{IJ} \nn\\
    &\qquad
    - U^{JK}_\ext
    \lb e_K^i \dow_i \psi_{IJ} 
    - \half e^i_I \dow_i\psi_{JK} \rb.
    \label{eq:source-fields}
\end{align}
We have massaged the two configuration equations so as to make them explicitly covariant under $\Diff(\phi)$. For a rigorous derivation, see the details in appendix \ref{app:background}. In the elastic limit $\ell\to 0$, the $\psi_{IJ}$ configuration equation \eqref{eq:config-bbh} is trivial and we just have \eqref{eq:config-phi} determining the configurations of $\phi^I$. For $\ell\neq 0$, however, one can check that \eqref{eq:config-phi} is completely dependent on \eqref{eq:config-bbh}. This makes sense because all the hydrostatic information in $\phi^I$ can be ``gauged away'' using the $\Diff(\phi)$ symmetry. This will no longer be the case out of equilibrium because $\phi^I$ can have independent physical information contained in $u^i_\phi$.

Normally, we are only interested in the effective theory arranged perturbatively around the strain-free configuration. For this purpose, we can consider the equation of state up to quadratic order in $\kappa_{IJ}$, giving us
\begin{align}  
    p
    &=  p_f + p_\ell\, \kappa^I_{~I}
     - \frac{1}{2} C^{IJKL} \kappa_{IJ} \kappa_{KL}~~,
    \label{eq:linearised-p}
\end{align}
where all the coefficients are understood to be arbitrary functions of $T$, $\mu$, and $\vec u^2$. Here $p_f$ denotes the ``fluid'' part of the thermodynamic pressure that is independent of the crystalline structure. We can use it to obtain the ``fluid'' thermodynamic densities $s_f$, $n_f$, $\rho_f$, and $\epsilon_f$ similarly to \eqref{eq:thermodynamics}. On the other hand, the \emph{elastic moduli tensor} $C^{IJKL}$ contains information about the bulk modulus $B$ and shear modulus $G$ via
\begin{align}
  C^{IJKL}  
  =  \left(B -  \frac{2}{d }  G \right) 
  h^{IJ} h^{KL}      
  + 2G\,  h^{I(K} h^{L)J} ~~.
\end{align}
Both $B$ and $G$ must be non-negative to ensure mechanical stability. Finally, $p_\ell$ denotes the \emph{lattice pressure}~\cite{Armas:2019sbe, Armas:2020bmo}. Mechanical stability requires that $p_\ell|_{\text{eq}} = 0$, when evaluated in thermodynamic equilibrium $T=T_0$, $\mu=\mu_0$, and $u^i=0$. However, its thermodynamic derivatives can generically be nonzero and define the crystal expansion coefficients 
\begin{align}
    \alpha_T = \frac{1}{B} \frac{\dow p_\ell}{\dow T}, \quad
    \alpha_m = \frac{1}{B} \frac{\dow p_\ell}{\dow \mu}, \quad
    \alpha_u = \frac{2}{B} \frac{\dow p_\ell}{\dow \vec u^2}~~,
\end{align}
related to temperature, chemical potential, and velocity fluctuations respectively. The thermodynamic densities from \eqref{eq:thermodynamics} can be obtained in terms of these as
\begin{align}
    s &= s_f + B\alpha_T \kappa^I_{~I} + {\cal O}(u^2), \nn\\
    n &= n_f + B\alpha_m \kappa^I_{~I} + {\cal O}(u^2), \nn\\
    \rho &= \rho_f + B\alpha_u \kappa^I_{~I} + {\cal O}(u^2), \nn\\
    r_{IJ} &= - p_\ell \bbh_{IJ} 
    + B\, h_{IJ} \kappa^K_{~K} 
    + 2G\,\kappa_{\langle IJ\rangle} + {\cal O}(u^2), \nn\\
    \bbr^{IJ} &= - p_\ell h^{IJ} 
    + B\, h^{IJ} \kappa^K_{~K} 
    + 2G\,\kappa^{\langle IJ\rangle}
    + {\cal O}(u^2)~,
    \label{eq:linearised-thermo}
\end{align}
where the angular brackets denote a symmetric-traceless combination.

Finally, we note that the free energy \eqref{eq:freeEnergy} can also admit higher-derivative corrections that we have ignored here for simplicity. Incidentally, assuming time-reversal invariance, no such corrections appear at first order in derivatives.

\subsection{Viscoplastic hydrodynamics}
\label{sec:viscoplastic-hydro}

Plastic deformations of a crystal are an out-of-equilibrium phenomenon. To study these, we will employ the framework of hydrodynamics~\cite{Glorioso:2018wxw, Kovtun:2012rj, landau1959fluid}, meaning that we model plastic deformations by small perturbations of a crystal near the equilibrium state discussed in the previous subsection.

Out of equilibrium, we do not have the liberty to derive the equations for the crystal fields $\phi^I$ and the reference metric $\psi_{IJ}$ from a hydrostatic free energy. But we know that such equations must exist and we take them to have the schematic form
\begin{subequations}
\begin{align}
    K_I + K^\ext_I = 0 \label{eq:josephson_phi} ~~, \\ 
    U^{I J} + U_\ext^{IJ} = 0 \label{eq:josephson_bbh} ~~.
 \end{align}
\label{eq:josehpson-together}%
\end{subequations}
Here $K_I$ and $U^{IJ}$ are some operators dual to $\phi_I$ and $\psi_{IJ}$ respectively. However, for the time being, we do not know much about these operators except that they must reduce to their hydrostatic form \eqref{eq:config-equations} in equilibrium. In foresight, the first one of these equations gives rise to the Josephson equation for the crystal fields, relating the crystal velocity $u^i_\phi$ to the thermodynamic velocity $u^i$. On the other hand, the second equation determines the time-evolution of the intrinsic metric $\dot\bbh_{IJ} = (\dow_t+u^i_\phi\dow_i)\bbh_{IJ}$, which is identically zero for an elastic crystal due to \eqref{eq:elastic-limit}.

Out of equilibrium, the conserved quantities: energy density $\epsilon$, momentum density $\pi_i$, and charge/particle density $n$, also become dynamical and are governed by their respective conservation equations
\begin{subequations} 
\begin{align}
    \dow_t\epsilon + \partial_i\epsilon^i
    &= 
    - K_{I } \dow_t{\phi}^I
    - \half U^{IJ}(\dow_t + u^k_\phi \dow_k)\psi_{IJ}, \\ 
    \dow_t\pi^i + \partial_j \tau^{ij }  
    &= K_{I } \partial^i  \phi^I, \label{momcons} \\ 
    \dow_t n + \partial_i j^{i}   &  =0 ~~ , \label{feieh09203209u}
\end{align}
\label{eq:conservation}%
\end{subequations}
where $\epsilon^i$ is the energy flux, $\tau^{ij}$ the symmetric stress tensor, and $j^i$ the particle flux. When the external sources are absent, the operators $K_I$ and $U^{IJ}$ are zero onshell due to \eqref{eq:josehpson-together}, and energy and momentum are both conserved.  In the presence of external sources, however, both the energy and momentum are sourced. The precise form of these couplings is derived in appendix \ref{app:background}. In particular, we note that the couplings are such that the conservation equations are invariant under the $\Diff(\phi)$ symmetry, provided that both $K_I$ and $U^{IJ}$ transform homogeneously.

It is also useful to introduce the interstitial density $n_\Delta$ and flux $j^i_\Delta$, defined as the total particle density/flux minus the particle density/flux of lattice sites
\begin{subequations}
\begin{align}
    n_\Delta &= n - m_0 v , \nn\\
    j^i_\Delta &= j^i - m_0 v\, u^i_\phi,
    \label{eq:interstitials}
\end{align}
where 
\begin{equation} \label{eq:local-volume-element}
    v =\sqrt{\det(e_i^Ie_j^J\bbh_{IJ})},
\end{equation}
is the local volume element of the lattice and $m_0$ denotes the (constant) number of particles per unit volume in the crystal.\footnote{In the context of Wigner crystals, the flux $m_0 v\, u^i_\phi$ is the electric current due to displacements of the collective electron state, and the density $m_0 v$ the electronic density caused by deformations of the electron state.} It measures the number density of interstitials or vacancies present in a crystal. For an elastic crystal, this quantity is conserved, while a plastic crystal can develop new interstitials or vacancies due to the change in the volume of the reference metric. To wit
\begin{equation} 
    \dow_t n_\Delta + \dow_i j^i_\Delta 
    = - m_0 (\dow_t + u^k_\phi\dow_k)\det\bbh~.
    \label{eq:interstitial-conservation}
\end{equation}
\label{eq:interstitial-stuff}
\end{subequations}
As we review in section~\ref{sec:dislocations}, changes in the volume of the reference metric corresponds to climb motion of dislocations \cite{1980PhRvB..22.2514Z, Beekman:2016szb, 2006PMag...86.2995C, 1989gfcm.book.....K, nabarro1967theory}. Climbing of dislocations is often neglected as it is energetically much more costly than glide motion (see figure~\ref{fig:dloc-glide}), which results in the interstitial density being conserved. For a Galilean crystal, we have $j^i = \pi^i = n u^i$. It is clear, therefore, that the diffusion of interstitials and vacancies is directly tied to the diffusion of the crystal fields responsible for the misalignment between $u^i$ and $u^i_\phi$. We shall see this explicitly in the course of our discussion.
  
To complete the hydrodynamic setup, we must specify a set of constitutive relations for all the unknown operators: $\epsilon^i$, $\tau^{ij}$, $j^i$, $K_I$, and $U^{IJ}$, arranged order-by-order in a derivative expansion. Note that we do not need to write down the constitutive relations for $j^i_\Delta$, which will be determined by the relations for $j^i$ along with $u^i_\phi$ obtained through the Josephson relation \eqref{eq:josephson_phi}. In general, however, the constitutive relations, cannot be arbitrary. They must satisfy the local second law of thermodynamics, which states that there must exist an entropy density $s^t$ and flux $s^i$ such that 
\begin{equation}
    \dow_t s^t + \dow_i s^i = \Delta \geq 0~,
\end{equation}
for all solutions of the equations of motion. The quantity $\Delta$ denotes the non-negative dissipation rate of the system. For the sake of simplicity, we shall assume that $s^t = s$ is the thermodynamic entropy density, although in general it can also admit higher-derivative corrections~\cite{Israel:1979wp, Banerjee:2015hra}. These corrections are directly related to the admissible derivative corrections in the hydrostatic free energy \eqref{eq:freeEnergy}, and are incidentally absent at one-derivative order for a time-reversal invariant crystal.

Let us take the following ansatz for the constitutive relations 
\begin{align}
    \epsilon^i 
    & = (\epsilon+p) u^i  
    + r_{IJ} e^{Ii} e^J_t
    + {\cal T}^{ij} u_j + \mathcal{E}^i  ~~, \nn\\
    \tau^{ij} 
    &= \rho\, u^i u^j  + p\,\delta^{ij} 
    - e_I^i e^j_J  r^{IJ} 
    + \mathcal{T}^{ij} ~~, \nn\\ 
    j^i  & = n u^i   + \mathcal{J}^i ~~, \nn\\
    K_I 
    &= -\partial_{i}\!\lb r_{IJ} e^{Ji}\rb
    + \frac{\ell}{2} \bbr^{JK} e^i_I \dow_i\psi_{JK}
    + \mathcal{K}_I ~~, \nn\\ 
    U^{IJ} 
    &= \ell\bbr^{IJ} + \mathcal{U}^{IJ} ~~,
    \label{eq:ideal-consti}
\end{align}
where ${\cal E}^i$, ${\cal T}^{ij}$, ${\cal J}^i$, ${\cal K}_I$, and ${\cal U}_{IJ}$ denote the respective dissipative corrections. The form of the constitutive relations is precisely picked so that entropy is conserved in the absence of these corrections. In detail, using the thermodynamic relations \eqref{eq:thermodynamics} together with the conservation laws \eqref{eq:conservation}, we find 
\begin{align}
    T\dow_t{s} 
    &= \dow_t{\epsilon} 
    - \mu \dow_t n
    - u^i  \dow_t{\pi}_i   
    + \frac{1}{2} r_{IJ} \dow_t h^{IJ}
    + \frac{1}{2} \bbr^{IJ} \dow_t\bbh_{IJ} \nn\\ 
    &= 
    - \frac{1}{T}  \mathcal{E}^i \partial_i T  
    - \mathcal{T}^{ij}  \partial_i u_j  
    - T{\cal J}^i \dow_i\frac{\mu}{T}
    \nn\\ 
    &\qquad
    - \mathcal{K}_I  (\dow_t + u^i\dow_i) \phi^I 
    - \half\mathcal{U}^{IJ} (\dow_t + u^i_\phi \dow_i) \psi_{IJ} \nn\\ 
    &\qquad
    -T\partial_i s^i~~,
    \label{eq:d_ts}
\end{align}
where the entropy (heat) flux is given as
\begin{align} \label{fehueiuwuiuiu}
    s^i  
    &= \frac1T \bigg( \epsilon^i + p u^i -  \mu j^i  - \tau^{ij} u_j   
    - r_{IJ} e^{Ii} (\dow_t + u^i\dow_i) \phi^J \bigg)~~ \nn\\
    &= s u^i
    + \frac1T {\cal E}^i - \frac{\mu}{T} {\cal J}^i.
\end{align}
Note that the ${\cal U}^{IJ}$ term in \eqref{eq:d_ts} comes with $ (\dow_t + u^i_\phi \dow_i) \psi_{IJ} $ and not $ (\dow_t + u^i \dow_i) \psi_{IJ} $. This ensures that the said term is $\Diff(\phi)$ covariant, because the operator $(\dow_t + u^i_\phi \dow_i)$ yields zero when acting on the $\Diff(\phi)$ parameters. To achieve this, we needed to add the $\dow_i\psi_{IJ}$ term in \eqref{eq:ideal-consti}, ensuring that ${\cal K}_I$ is also $\Diff(\phi)$ covariant. We can check that the form of $K_I$ and $U^{IJ}$ in \eqref{eq:ideal-consti} matches their hydrostatic expectation from \eqref{eq:config-equations}.

Moving on, from \eqref{eq:d_ts}, we infer that the dissipation rate is given as 
\begin{align}
    T\Delta 
    &=
    - \frac{1}{T}  \mathcal{E}^i \partial_i T  
    - \mathcal{T}^{ij}  \partial_i u_j  
    - T{\cal J}^i\dow_i\frac{\mu}{T}
    - \mathcal{K}_I  (\dow_t + u^i\dow_i) \phi^I 
    \nn\\ 
    &\qquad
    - \half\mathcal{U}^{IJ} (\dow_t + u^i_\phi \dow_i) \psi_{IJ}
    \geq 0~,
    \label{eq:dissipationrate}
\end{align}
which, as promised, is trivially zero in the absence of dissipative corrections. Furthermore, the dissipative corrections themselves must be constrained so as to ensure that $\Delta \geq 0$.

\subsection{Constitutive relations}
\label{sec:constitutive}

To obtain all the dissipative corrections permitted by the second law of thermodynamics, we will need to solve \eqref{eq:dissipationrate} order-by-order in the derivative expansion. Let us outline our derivative counting scheme. The thermodynamic variables $T$, $\mu$, and $u^i$ are treated as ${\cal O}(\dow^0)$.  The crystal fields $\phi^I$ are counted as ${\cal O}(\dow^{-1})$, ensuring that the frame fields $e^I_i$ and the crystal velocity $u^i_\phi$ are both ${\cal O}(\dow^0)$. We shall treat $\psi_{IJ}$ also as ${\cal O}(\dow^0)$, with the suppression of plasticity controlled by $\ell\sim{\cal O}(\dow)$.

Looking at the form of \eqref{eq:dissipationrate} and the derivative counting scheme above, we notice that we can write down a term in ${\cal K}_I$ which contributes even before the thermodynamic contributions in \eqref{eq:ideal-consti}. These are given by
\begin{equation}
    {\cal K}_I = - \sigma_\phi h_{IJ} (\dow_t+ u^i\dow_i)\phi^I + {\cal O}(\dow),
\end{equation}
for some non-negative coefficient $\sigma_\phi$.
Noting that all the other terms in $K_I$ in \eqref{eq:ideal-consti}, as well as $K_I^\ext$, are already ${\cal O}(\dow)$, the Josephson equation \eqref{eq:josephson_phi} at leading order tells us that $(\dow_t+ u^i\dow_i)\phi^I = {\cal O}(\dow)$ or that the crystal velocity is the same as the thermodynamic fluid velocity up to derivative corrections: $u^i_\phi = u^i + {\cal O}(\dow)$. In view of this, we will count the combination $(\dow_t+ u^i\dow_i)\phi^I$ as ${\cal O}(\dow)$ in the rest of the discussion. This general structure occurs whenever we include massless degrees of freedom, like Goldstones, in the hydrodynamic description~\cite{Jain:2018jxj}.



We want to obtain the dissipative corrections to the constitutive relations of viscoplastic hydrodynamics up to first order in derivatives. For simplicity, we shall only look at the dissipative corrections that affect the constitutive relations at the level of linearised fluctuations. This, in particular, means that we will ignore any terms that are non-linear in strain $\kappa_{IJ}$ or fluid velocity $u^i$. In this regime, we can split the solutions of \eqref{eq:dissipationrate} into scalar, vector, and tensor sectors. Let us first consider the vector sector 
\begin{subequations}
\begin{align}
    \begin{pmatrix}
        \mathcal{E}^I \\
        \mathcal{J}^I  \\
        \mathcal{K}^I
    \end{pmatrix}    
    = -
    \begin{pmatrix}
        \sigma_\epsilon & \gamma_{\epsilon n}  &  \gamma_{\epsilon\phi}   \\
        \gamma'_{\epsilon n} & \sigma_n & \gamma_{n\phi}  \\
        \gamma'_{\epsilon\phi}  & \gamma'_{n\phi}   & \sigma_\phi  \\
    \end{pmatrix} \begin{pmatrix}
        e^I_i \frac{1}{T} \partial^i T   \\ 
        e^I_i T\dow^i\frac{\mu}{T}   \\ 
        (\dow_t + u^i\dow_i)\phi^I
    \end{pmatrix}, 
    \label{eq:consti-vector}
\end{align}
where we have defined ${\cal E}^i = e^i_I{\cal E}^I$, ${\cal J}^i = e^i_I{\cal J}^I$. These constitutive relations are the same as obtained for elastic crystals in~\cite{Armas:2019sbe, Armas:2020bmo}, generalised to systems without a boost symmetry. The coefficients in the first $2\times 2$ block are the thermo-electric conductivities. The very last entry $\sigma_\phi$ is the crystal conductivity, while the remaining off-diagonal entries $\gamma_{\epsilon\phi}$, $\gamma_{n\phi}$ capture the response of Goldstone fields to thermal and particle number fluctuations.
The plasticity effects only enter the scalar and symmetric-traceless tensor sector, where we find
\begin{align}
    \begin{pmatrix}
        {\cal T}^{IJ} \nn\\
        {\cal U}^{IJ}
    \end{pmatrix}
    &= 
    - h^{IJ}\begin{pmatrix}
        \zeta_\tau & \zeta_{\tau\bbh} \\
        \zeta'_{\tau\bbh} & \zeta_\bbh
    \end{pmatrix}
    \begin{pmatrix}
        \dow_k u^k \\
        \half h^{KL}(\dow_t + \bar u^i \dow_i) \psi_{KL}
    \end{pmatrix} \nn\\
    &\hspace{-2em}
    - \begin{pmatrix}
        \eta_\tau & \eta_{\tau\bbh} \\
        \eta'_{\tau\bbh} & \eta_\bbh
    \end{pmatrix}
    \begin{pmatrix}
        2 e^{\langle I}_i e^{J\rangle}_j \dow^i u^j \\
        h^{K\langle I}h^{J\rangle L}(\dow_t + \bar u^i \dow_i) \psi_{KL}
    \end{pmatrix}.
    \label{eq:consti-scalar-tensor}
\end{align}
\end{subequations}
We have defined ${\cal T}^{ij} = e^i_I e^j_J {\cal T}^{IJ}$. The first entries in the respective matrices $\zeta_\tau$ and $\eta_\tau$ are the well-known bulk viscosity and shear viscosity terms. The respective last entries $\zeta_\bbh$ and $\eta_\bbh$ will be related to the relaxation rates of crystal strain tensor due to plasticity. We will return to the remaining off-diagonal entries in the next section.

The off-diagonal primed and unprimed coefficients in the expressions above are related by Onsager's relations~\cite{1931PhRv...37..405O, 1931PhRv...38.2265O, 1945RvMP...17..343C}
\begin{subequations}
\begin{gather}
    \gamma'_{\epsilon n}  = \gamma_{\epsilon n}, \qquad 
    \gamma'_{\epsilon\phi}  =  - \gamma_{\epsilon\phi}, \qquad 
    \gamma'_{n\phi}  =  - \gamma_{n\phi}, \nn\\
    \eta'_{\tau\bbh} = \eta_{\tau\bbh}, \qquad 
    \zeta'_{\tau\bbh} = \zeta_{\tau\bbh}.
    \label{eq:onsager} 
\end{gather}
Demanding that the dissipation rate is positive-semidefinite further  results in the sign constraints 
\begin{gather}
    \eta_\tau \geq 0, \qquad 
    \zeta_\tau \geq 0, \qquad 
    \sigma_\epsilon \geq 0, \qquad 
    \sigma_\phi \geq 0, \nn\\
    \sigma_\epsilon \sigma_n \geq \gamma_{\epsilon n}^2, \qquad 
    \eta_\tau\eta_\bbh \geq \eta_{\tau\bbh}^2, \qquad 
    \zeta_\tau\zeta_\bbh \geq \zeta_{\tau\bbh}^2.
\end{gather}
\end{subequations}
          
We note that if Galilean boost symmetry were imposed, the particle flux does not receive any derivative corrections, leading to the coefficients $\gamma_{\epsilon n}$, $\gamma'_{\epsilon n}$, $\gamma_{n\phi}$, $\gamma'_{n\phi}$, and $\sigma_n$ being zero. However, we will keep these coefficients non-zero for now because some of these are relevant for relativistic crystals. In the relativistic case, the energy flux does not receive any derivative corrections and we instead must set $\gamma_{\epsilon n}$, $\gamma'_{\epsilon n}$, $\gamma_{\epsilon\phi}$, $\gamma'_{\epsilon\phi}$, $\sigma_\epsilon$ to zero. See appendix \ref{app:relativistic} for more details on relativistic crystals.

\section{Linear viscoplasticity}
\label{sec:linear-viscoplasticity}

We devote this section to understanding the physical implications of the hydrodynamic model we developed above. For simplicity, we shall assume the crystal to evolve isothermally, i.e. we will fix $T$ to the constant equilibrium temperature $T_0$. This has the consequence that energy conservation decouples from the rest of the system and we will not be able to probe the respective coefficients $\sigma_\epsilon$, $\gamma_{\epsilon n}$, and $\gamma_{\epsilon\phi}$. Generalising our discussion to restore the effects of temperature variations is straight-forward, albeit the explicit manipulations become more involved. Furthermore, we will turn off the external stress source $T^{ij}_\ext$ for simplicity, which in turn sets $K_I^\ext$ and $U^{IJ}_\ext$ to zero, and focus on linearised fluctuations in
\begin{gather}
    u^i, \qquad 
    \mu = \mu_0 + \delta\mu, \nn\\
    \phi^I = x^I - \delta^I_i \delta\phi^i, \qquad 
    \psi_{IJ}.
\end{gather}
In particular, this means that 
\begin{equation}
    r^{IJ} = \bbr^{IJ} 
    = B\lb \kappa^K_{~K} - \alpha_m \delta\mu \rb h^{IJ} 
    + 2G\,\kappa^{\langle IJ\rangle}
    + \ldots.
\end{equation}

\subsection{Hydrodynamic equations}

First, let us look at the Josephson equation for the crystal fields $\phi^I$ given by \eqref{eq:josephson_phi}, which determines the crystal velocity $u^i_\phi = \dow_t\delta\phi^i$. Plugging in the constitutive relations, we find 
\begin{subequations}
\begin{align}
    u^i_\phi
    &= u^i 
    + D_\phi^\| \dow^i\kappa^k_{~k} 
    + 2D_\phi^\perp 
    \lb  \dow_j\kappa^{ij} {\,-\,} \dow^i\kappa^k_{~k}  \rb  \nn\\
    &\qquad 
    - \lb \gamma_n + \frac{B\alpha_m}{\sigma_\phi} \rb \dow^i\mu~,
    \label{eq:josephson_phi_linear}%
\end{align}
where we have defined 
\begin{gather}
    D_\phi^\| = \frac{B+2\frac{d-1}{d}G}{\sigma_\phi}, \quad 
    D_\phi^\perp = \frac{G}{\sigma_\phi}, \quad 
    \gamma_n = \frac{\gamma_{n\phi}}{\sigma_\phi}~.
    \label{eq:Dphi-def}
\end{gather}
The coefficients $D_\phi^\|$ and $D_\phi^\perp$ are the diffusion rates of the Goldstone field $\delta\phi^i$ longitudinal and transverse to the wavevector respectively, while $\gamma_n$ denotes its response to the chemical potential fluctuations; see~\cite{Armas:2021vku}. For a plastic crystal, we see that $D_\phi^\|$ and $D_\phi^\perp$ also capture the response of $\delta\phi^i$ to the fluctuations in the reference metric via the strain tensor. We can use \eqref{eq:josephson_phi_linear}  to obtain the evolution of the distortion strain tensor $\varepsilon_{ij} = e^I_i e^J_j \varepsilon_{IJ}$ defined in \eqref{eq:distortion-strain}. In the rest frame of the crystal, this is given by the gradient of the crystal velocity
\begin{align} \label{eq:distortionstrain}
    \dot\varepsilon_{ij}
    &= \dow_{(i} u^\phi_{j)} \nn\\
    &= \dow_{(i} u_{j)}
    - \lb \gamma_n + \frac{B\alpha_m}{\sigma_\phi} \rb  \dow_i\dow_j\mu 
    + D_\phi^\| \dow_i\dow_j \kappa^k_{~k} \nn\\
    &\qquad 
    + 2D_\phi^\perp \lb \dow_{(i}\dow^k \kappa_{j)k} 
    - \dow_i\dow_{j} \kappa^k_{~k}\rb .
\end{align}
\end{subequations}
In \eqref{eq:distortionstrain} the ``dot'' denotes the Lie derivative operator with respect to $\dow_t + u^i_\phi \dow_i$.\footnote{We use the convention that traceless and trace combinations are computed after the ``dot'' derivative. This means that $\dot\kappa^K_{~K}$ equals $h^{IJ}(\dow_t + u^i_\phi \dow_i)\kappa_{IJ}$ and not 
$(\dow_t + u^i_\phi \dow_i) (h^{IJ}\kappa_{IJ})$. This distinction is trivial at linear order as long as the field under consideration is at least linear in fluctuations.} Importantly, we see that the distortion strain diffuses but does not relax to zero. This is in contrast to the plastic crystal strain as we will see below.

Next, we have the dynamical equation for the reference metric $\bbh_{ij} = e^I_i e^J_j\bbh_{IJ}$ given by \eqref{eq:josephson_bbh}. Using the constitutive relations, we find 
\begin{subequations}
\begin{align} \label{eq:bbh-equation}
    \dot\bbh_{ij}
    &= 
    - \frac{2}{d}\delta_{ij} (\lambda_B-1) \dow_k u^k
    - 2(\lambda_G-1) \dow_{\langle i}u_{j\rangle}\nn\\
    &\hspace{1em}
    + \frac{2}{d} \delta_{ij} \Omega_B\lb \kappa^k_{~k} - \alpha_m \delta\mu \rb
    + 2\Omega_G \kappa_{\langle ij\rangle},
\end{align}
where we have defined 
\begin{gather} \label{redefinition2}
    \lambda_G = 1 + \frac{\ell\eta_{\tau\bbh}}{\eta_{\bbh}}, \qquad 
    \lambda_B = 1 + \frac{\ell\zeta_{\tau\bbh}}{\zeta_\bbh}, \nn\\
    \Omega_G = \frac{\ell^2G}{\eta_{\bbh}}, \qquad 
    \Omega_B = \frac{\ell^2B}{\zeta_{\bbh}}.
\end{gather}
In the limit $\ell\to0$, the right-hand side vanishes and the reference metric does not evolve with respect to the crystal, as expected for a purely elastic material. The coefficients defined here find physical meaning in the evolution of the crystal strain tensor; we get 
\begin{align}   
    \dot\kappa_{ij}
    &= \frac{1}{d} \lambda_B\delta_{ij} \dow_k u^k
    + \lambda_G \dow_{\langle i}u_{j\rangle}
    - \lb \gamma_n + \frac{B\alpha_m}{\sigma_\phi} \rb \dow_i\dow_j\mu  \nn\\
    &\qquad 
    + D_\phi^\| \dow_i\dow_j\kappa^k_{~k}
    + 2D_\phi^\perp \lb \dow_{(i}\dow^k\kappa_{j)k} 
    - \dow_i\dow_{j}\kappa^k_{~k}\rb \nn\\
    &\qquad
    - \frac{1}{d} \delta_{ij} \Omega_B\lb \kappa^k_{~k} - \alpha_m \delta\mu \rb
    - \Omega_G \kappa_{\langle ij\rangle}~.
    \label{eq:strain-evolution-linear}
\end{align}
\end{subequations}
The first thing we notice is that the leading order term is no-longer just the gradient of the fluid velocity. We have new coefficients $\lambda_B,\lambda_G\neq 1$ that modify these relations in the expansion and shear channels respectively. Similar phenomena was also observed in the presence of pinning in our previous paper~\cite{Armas:2021vku}. The diffusion coefficients $D_\phi^\|$, $D_\phi^\perp$ and the chemical potential response $\gamma_n$ are the same as in the distortion strain. Importantly, however, the crystal strain also relaxes with independent rates $\Omega_B$, $\Omega_G$ in the expansion and shear channels respectively. Such relaxation processes in the crystal strain are expected to be found in generic plastic materials. We also see a relaxation of strain due to chemical potential fluctuations via a nonzero chemical expansion coefficient $\alpha_m$.

Let us now look at the conserved fluxes. We have the stress tensor and the particle flux 
\begin{subequations}  \label{eq:stress-flux-linear}
\begin{align} \label{eq:stressonly}
    \tau^{ij} 
    &= \lb p + \lambda_B B\alpha_m \delta\mu \rb \delta^{ij} 
    - \lambda_B B \delta^{ij} \kappa^k_{~k} 
    - 2\lambda_G G \kappa^{\langle ij\rangle}
    \nn\\
    &\qquad 
    - \zeta \delta^{ij} \dow_k u^k
    - 2\eta \dow^{\langle i} u^{j\rangle}, \\
    j^i
    &= n\, u^i
    - \lb \sigma + B\alpha_m\gamma_n \rb \dow^i\mu \nn\\
    &\qquad 
    - n D^\|_{n} \dow^i \kappa^k_{~k} 
    - 2 n D_n^\perp 
    \lb  \dow_j \kappa^{ij} {\,-\,} \dow^i \kappa^k_{~k}  \rb ,
\end{align}
where 
\begin{gather} \label{redefinition1}
    \eta = \eta_\tau - \frac{\eta_{\tau\bbh}^2}{\eta_\bbh}, \qquad 
    \zeta = \zeta_\tau - \frac{\zeta_{\tau\bbh}^2}{\zeta_\bbh}, \nn\\
    \sigma = \sigma_n + \frac{\gamma_{n\phi}^2}{\sigma_\phi},
\end{gather}
are the true shear viscosity, bulk viscosity, conductivity of the crystal and 
\begin{gather}
    D_{n}^\| = -\frac{\gamma_{n\phi}}{n} D_\phi^\|, \qquad 
    D_{n}^\perp = -\frac{\gamma_{n\phi}}{n} D_\phi^\perp~,
\end{gather}
\end{subequations}
are the longitudinal and transverse diffusion coefficients for the particle flux. For a Galilean crystal, the coefficients $\sigma$, $\gamma_n$, $D_n^\|$, and $D_n^\perp$ all vanish and the particle flux $j^i$ just becomes $nu^i$.

We can also obtain the interstitial flux using \eqref{eq:interstitials}. We can always choose $m_0$ to be the equilibrium number density $n_0$, so that the interstitial density $n_\Delta$ vanishes in equilibrium. We find 
\begin{subequations}
\begin{align} \label{eq:diffinter}
    n_\Delta 
    &= \delta n + \kappa^k_{~k}, \nn\\
    j^i_\Delta 
    &= - \sigma_\Delta  \dow^i\mu \nn\\
    &\qquad 
    - nD_\Delta^\| \dow^i \kappa^k_{~k} 
    - 2 D_\Delta^\perp\!
    \lb  \dow_j \kappa^{ij} {\,-\,} \dow^i \kappa^k_{~k}  \rb,
\end{align}
where 
\begin{gather}
    \sigma_\Delta = \sigma - n\gamma_n 
    - \lb 1 - \frac{\gamma_{n\phi}}{n}\rb \frac{n B\alpha_m}{\sigma_\phi}, \nn\\
    D_\Delta^\| = \lb 1 - \frac{\gamma_{n\phi}}{n} \rb D_\phi^\|, \quad 
    D_\Delta^\perp = \lb 1 - \frac{\gamma_{n\phi}}{n} \rb D_\phi^\perp.
\end{gather}
\end{subequations}
Notice from \eqref{eq:diffinter} that the diffusion of interstitials and vacancies is controlled by the Goldstone diffusion coefficients appearing in \eqref{eq:josephson_phi_linear}. The pre-factor appearing in the relation is unity for Galilean crystals.

\subsection{Field redefinitions of crystal strain}
\label{eq:strain-redef}

We have already discussed the physical distinction between the distortion strain $\varepsilon_{ij}$ representing the mechanical distortions of the crystal with respect to an original configuration, and the crystal strain $\kappa_{ij}$ representing the strain felt by the crystal. Since $\kappa_{ij}$ is only really meaningful in the context of the effective field theory description, we can arbitrarily redefine it as long as the redefinition scales as $\ell$ and drops out in the elastic limit. For example, we can take
\begin{subequations}
\begin{equation}
    \kappa_{ij} \to \kappa_{ij} 
    + \ell a_G \kappa_{\langle ij \rangle}
    + \ell a_B \delta_{ij} \kappa^k_{~k}~.
\end{equation}
This is equivalent to redefining the plastic part of the reference metric as
\begin{equation}
    \psi_{IJ} \to \psi_{IJ} 
    - 2a_G e^i_I e^j_J \kappa_{\langle ij \rangle}
    - 2a_B h_{IJ} \kappa^k_{~k}~.
\end{equation}
\end{subequations}
The upshot of this procedure is that we can choose
\begin{subequations}
\begin{equation}
    a_G = \frac{\eta_{\tau\bbh}}{\eta_\bbh}, \qquad 
    a_B = \frac1d \frac{\zeta_{\tau\bbh}}{\zeta_\bbh}, 
\end{equation}
and rescale the thermodynamic coefficients 
\begin{gather}
    G \to \frac{1}{\lambda_G^2} G, \qquad
    B \to \frac{1}{\lambda_B^2} B , \qquad
    \alpha_m \to \lambda_B\alpha_m,
\end{gather}
\end{subequations}
to get rid of $\lambda_B$ and $\lambda_G$ entirely from the hydrodynamic equations, up to the one-derivative order terms in the constitutive relations. In view of this, we will set $\lambda_B=\lambda_G=1$ in the remainder of this section.

We note that the situation is qualitatively different for the analogous $\lambda$ coefficient appearing in pinned crystals~\cite{Armas:2021vku}. In that context, we could also perform a field redefinition to remove the $\lambda$ coefficient from the hydrodynamic equations. However, this procedure required redefining the crystal phase fields $\phi^I\to\lambda\phi^I$, which are charged under translations and hence modified the correlation functions involving the stress tensor.



\subsection{Rheology equations}
\label{sec:rheology}

To understand the physical materials our hydrodynamic model is describing, let us further freeze the chemical potential fluctuations by setting $\mu = \mu_0$ where $\mu_0$ is a constant chemical potential and focus on just the mechanical fluctuations of the crystal. In addition, we assume that the relevant observable to probe the material with is the time derivative of the distortion strain \eqref{eq:distortionstrain}. Under these conditions, we can  write the rheology equations by eliminating the shear tensor $\dow_{(i}u_{j)}$ in favour of the distortion strain $\varepsilon_{ij}$ using \eqref{eq:distortionstrain}, i.e.
\begin{subequations} \label{eq:rheocrystal}
    \begin{align}
    \tau_{ij} 
    &= - \delta_{ij} \lb B \kappa^k_{~k} 
    + \zeta \dot\varepsilon^k_{~k}  \rb
    \nn\\
    &\qquad 
    - 2G \kappa_{\langle ij\rangle}
    - 2\eta \dot\varepsilon_{\langle ij\rangle}, 
    \label{eq:rheo-1} \\
    \dot\kappa_{ij}
    &= \frac{1}{d} \delta_ {ij}\lb \dot\varepsilon^k_{~k}
    - \Omega_B \kappa^k_{~k} \rb \nn\\
    &\qquad 
    + \dot\varepsilon_{\langle ij\rangle} 
    - \Omega_G \kappa_{\langle ij\rangle}~~.
    \label{eq:rheo-2}
\end{align}
\end{subequations}
In writing these equations, we have only considered terms that contribute to the hydrodynamic equations up to second derivative order, meaning that we consider terms up to $\mathcal{O}(\partial^2)$ for $ \dot\kappa_{ij}$ and terms up to $\mathcal{O}(\partial)$ for $\tau_{ij}$. We have also ignored the constant pressure term in the stress tensor. Due to this power counting scheme and the specific observable under consideration $\dot\varepsilon_{ij}$, we see that all the diffusive corrections disappear and the rheology equations neatly split into the bulk and shear channels.

\begin{figure}[t]
    {\includegraphics[width=0.8\linewidth]{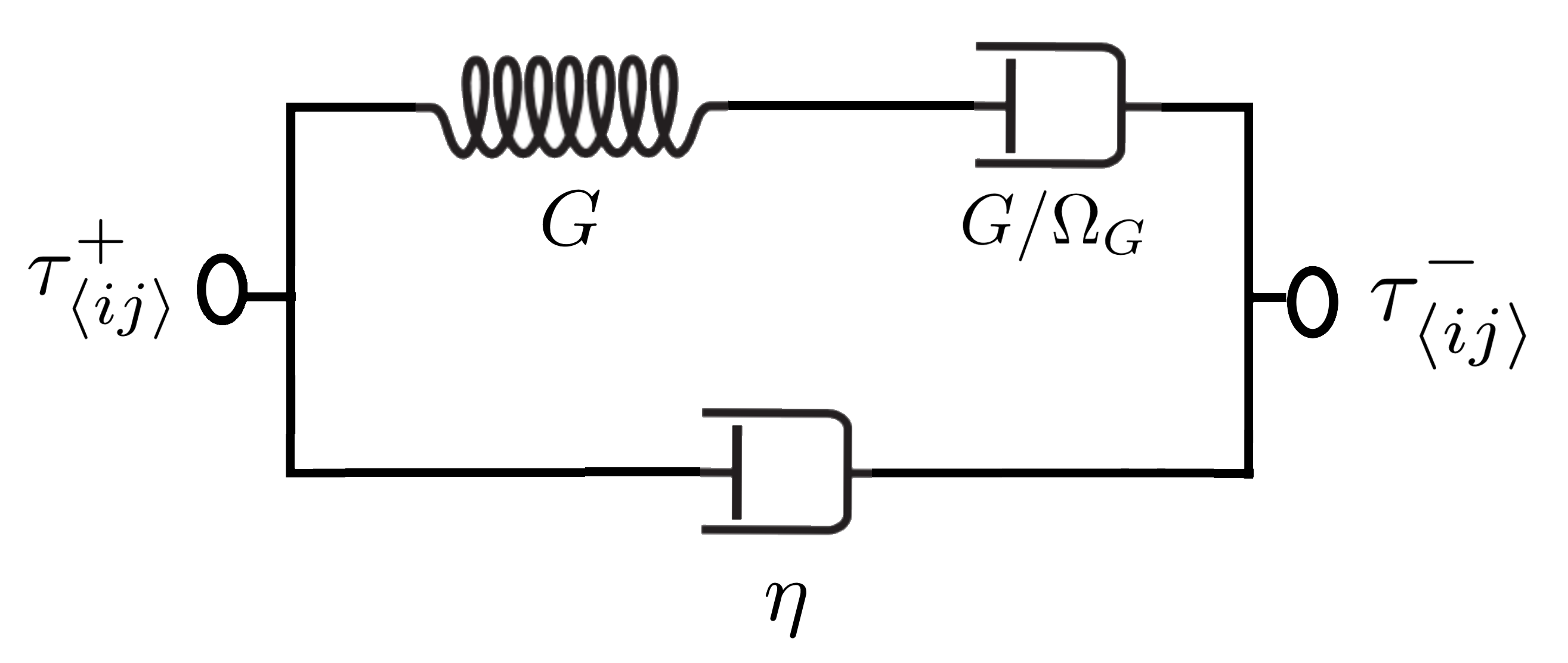}}
     \caption{Circuit representation of the rheology equations for a plastic crystal in the shear sector. The circuit describes a Jeffrey material. In the limit $\eta\to 0$, the dashpot in the lower arm disappears and we get a Maxwell material, whereas in the limit $\Omega_G\to 0$, the dashpot in the upper arm becomes infinitely rigid and we get a Kelvin-Voigt material. The bulk sector of the rheology equations behaves similarly, with $G$, $\Omega_G$, $\eta$ replaced with $B$, $\Omega_B$, $\zeta$ respectively.
     \label{fig:jeffrey}}
\end{figure}

If we momentarily ignore the plasticity effects by setting $\ell\to 0$, we get $\Omega_B=\Omega_G =0$. In this case, \eqref{eq:rheo-2} implies that the crystal strain tensor $\kappa_{ij}$ and the distortion strain tensor $\varepsilon_{ij}$ are the same objects and we arrive at the Kelvin-Voigt model 
\begin{align}
    \tau^k_{~k}
    &= - d B \varepsilon^k_{~k} 
    - d\zeta \dot\varepsilon^k_{~k}, \nn\\
    \tau_{\langle ij\rangle} 
    &= - 2G \varepsilon_{\langle ij\rangle}
    - 2\eta \dot\varepsilon_{\langle ij\rangle}.
\end{align}
If we keep the plasticity effects, however, we end up with the  Jeffrey model~\cite{Fukuma:2011pr, Azeyanagi:2009zd, Lier:2021wxd}
\begin{align} \label{eq:Jeffrey}
    \dot\tau^k_{~k} + \Omega_B\tau^k_{~k}
    &= - d \lb B 
    + \zeta\Omega_B \rb \dot\varepsilon^k_{~k}
    - d\zeta \ddot\varepsilon^k_{~k}, \nn\\
    \dot\tau_{\langle ij\rangle} + \Omega_G \tau_{\langle ij\rangle}
    &= 
    - 2 \lb G + \eta\Omega_G \rb \dot\varepsilon_{\langle ij\rangle}
    - 2\eta \ddot\varepsilon_{\langle ij\rangle}~.
\end{align}
The respective material diagram in given in figure \ref{fig:jeffrey}.
Therefore, our hydrodynamic theory describes the non-equilibrium fluctuations of a Jeffrey material. The special case $\eta = \zeta =0$ describes a Maxwell material. 

As we see from \eqref{eq:rheocrystal}, there are no diffusive corrections in the rheology equations from the perspective of the crystal evolution. The first such corrections appear at ${\cal O}(\ell\dow^2)$ in \eqref{eq:rheo-2}, which are suppressed due to our weak-plasticity assumption $\ell\sim{\cal O}(\dow)$. The diffusive effects in our viscoplastic model relate to the diffusion of interstitials relative to the crystal. In appendix~\ref{app:diagramsprocess}, we consider the rheology equations from the point of view of total matter displacement, including both crystalline and interstitial matter. In this context, the rheology equations do admit diffusive corrections and we find a rich structure akin to generalised Maxwell materials.

\subsection{Mode spectrum}
\label{sec:plastic-modes}

Let us now compute the linearised mode spectrum of plastic hydrodynamics. The behaviour of the spectrum is qualitatively different based on the interplay between the plasticity scale\footnote{The use of $\ell$ as the scale of plasticity should only be understood qualitatively. The precise value of the scale depends on the observable under consideration. For example, in the transverse sector this scale is $\Omega_G/v_\perp$.} $\ell$ and the momentum (wave-number) scale $k$ of fluctuations. To stay within the hydrodynamic regime, we must require that $k \ll L^{-1}_{T}$, where $L_T$ is the thermal length scale. To stay within the weak plasticity regime, we also require that $\ell\ll L_{T}^{-1}$, however the relative ordering of $k$ and $\ell$ may differ. If we focus on fluctuations at momentum scales larger than the scale of plasticity, but still within the hydrodynamic regime, i.e. $\ell \ll k\ll L_{T}^{-1}$, the modes behave like that of a solid with small ``softening'' corrections due to plasticity. On the other hand, if we look at fluctuations at momentum-scales smaller than the plasticity scale, i.e. $k \ll \ell \ll L_{T}^{-1}$, the modes behave like that of a liquid with ``rigidity'' corrections.

To begin with it, is useful to make a wavevector decomposition of the strain such that
\begin{subequations}
\begin{equation} \label{eq:kappa-k-decomposition}
\kappa_{ij}=\frac{k_ik_j}{k^2} \kappa^\| + \frac{k_{(i}}{k} \kappa_{j)}^\perp
    + \lb\delta_{ij} - \frac{k_ik_j}{k^2}\rb \kappa^{\sfT}
    + \kappa^{\sfTT}_{ij},
\end{equation}
where $\kappa_{i}^\perp$ is transverse to $k^i$, $\kappa_{ij}^\sfTT$ is transverse and traceless, $\kappa^\|$ is the longitudinal component and $\kappa^{\sfT}$ is the transverse trace. We can correspondingly decompose the fluid velocity as
\begin{equation}
    u^i = \frac{k^i}{k} u_\| + u^i_\perp,
\end{equation}
\end{subequations}
where $u_\|$ is the longitudinal component and $u^i_\perp$ is transverse to $k_i$. Given this decomposition, we note that the transverse-traceless fluctuations $\kappa_{ij}^\sfTT$ of the strain tensor completely decouple from the rest of the system and lead to a momentum-independent gapped mode $\omega = -i\Omega_G$. This mode is not important for our linearised analysis in this section, thought might play a significant role upon taking non-linear interactions into account. Note that this sector only exists in $d>2$.

Let us now focus on the transverse vector sector spanned by the transverse velocity $u^i_\perp$ and the transverse strain $\kappa^i_\perp$. In the solid regime $\ell\ll k  \ll L_T^{-1}$, we find the well-known transverse sound mode characteristic of a crystal, but damped due to plastic effects\footnote{Note that the ``solid regime'' modes are only valid for $\ell\ll k$. This implies, in particular, that these expressions are not valid if we take $k\to 0$ without taking $\ell\to 0$ first. For $k\to 0$ but $\ell\neq 0$, the ``liquid regime'' modes apply.}
\begin{subequations}
\begin{equation}
    \omega = \pm v_\perp k - \frac{i}{2} \lb \Gamma_\perp k^2 + \Omega_\perp \rb
    + \ldots,
\end{equation}
where we have isolated the speed of transverse sound $v_\perp$, attenuation rate $\Gamma_\perp$, and damping rate $\Omega_\perp$ as
\begin{equation}
    v_\perp^2 = \frac{G}{\rho}, \qquad 
    \Gamma_\perp = \frac{\eta}{\rho} + \frac{G}{\sigma_\phi}, \qquad 
    \Omega_\perp = \Omega_G.
\end{equation}
\label{eq:transverse-modes}%
\end{subequations}
The longitudinal sector is considerably more involved and is spanned by the temperature $T$, chemical potential $\mu$, longitudinal velocity $u_\|$, longitudinal strain $\kappa_\|$, and the transverse trace strain $\kappa_\sfT$. As with the earlier discussion, we ignore the temperature fluctuations which removes the energy diffusion mode. Other than this, focusing on the solid regime $\ell\ll k  \ll L_T^{-1}$, we find a damped longitudinal sound mode and a damped crystal diffusion mode
\begin{subequations}
\begin{align}
    \omega 
    &= \pm v_\| k
    - \frac{i}{2} \lb\Gamma_\| k^2 + \Omega_\| \rb + \ldots, \nn\\
    \omega
    &= -iD_\| k^2 - i\Omega_D + \ldots.
\end{align}
The speed of longitudinal sound $v_\|$, attenuation rate $\Gamma_\|$, and damping rate $\Omega_\|$ are given as
\begin{align}
    v_\|^2 
    &= \frac{n_m^2}{\rho\chi} + \frac{B + 2\frac{d-1}{d}G}{\rho}
    , \nn\\
    \Gamma_\|
    &= \frac{\zeta + 2\frac{d-1}{d}\eta}{\rho} 
    + \frac{n_m^2\sigma}{\rho\chi^2v_\|^2} \nn\\
    &\qquad 
    + \frac{\rho\lb v_\|^2 - \frac{n n_m}{\rho\chi} \rb^2}{\sigma_\phi v_\|^2}
    + \frac{2n_m\gamma_n \lb v_\|^2 - \frac{n n_m}{\rho\chi} \rb}{\chi v_\|^2}, \nn\\
    \Omega_\|
    &= \frac{\lb 1 + n_m\alpha_m/\chi \rb^2 B\Omega_B + 2\frac{d-1}{d}G\Omega_G}{\rho v_\|^2},
\end{align}
while the diffusion constant $D_\|$ and diffusion damping rate $\Omega_D$ are given as
\begin{align}
    D_\|
    &= \frac{B+2\frac{d-1}{d}G}{\rho\chi v_\|^2} \lb \sigma - 2n\gamma_n + \frac{n^2}{\sigma_\phi} \rb, \nn\\
    \Omega_D
    &= \frac{B\Omega_B (n- 2\frac{d-1}{d} G \alpha_m)^2+ 2\frac{d-1}{d} G \Omega_G n_m^2}{\rho v_\|^2 \chi  (B+ 2\frac{d-1}{d}G)}~,
\end{align}
\label{eq:parr-modes}%
\end{subequations}
along with $\chi = \dow n/\dow\mu$ and $n_m = n+B\alpha_m$. We also find a gapped non-hydrodynamic mode 
\begin{equation}
    \omega
    = -i\frac{B\Omega_G+ 2\frac{d-1}{d} G\Omega_B}{B+ 2\frac{d-1}{d} G} + \ldots,
    \label{eq:non-hydro-mode}
\end{equation}
which is absent in an elastic crystal, arising from the fact that we now have an additional degree of freedom $\kappa_\sfT$. Upon turning off the plasticity effects, i.e. $\Omega_B=\Omega_G=0$, both the sound and diffusion modes become undamped, while the last non-hydrodynamic mode disappears altogether. This mode spectrum is a direct generalisation of the mode spectrum reported in~\cite{Armas:2019sbe, Armas:2020bmo} to plastic crystals. Compared to these references, we have one less diffusion mode because we have ignored thermal fluctuations for simplicity.

Passing onto the liquid regime $k\ll\ell \ll L_T^{-1}$, the two branches of the transverse sound mode \eqref{eq:transverse-modes} decouple into the liquid shear mode and a gapped mode
\begin{subequations}
\begin{align}
    \omega 
    &= - i\frac{\eta_l}{\rho} k^2 + \ldots, \nn\\
    \omega 
    &= - i\Omega_G + \ldots,
    \label{eq:modes-transverse-fluid}
\end{align}
while the four longitudinal modes mix to give the liquid sound mode and two other gapped modes
\begin{align}
    \omega 
    &= \pm \sqrt{\frac{n^2}{\rho\chi_l}}\, k 
    - \frac{i}{2} \lb \frac{\zeta_l + 2\frac{d-1}{d}\eta_l}{\rho} 
    + \frac{\sigma }{\chi_l} \rb k^2 + \ldots, \nn\\
    \omega 
    &= -i\Omega_B \lb 1 + \frac{B\alpha_m^2}{\chi} \rb + \ldots, \nn\\
    \omega 
    &= - i\Omega_G + \ldots.
    \label{eq:modes-long-fluid}
\end{align}
Here we have defined the effective ``liquid coefficients'' corrected due to rigidity effects as
\begin{align}
    \chi_l &=  \chi + B\alpha_m^2, \nn\\
    \eta_l &= \eta + \frac{G}{\Omega_G}, \nn\\
    \zeta_l &=  \zeta
    + \lb 1 + \frac{n\alpha_m}{\chi + B\alpha_m^2} \rb^2
    \frac{B}{\Omega_B}.
    \label{eq:fluid-viscosities}
\end{align}
\end{subequations}
These can be compared directly with the modes in a relativistic fluid~\cite{Kovtun:2012rj} or a boost-agnostic fluid~\cite{Armas:2020mpr}. Again, we do not find a charge/energy diffusion mode because we have ignored temperature fluctuations.

We note that we have only focused on modes near the zero velocity equilibrium state $u^i=0$. If the system has some boost symmetry, Galilean or relativistic, all boosted equilibrium states can be related to this one via symmetry transformations. However, for systems without a boost symmetry, the finite-velocity equilibrium states can carry qualitatively new information; see e.g.~\cite{Armas:2020mpr}. We do not analyse this scenario here.

\subsection{Correlation functions}
\label{sec:correlations}

We can use this hydrodynamic framework to compute the retarded correlation functions, also called response functions, of conserved operators and fluxes. To this end, we use the variational approach described in~\cite{Kovtun:2012rj}; see~\cite{Armas:2020mpr} for discussions specific to the boost-agnostic construction. We consider coupling the hydrodynamic equations to background sources: frame velocity $v^i$ for momentum density $\pi_i$, spatial metric $g_{ij}$ for stress tensor $\tau^{ij}$, gauge potential $A_t$ for number density $n$, and gauge field $A_i$ for number flux $j^i$, along with the external stress tensor $T^{ij}_\ext$ already introduced for the strain tensor $\kappa_{ij}$. The precise form of these couplings has been derived in appendix \ref{app:background}. Having done that, schematically, the correlation functions $G^R_{{\cal O}_1{\cal O}_2}$ can be obtained by observing how the operator ${\cal O}_1$ responds to the source of the operator ${\cal O}_2$; see appendix \ref{app:correlation} for the precise formulae.

The following results apply to Galilean crystals upon setting $\rho = n$ along with $\gamma_{n}$, $\sigma$ set to zero. On the other hand, they directly apply to relativistic crystals upon setting $\rho = (\epsilon+p)/c^2$.

For simplicity, we will focus only on the zero-wavevector correlation functions. In this limit, all the number density and momentum density correlators are trivial on account of the conservation laws (Ward identities), while all the number flux and stress tensor correlators become isotropic. Let us start with the flux and crystal velocity: all the non-trivial $\omega$-dependent correlators involving the flux $j^i$ and crystal velocity $u_\phi^i$ are given by
\begin{align}
    G^R_{j^ij^j}(\omega)
    &= \delta^{ij} \lb \frac{n^2}{\rho } - i\omega \sigma \rb, \nn\\
    G^R_{u^i_\phi u^j_\phi}(\omega)
    &= \delta^{ij} \lb \frac{1}{\rho }- \frac{i\omega}{\sigma_\phi}\rb, \nn\\
    G^R_{j^iu^j_\phi}(\omega)
    &= \delta^{ij} \lb \frac{n}{\rho }
    -i\omega \gamma_n \rb~.
    \label{eq:current-correlators}
\end{align}
Interestingly, these results are the same as found for pure elasticity in \cite{Armas:2019sbe, Armas:2020bmo}, meaning that they are not sensitive to the plastic nature of the crystal. This is not really surprising as we already noted that there are no plasticity-dependent dissipative transport coefficients in the vector sector; see \eqref{eq:consti-vector}. 

Notably, the flux-flux correlator is typically the only available observable in experiments, so an important question to consider is whether there are ways to still see the effects of plasticity in this correlator. As it turns out, one can get some mileage by looking at nonzero wavevectors. For example, turning on the wavevector $k_i = k\delta^x_i$ and looking at the transverse sector we find
\begin{align} \label{eq:optkd}
   G^R_{j^yj^y}(\omega,k)  
   &= \frac{n^2}{\rho} -i\omega\sigma
   \nn\\
   & 
   -\frac{G k^2 \left(n/\rho
   - i\omega\gamma_n \right)^2}
   {i \omega  (i\omega - \Omega_G)
   + k^2 \left(v_\perp^2 - i\omega D_\phi^\perp \right)
   }~.
\end{align}
Here we have switched off the viscosity $\eta$ to isolate the plasticity effects on conductivities, which we clearly see appearing with $k^2$ terms. Further turning off $\sigma$ and $\gamma_n$, one can then obtain the expression \eqref{eq:omega-k-conductivity} reported in the introduction, specialised to no-pinning case. 

Next we consider the stress and strain correlators at zero wavevector. As one would expect, these are sensitive to plasticity and we find
\begin{align} \label{eq:correlationfunctions}
    G^R_{\tau^{ij}\tau^{kl}}(\omega)
    &= p \lb 2\delta^{i(k}\delta^{l)j} - \delta^{ij}\delta^{kl} \rb
    + \frac{n^2}{\chi + B\alpha_m^2 } \delta^{ij}\delta^{kl}
    \nn\\
    &\hspace{-2em}
    -i \omega \delta^{ij}\delta^{kl} 
     \left(\zeta
     + \frac{\lb 1+\frac{B\alpha_m^2}{\chi}\rb
     \lb 1 + \frac{n\alpha_m}{\chi+B \alpha_m^2}\rb^2 B}
     {\lb 1+\frac{B\alpha_m^2}{\chi}\rb \Omega_B -i \omega } \right) \nn\\
    &\hspace{-2em}
    - 2i\omega  \delta^{i\langle k}\delta^{l\rangle j} 
    \lb \eta +\frac{G}{\Omega_G -i \omega } \rb,
    \nn\\[0.5em]
    G^R_{\kappa^{ij}\kappa^{kl}}(\omega)
    &= - \delta^{ij}\delta^{kl}  \frac{1}{4B}
    \frac{\Omega_B}
    {\lb 1+\frac{B\alpha_m^2}{\chi}\rb \Omega_B -i \omega }
    \nn\\
    &\qquad 
    - 2\delta^{i\langle k}\delta^{l\rangle j}
    \frac{1}{4G}\frac{\Omega_G}{\Omega_G-i\omega}, \nn\\[0.5em]
    G^R_{\tau^{ij}\kappa^{kl}}(\omega)
    &= \half \delta^{ij}\delta^{kl} \frac{
    \frac{n\alpha_m}{\chi } \Omega_B + i\omega}
    {\lb 1+\frac{B\alpha_m^2}{\chi}\rb \Omega_B -i \omega } \nn\\
    &\qquad 
    + \delta^{i\langle k}\delta^{l\rangle j}\frac{i \omega}{\Omega_G-i \omega}~.
\end{align}
The imaginary part of the stress-stress correlator can be used to obtain the $\omega$-dependent viscosities, i.e.\footnote{Our result for $\zeta(\omega)$ differs from~\cite{Delacretaz:2017zxd} in its functional form. The difference can be traced back to the fact that the hydrodynamic theory presented in~\cite{Delacretaz:2017zxd} only works with two components of the strain tensor in $d=2$. In the presence of plasticity effects arising from crystal dislocations, all three components of the symmetric strain tensor can evolve independently, and couple non-trivially in the hydrodynamic equations.}
\begin{align} \label{eq:viscocorrelators}
    \zeta(\omega)
    &=
    - \frac{1}{\omega}
    \mathrm{Im}\bigg( G^R_{\tau^{xx}\tau^{xx}}(\omega)
    - 2\frac{d-1}{d} G^R_{\tau^{xy}\tau^{xy}}(\omega) \bigg) \nn\\
    &= \zeta
     + \frac{\lb 1+\frac{B\alpha_m^2}{\chi}\rb^2
     \lb 1 + \frac{n\alpha_m}{\chi+B \alpha_m^2}\rb^2 B\Omega_B}
     {\lb 1+\frac{B\alpha_m^2}{\chi}\rb^2 \Omega_B^2 + \omega^2} \nn\\
    \eta(\omega)
    &=
    - \frac{1}{\omega}
    \mathrm{Im}\,G^R_{\tau^{xy}\tau^{xy}}(\omega) \nn\\
    &= \eta + \frac{G \Omega_G}{\Omega_G^2 + \omega^2} ~.     
\end{align}
In the solid regime $\omega \gg \Omega_B,\Omega_G$, these objects reduce to the crystal viscosities $\zeta$, $\eta$. Whereas, in the liquid regime $\omega \ll \Omega_B, \Omega_G$, these give rise to the effective liquid viscosities $\zeta_l$, $\eta_l$ defined in \eqref{eq:fluid-viscosities}.

\section{Plastic deformations from lattice dislocations}
\label{sec:dislocations}

In this section we discuss how dislocations in a crystal give rise to plasticity. We started our effective description of crystals in section \ref{eq:non-rel-plasticity} with the crystal fields $\phi^I$. These are arbitrary labels that we assigned to each lattice site, however, these labels might not correspond to how the sites are actually organised in the crystal. In fact, if the crystal has topological defects, it might not be possible at all to assign a smooth labelling $\phi^I$ throughout the crystal. For this purpose, let us introduce another crystal field $\tilde\varphi^I$ that does correspond to the actual structure of the crystal. As a downside, $\tilde\varphi^I$ is no longer smooth and cannot be directly used in the effective field theory description. For simplicity, we will assume that the associated frame fields $\tilde e^I_i = \dow_i\tilde\varphi^I$ and $\tilde e^I_t = \dow_t\tilde\varphi^I$ are smooth, which means that the crystal only has translational defects called \emph{dislocations}, but no rotational defects called \emph{disclinations}.

Even though the frame fields $\tilde e_i^I$ are smooth, they are no longer curl-free, i.e.
\begin{equation}
    \epsilon^{ij}\dow_i \tilde e^I_j = \ell n^I_\dloc \neq 0~.
    \label{eq:dloc-density}
\end{equation}
The object $n^I_\dloc$ is known as the \emph{dislocation density} and the small parameter $\ell$ controls the strength of dislocations. We have specialised to $d=2$ spatial dimensions for simplicity, though the notion of dislocations can also be generalised to higher-dimensions; see e.g.~\cite{Beekman:2017brx}. We can integrate the dislocation density in some volume $U$ to define the \emph{Burgers vector}
\begin{align}
    B^I 
    &= \int_U \df^2 x\, \ell n^I_\dloc \nn\\
    &= \oint_{\dow U} \df x^i \dow_i \tilde\varphi^I~.
\end{align}
It measures the displacement, in lattice units, as we circle a loop around $U$. If the region $U$ contains no defects, the net displacement is zero. However, when $U$ contains a dislocation, going around the loop we pick up a net displacement. Dislocations are topologically conserved and satisfy the conservation law
\begin{equation}
    \dow_t n^I_\dloc + \dow_i j^I_\dloc = 0~,
\end{equation}
where the dislocation flux is defined as
\begin{equation}
    \ell j^{Ii}_\dloc = 
    - \epsilon^{ij}\dow_t\tilde e^I_j
    + \epsilon^{ij}\dow_j\tilde e^I_t~~,
    \label{eq:dloc-flux}
\end{equation}
which follows from the definition of dislocation density in \eqref{eq:dloc-density}.
This means that once excited, dislocations cannot decay locally and must interact with anti-dislocations to annihilate. Furthermore, creating dislocation/anti-dislocation pairs only requires restructuring the bond-structure of a crystal and can be achieved with comparatively little energy than the creation of interstitials/vacancies that require displacement of lattice sites. Thus dislocations can be found even in the purest of crystals. See~\cite{Beekman:2016szb, 2006PMag...86.2995C} for a more detailed discussion.

We wish to see how dislocations in a crystal give rise to plasticity. For this purpose, we decompose the frame fields $\tilde e^I_i = \dow_i\tilde\varphi^I$ into an ``undefected'' part expressed as the gradient of a smooth crystal field $\phi^I$ and a part arising from dislocations
\begin{align}
    \dow_i \tilde\varphi^I 
    &= \dow_i \phi^I + \ell V^I_i~, \nn\\
    \dow_t \tilde\varphi^I 
    &= \dow_t \phi^I + \ell V^I_t~.
    \label{eq:phi-V-decomposition}
\end{align}
It is trivial to see that the $\phi^I$ part does not contribute to the dislocation density. We can think of $\phi^I$, $V^I_i$, and $V^I_t$ as independent degrees of freedom describing a crystal with dislocations. However, the decomposition in \eqref{eq:phi-V-decomposition} is not unique and must be accompanied by a local ${\rm U}(1)^d$ gauge symmetry, with $V^I_i$, $V^I_t$ acting as the respective gauge fields
\begin{align}
    \phi^I 
    &\to \phi^I - \ell \lambda^I~, \nn\\
    V^I_i 
    &\to V^I_i + \dow_i\lambda^I~, \nn\\
    V^I_t
    &\to V^I_t + \dow_t\lambda^I~.
\end{align}
We can partially gauge-fix this symmetry by demanding that the crystal velocity $u^i_\phi$ defined using the $\phi^I$ fields in \eqref{eq:crystal-velocity}, also defines the local rest frame of the $\tilde\varphi^I$ fields, i.e.
\begin{equation}
    \dow_t\tilde\varphi^I + u^i_\phi \dow_i\tilde\varphi^I = 0~.
\end{equation}
This can be seen as determining $V^I_t$ in terms of the other fields. Note that  this is not a complete gauge-fixing. We are still left with the residual $\Diff(\phi)$ symmetry in the crystal fields that leaves $u^i_\phi$ invariant. This can be understood as the origin of the $\Diff(\phi)$ symmetry introduced in \eqref{sec:elastic-plastic} in the context of dislocations.

Since the crystal fields $\tilde\varphi^I$ are allowed to be singular, we can always choose them such that the reference metric of the crystal is just $\delta_{IJ}$. This allows us to define the strain tensor of the crystal
\begin{equation}
    \tilde\kappa_{IJ} = \half \lb \tilde h_{IJ} - \delta_{IJ} \rb~,
\end{equation}
where $\tilde h_{IJ}$ is the inverse of $\tilde h^{IJ} = \tilde e^{Ii}\tilde e^J_j$. Given that $\phi^I$ and $\tilde\varphi^I$ describe the same physical crystal, the pullbacks of the respective strain tensors to the physical space must be identical
\begin{equation}
    \kappa_{ij} = e^I_ie^J_j \kappa_{IJ}
    = \tilde e^I_i \tilde e^J_j \tilde\kappa_{IJ} ~.
\end{equation}
This gives us a non-linear relation between the plastic part of the reference metric $\psi_{IJ}$ and the dislocation part of the frame fields $V^I_i$, i.e.
\begin{equation}
    e^I_ie^J_j \psi_{IJ}
    = 2\delta_{IJ} e^I_{(i} V^J_{j)} + \ell^2 \delta_{IJ} V^I_i V^J_j~.
    \label{eq:plasticity-dislocation-map}
\end{equation}
We have therefore established that a crystal with dislocations gives rise to plasticity.

It is instructive to do a counting exercise at this point. Note that $V^I_i$ has $d^2$ degrees of freedom in $d$ spatial dimensions as opposed to $\psi_{IJ}$, which only has $d(d+1)/2$ components. It suggests that dislocations do a bit more than just making the crystal plastic. Normally, in an undefected crystal, the bond-angles among the lattice sites are entirely determined by the respective bond-lengths. This, however, is no longer the case in the vicinity of dislocations and the crystal can carry independent local bond-angles in addition to the reference metric. This accounts for the missing $d(d-1)/2$ degree of freedom. If we assume that the dislocation density is weak enough so that the orientations of nearby dislocations is effectively uncorrelated, we can ignore the angular degrees of freedom. In this regime, the effective dynamics of a crystal with dislocations is exactly captured by the theory of plasticity constructed above.

\subsection*{Glide constraint}

\begin{figure}[t]
\center
\includegraphics[width=7cm]{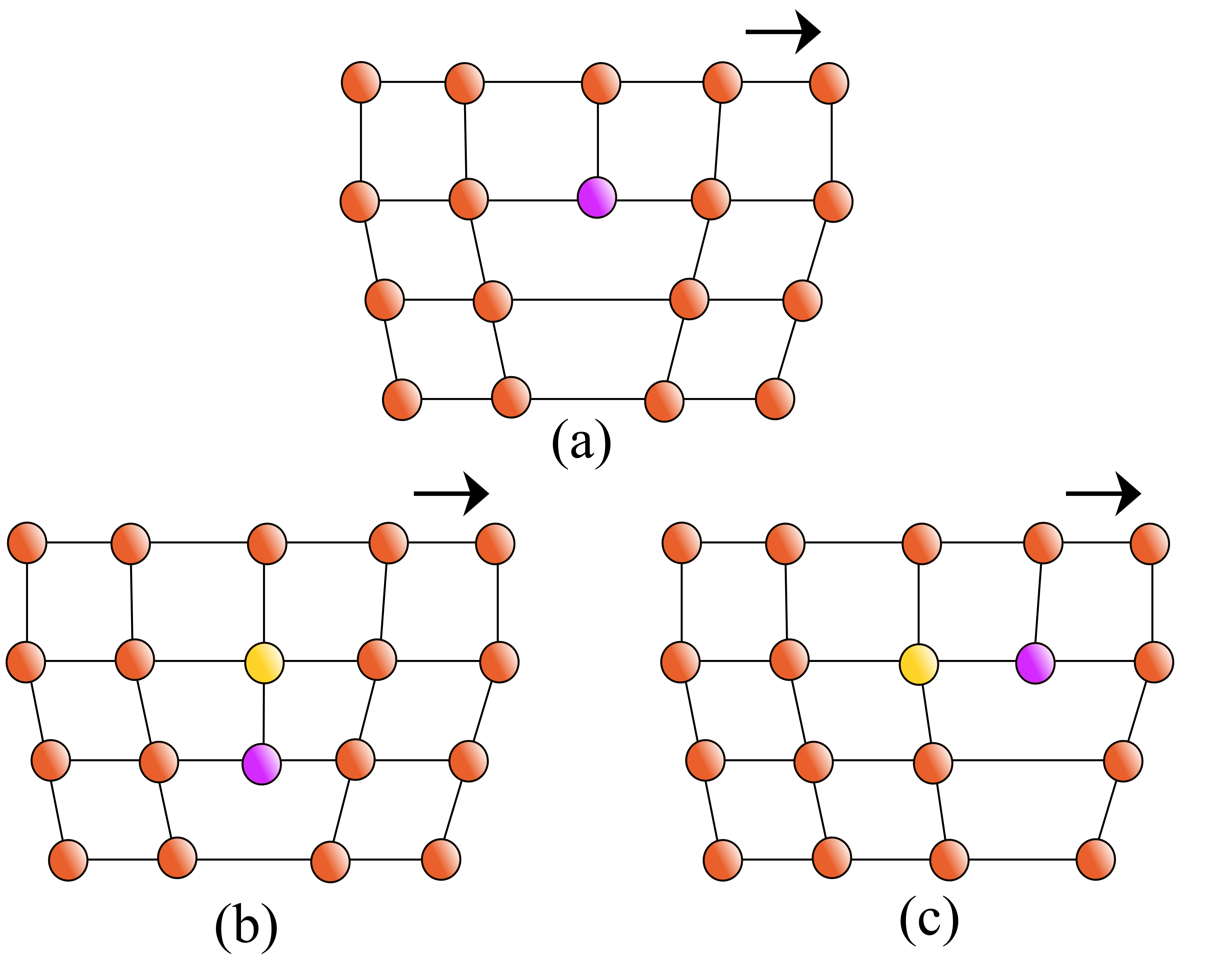}
\caption{Climb motion of a dislocation perpendicular to its Burgers vector (given in black) takes the lattice configuration (a) to (b), creating a new lattice site in the process. On the other hand, glide motion of a dislocation parallel to its Burgers vector takes the lattice configuration (a) to (c), without the need to create or remove a lattice site.}
\label{fig:dloc-glide}
\end{figure}

The motion of dislocations can be categorised into either ``glide'' (movement parallel to the Burgers vector) or ``climb'' (movement perpendicular to the Burgers vector). Glide motion merely requires a reconfiguration of the bond structure of the lattice, while climb motion requires creation or removal of lattice sites and is thus energetically much costlier; see figure \ref{fig:dloc-glide}.
Creation or removal of lattice sites result in permanent expansion or compression of the local volume element of a material and is hard to achieve except in extreme cases, such as during the process of densification; see e.g. \cite{densification}. As such, for most practical purposes, climb motion of dislocations can be neglected, resulting in the so-called \emph{glide constraint}.

Since climb motion acts as a local source or sink for lattice sites, the glide constraint can simply be formulated as the statement that the interstitial density (total particle density minus the density of lattice sites) given in \eqref{eq:interstitial-stuff} is conserved.  Due to~\eqref{eq:interstitial-conservation}, this is equivalent to the statement that the local volume element associated with the reference metric of the crystal remains fixed in the rest frame of the crystal
\begin{subequations}
\begin{equation}
    (\dow_t + u^k_\phi\dow_k) \det\bbh=0 ~~.
    \label{eq:glide-constraint-plastic}
\end{equation}
Using relation \eqref{eq:plasticity-dislocation-map} between plasticity and dislocations, together with the definitions of dislocation density \eqref{eq:dloc-density} and flux \eqref{eq:dloc-flux}, we can obtain the equivalent statement 
\begin{gather}
    \tilde e_I^i \epsilon_{ik} \lb j^{Ik}_\dloc
     - n^I_\dloc u^k_\phi \rb
     = 0,
\end{gather}
\end{subequations}
which implies that there is no flux of dislocations with respect to the crystal velocity perpendicular to the local Burgers vector. Imposing the glide constraint \eqref{eq:glide-constraint-plastic} on the evolution equation of the reference metric~\eqref{eq:bbh-equation}, we find $\lambda_B=1$ and $\Omega_B=0$, or equivalently $\zeta_\bbh \to\infty$. We have already set $\lambda_B=1$ following our discussion in section \ref{eq:strain-redef} using the redefinition freedom of the reference metric. However, we have kept $\Omega_B\ne0$ throughout the paper for generality.


\section{Pinned plastic crystals}
\label{sec:pinning}

In this section, we consider viscoplastic crystals with pinning due to possible point-like impurities and inhomogeneities in the lattice structure, thereby breaking the translational invariance of the theory. This is particularly relevant for electronic crystals because the background lattice of ions serves as a natural source of pinning in these phases. Notably, just like plasticity, pinning also causes the strain tensor of a crystal to relax. Considering both pinning and plasticity in the same hydrodynamic framework will also allow us to clearly distinguish between the physical signatures of the two phenomena.

\subsection{Pinned viscoplastic hydrodynamics}
\label{sec:pinned-consti}

In a recent paper~\cite{Armas:2021vku}, we formulated the hydrodynamic theory of viscoelastic crystals with pinning. The main ingredient of this construction was background crystal fields $\Phi^I(x)$ that explicitly break the spatial translational symmetry of the crystal. In the presence of these fields, the equation of state of the crystal \eqref{eq:linearised-p} additionally contains a mass term dependent on the difference between the dynamical and background crystal fields
\begin{equation}
    p \sim -\half \ell'^2 m^2 h_{IJ} \lb \phi^I - \Phi^I\rb \lb \phi^J - \Phi^J \rb~.
    \label{eq:pinning-pressure}
\end{equation}
Here we have introduced another small parameter $\ell'$ to control the strength of pinning. Just like the plasticity parameter $\ell$, we choose $\ell'$ to also scale as ${\cal O}(\dow)$ to keep the pinning effects weak.

The consequence of pinning for the hydrodynamic equations is that energy and momentum can now also be sourced by the background crystal fields $\Phi^I$, leading to a modification of the conservation laws \eqref{eq:conservation} to 
\begin{align}
    \dow_t\epsilon + \partial_i\epsilon^i
    &= 
    - K_{I } \dow_t{\phi}^I
    - \half U^{IJ}(\dow_t + \bar u^k \dow_k)\psi_{IJ} \nn\\
    &\qquad 
    - \ell' L_I \dow_t\Phi^I
    , \nn\\ 
    \dow_t\pi^i + \partial_j \tau^{ij }  
    &= K_{I } \partial^i  \phi^I
    + \ell' L_I \dow^i\Phi^I , \nn\\ 
    \dow_t n + \partial_i j^{i}   &  =0 ~~ . 
    \label{eq:conservation-pinning}
\end{align}
See appendix \ref{app:background} for the derivation.
Here $L_I$ is a new operator dual to $\Phi^I$, which has to be fixed using the second law of thermodynamics. We have introduced the parameter $\ell'$ in these equations so that the pinning-induced modifications vanish when we tune $\ell'\to 0$. Note that the particle continuity equation does not get affected by pinning. The schematic form of the Josephson equations also remains the same as \eqref{eq:josehpson-together}, however the constitutive relations for the operators $K_I$ and $U^{IJ}$ can now contain new terms due to pinning.

To implement the constraints coming from the second law of thermodynamics, let us take the same parametrisation for the constitutive relations as \eqref{eq:ideal-consti} except
\begin{align}
    K_I 
    &= -\partial_{i}\!\lb r_{IJ} e^{Ji}\rb
    + \frac{\ell}{2} \bbr^{JK} e^i_I \dow_i\psi_{JK} \nn\\
    &\qquad 
    - \ell'^2 m^2 h_{IJ}\lb\phi^I-\Phi^I\rb
    + \mathcal{K}_I, \nn\\
    L_I 
    &= \ell' m^2 h_{IJ}\lb\phi^I-\Phi^I\rb
    + {\cal L}_I~.
    \label{eq:consti-pinned}
\end{align}
The pinning contributions to these operators are the same as found in~\cite{Armas:2021vku} for pinned elastic crystals.
Proceeding with the second law calculation as before, we find that
the entropy production rate \eqref{eq:dissipationrate} receives an additional contribution from pinning 
\begin{align}
    T\Delta 
    &=
    - \frac{1}{T}  \mathcal{E}^i \partial_i T  
    - \mathcal{T}^{ij}  \partial_i u_j  
    - T{\cal J}^i\dow_i\frac{\mu}{T}
    \nn\\ 
    &\qquad
    - \mathcal{K}_I  (\dow_t + u^i\dow_i) \phi^I 
    - \half\mathcal{U}^{IJ} (\dow_t + u^i_\phi \dow_i) \psi_{IJ} \nn\\
    &\qquad 
    - \ell'\mathcal{L}_I  (\dow_t + u^i\dow_i) \Phi^I 
    \geq 0.
    \label{eq:entropy-production-pinning}
\end{align}
The details of this derivation in the presence of background sources can be found in appendix \ref{app:background}.

The one-derivative corrections to the constitutive relations allowed by the second law can be found similarly to section \ref{sec:constitutive}. The tensor sector \eqref{eq:consti-scalar-tensor} remains unchanged, while the vector sector receives an new row and column 
\begin{equation}
    \begin{pmatrix}
        \mathcal{E}^I \\
        \mathcal{J}^I  \\
        \mathcal{K}^I \\
        \mathcal{L}^I
    \end{pmatrix}    
    = -
    \begin{pmatrix}
        \sigma_\epsilon & \gamma_{\epsilon n}  &  \gamma_{\epsilon\phi} & \gamma_{\epsilon\Phi}  \\
        \gamma'_{\epsilon n} & \sigma_n & \gamma_{n\phi} & \gamma_{n\Phi}  \\
        \gamma'_{\epsilon\phi}  & \gamma'_{n\phi}   & \sigma_\phi & \sigma_\times  \\
        \gamma'_{\epsilon\Phi} & \gamma'_{n\Phi} & \sigma'_\times & \sigma_\Phi
    \end{pmatrix} \begin{pmatrix}
        e^I_i \frac{1}{T} \partial^i T   \\ 
        e^I_i T\dow^i\frac{\mu}{T}   \\ 
        (\dow_t + u^i\dow_i)\phi^I \\
        \ell'(\dow_t + u^i\dow_i)\Phi^I
    \end{pmatrix}.
    \label{eq:consti-vector-pinning}
\end{equation}
Onsager's relations imply that the new off-diagonal terms are related as
\begin{subequations}
\begin{equation}
    \gamma'_{\epsilon\Phi} = - \gamma_{\epsilon\Phi}, \qquad 
    \gamma'_{n\Phi} = - \gamma_{n\Phi}, \qquad 
    \sigma'_\times = \sigma_\times~,
    \label{eq:onsager-pinning} 
\end{equation}
whereas the positivity of entropy production leads to the sign constraint
\begin{equation}
    \sigma_\Phi\sigma_\phi \geq \sigma_\times^2~.
\end{equation}
\end{subequations}

\subsection{Linearised hydrodynamic equations}

Pinning leads to some qualitative changes in the dynamical properties of the crystal. Notably, the Josephson equation for $u^i_\phi$ modifies from \eqref{eq:josephson_phi_linear} to
\begin{align}
    u^i_\phi
    &= \lambda u^i 
    + D_\phi^\| \dow^i\kappa^k_{~k} 
    + 2D_\phi^\perp 
    \lb  \dow_j\kappa^{ij} {\,-\,} \dow^i\kappa^k_{~k}  \rb  \nn\\
    &\qquad 
    - \lb \gamma_n + \frac{B\alpha_m}{\sigma_\phi} \rb \dow^i\mu
    - \Omega_\phi \delta\phi^i
    ~,
\end{align}
where we have defined
\begin{equation}
    \Omega_\phi = \frac{\ell'^2m^2}{\sigma_\phi}, \qquad 
    \lambda = 1 + \frac{\ell'\sigma_\times}{\sigma_\phi}~.
\end{equation}
The coefficient $\lambda$ modifies the leading order Josephson relation and controls the screening (or enhancement) of the speed of sound due to impurities; see \cite{Armas:2021vku}. We note that, unlike its analogue plastic coefficients $\lambda_B$, $\lambda_G$, the coefficient $\lambda$ is physical and cannot be rescaled away from the theory; see our discussion at the end of section \ref{eq:strain-redef} or the appendix of~\cite{Armas:2021vku}. More importantly, there is a relaxation term in the Josephson equation, which was absent in the case of pure plasticity. The physical consequence of this is that while previously only the crystal strain underwent relaxation, see \eqref{eq:distortionstrain}, the distortion strain now also relaxes
\begin{align} 
    \dot\varepsilon_{ij}
    &= \lambda\dow_{(i} u_{j)}
    - \lb \gamma_n + \frac{B\alpha_m}{\sigma_\phi} \rb  \dow_i\dow_j\mu 
    + D_\phi^\| \dow_i\dow_j \kappa^k_{~k} \nn\\
    &\qquad 
    + 2D_\phi^\perp \lb \dow_{(i}\dow^k \kappa_{j)k} 
    - \dow_i\dow_{j} \kappa^k_{~k}\rb 
    - \Omega_\phi \varepsilon_{ij}~.
\end{align}

The equation \eqref{eq:bbh-equation} for the evolution of the reference metric $\bbh_{IJ}$ is unaffected by pinning because there are no pinning induced transport coefficients in the tensor sector. However, the evolution of the crystal strain tensor does get pinning contributions coming from the distortion strain and \eqref{eq:strain-evolution-linear} modifies to
\begin{align}   
    \dot\kappa_{ij}
    &= \frac{1}{d} \lambda'_B\delta_{ij} \dow_k u^k
    + \lambda'_G \dow_{\langle i}u_{j\rangle}
    - \lb \gamma_n + \frac{B\alpha_m}{\sigma_\phi} \rb \dow_i\dow_j\mu  \nn\\
    &\quad 
    + D_\phi^\| \dow_i\dow_j\kappa^k_{~k}
    + 2D_\phi^\perp \lb \dow_{(i}\dow^k\kappa_{j)k} 
    - \dow_i\dow_{j}\kappa^k_{~k}\rb \nn\\
    &\quad
    - \frac{1}{d} \delta_{ij} \Omega_B \lb \kappa^k_{~k} - \alpha_m \delta\mu \rb
    - \Omega_G \kappa_{\langle ij\rangle}
    - \Omega_\phi \varepsilon_{ij},
\end{align}
where we have defined the new pinning-induced versions of the coefficients
\begin{gather}
    \lambda'_B = 1
    + \frac{\ell\eta_{\tau\bbh}}{\eta_{\bbh}}
    + \frac{\ell'\sigma_\times}{\sigma_\phi}, \quad 
    \lambda'_G = 1
    + \frac{\ell\zeta_{\tau\bbh}}{\zeta_{\bbh}}
    + \frac{\ell'\sigma_\times}{\sigma_\phi},
    \label{eq:pinning-defs-kappa}
\end{gather}
We have kept $\lambda_B$, $\lambda_G$ non-unity here for generality; if were to implement the discussion from section \ref{eq:strain-redef} and set $\lambda_B=\lambda_G=1$, we would have $\lambda'_B=\lambda'_G = \lambda$.

The expression for the particle flux modifies compared to \eqref{eq:stress-flux-linear} due to pinning
\begin{align}
    j^i
    &= (n+\lambda_n)\, u^i 
    + \Omega_n \delta\phi^i
    - \lb \sigma + B\alpha_m\gamma_n \rb \dow^i\mu \nn\\
    &\qquad 
    - n D^\|_{n} \dow^i \kappa^k_{~k} 
    - 2 n D_n^\perp 
    \lb  \dow_j u^{ij} {\,-\,} \dow^i \kappa^k_{~k}  \rb~,
\end{align}
where 
\begin{equation}
    \lambda_n = -\ell'
    \lb \gamma_{n\Phi} 
    - \gamma_{n\phi}\frac{\sigma_\times}{\sigma_\phi} \rb, \quad 
    \Omega_n = -\ell'^2m^2\gamma_{n\phi}~,
\end{equation}
are new coefficients related to the screening of particle flux and pinning due to impurities. The pinning corrections to interstitial flux can be obtained from here by combining $j^i$ and $u^i_\phi$. The expression for the stress tensor in \eqref{eq:stress-flux-linear} does not change due to the absence of any plasticity-induced transport in the tensor sector. However, due to broken translations, momentum is now sourced by 
\begin{align}
    \ell L^i
    &= - \frac{\rho}{\lambda} \omega_0^2\delta\phi^i
    - \Gamma \pi^i
    + \lb \lambda_n
    - \ell\sigma_\times \frac{B\alpha_m}{\sigma_\phi} \rb \dow^i\mu
    \nn\\
    &\quad 
    + \ell'\sigma_\times D_\phi^\| \dow^i u^k_{~k} 
    + 2\ell'\sigma_\times D_\phi^\perp 
    \lb  \dow_j u^{ij} {\,-\,} \dow^i u^k_{~k}  \rb~,
\end{align}
where 
\begin{equation}
    \omega_0^2 = \frac{\lambda^2\ell'^2 m^2}{\rho}, \qquad 
    \Gamma = \frac{\ell'^2}{\rho} 
    \lb \sigma_\Phi - \frac{\sigma_\times^2}{\sigma_\phi} \rb~,
\end{equation}
are the pinning frequency and the rate of momentum relaxation respectively. 

The pinning-induced phase relaxation rate $\Omega_\phi$, crystal conductivity $\sigma_\phi$, and pinning frequency $\omega_0$ are related via the damping-attenuation relation~\cite{Armas:2021vku, Delacretaz:2021qqu}
\begin{subequations}
\begin{equation}
    \Omega_\phi 
    = \frac{\rho\, \omega_0^2}{\lambda^2\sigma_\phi}~~.
\end{equation}
Using the speed of transverse sound $v_\perp^2 = \lambda^2G/\rho$ and the elastic contribution to the speed of longitudinal sound $v_{\|\phi}^2 = \lambda^2(B+2\frac{d-1}{d}G)/\rho$ (see section \ref{sec:pinned-modes}), the same relation can also be recast as
\begin{equation}
    \Omega_\phi 
    = D_\phi^\perp \omega_0^2/v_\perp^2 
    = D_\phi^\| \omega_0^2/v_{\|\phi}^2~~.
\end{equation}
\label{eq:damping-attentuation}%
\end{subequations}
In particular, this tells us that the phase or distortion strain in a pinned crystal cannot relax without diffusing or cannot diffuse without relaxing. Note that no such relation applies for the plasticity-induced strain relaxation rates $\Omega_B$, $\Omega_G$.

\subsection{Rheology equations}
\label{sec:rheology-pinning}

Pinning results in modifications of the rheology of the materials under consideration. Following the same procedure as section \ref{sec:rheology}, we can obtain the rheology equations for pinned plastic crystals
\begin{subequations} \label{eq:rheocrystal-pinned}
    \begin{align}
    \tau_{ij} 
    &= - \delta_{ij} \lb B \kappa^k_{~k} 
    + \frac{\zeta}{\lambda} \lb \dot\varepsilon^k_{~k}
    + \Omega_\phi\, \varepsilon^k_{~k}\rb  \rb
    \nn\\
    &\qquad 
    - 2G \kappa_{\langle ij\rangle}
    - \frac{2\eta}{\lambda} 
    \lb \dot\varepsilon_{\langle ij\rangle} 
    + \Omega_\phi\,\varepsilon_{\langle ij\rangle}\rb,  \\
    \dot\kappa_{ij}
    &= \frac{1}{d} \delta_ {ij}\lb \dot\varepsilon^k_{~k}
    - \Omega_B \kappa^k_{~k} \rb \nn\\
    &\qquad 
    + \dot\varepsilon_{\langle ij\rangle} 
    - \Omega_G \kappa_{\langle ij\rangle}~~.
\end{align}
\end{subequations}
The relation between the distortion strain and the crystal strain is only sensitive to plasticity and does not receive any corrections due to pinning. However, we do generate some pinning-dependence in the stress tensor after eliminating $\dow_{(i}u_{j)}$ in favor of $\varepsilon_{ij}$. With our choice of order counting $\ell'\sim{\cal O}(\dow)$, these new terms are technically suppressed and their effects on the rheology of the material will only become important when the strength of pinning is increased. However, given their qualitatively distinct nature, it is interesting to consider the effects of these pinning-induced phenomena on the material diagrams. 

\begin{figure}[t]
    {\includegraphics[width=\linewidth]{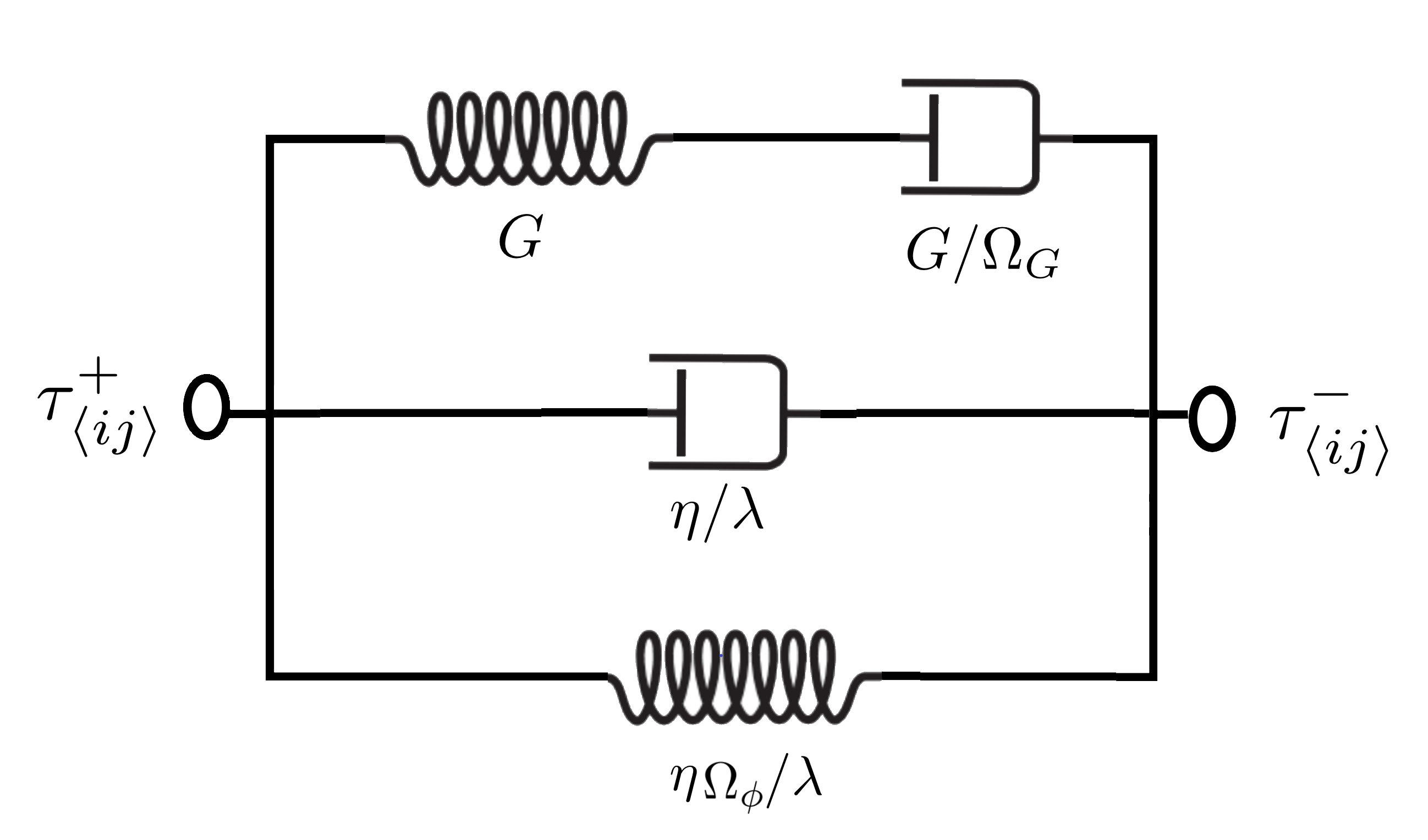}}
     \caption{Circuit representation of the rheology equations for a pinned plastic crystal in the shear sector. Pinning introduces a weak spring component parallel to the Jeffrey circuit, causing the material to return to its original state at very late time scales similar to Zener materials. The bulk sector behaves similarly, with $G$, $\Omega_G$, $\eta$ replaced with $B$, $\Omega_B$, $\zeta$ respectively.
     \label{fig:zener}}
\end{figure}

Eliminating $\kappa_{ij}$, we can write down the stress-strain relations associated with \eqref{eq:rheocrystal-pinned}, i.e.
\begin{align} \label{eq:zener}
    \dot\tau^k_{~k} + \Omega_B\tau^k_{~k} 
    &= - d\lb B + \frac{\zeta}{\lambda} (\Omega_B+\Omega_\phi) \rb
    \dot\varepsilon^k_{~k}  
    - \frac{d\zeta}{\lambda} \ddot\varepsilon^k_{~k} \nn\\
    &\qquad 
    - \frac{d\zeta}{\lambda}\Omega_\phi\Omega_B\varepsilon^k_{~k}, \nn\\
    \dot\tau_{\langle ij\rangle} + \Omega_G\tau_{\langle ij\rangle}
    &= - 2\lb G + \frac{\eta}{\lambda} (\Omega_G+\Omega_\phi) \rb
    \dot\varepsilon_{\langle ij\rangle}
    - \frac{2\eta}{\lambda} \ddot\varepsilon_{\langle ij\rangle} \nn\\
    &\qquad 
    - \frac{2\eta}{\lambda}\Omega_\phi\Omega_G\varepsilon_{\langle ij\rangle}.
\end{align}
The respective last terms in these relations are new compared to the pure plasticity case in \eqref{eq:Jeffrey} and result in the system eventually returning to its original state at very late time scales, similar to a Zener material \cite{banks_hu_kenz_2011}; see the material diagram in figure \ref{fig:zener}. Note that the last term in both the sectors in \eqref{eq:zener} combines viscosities, plasticity-induced relaxation, and pinning-induced relaxation, and is truly a concoction of the three physical phenomena.

\subsection{Mode spectrum} 
\label{sec:pinned-modes}

We can use our hydrodynamic framework to compute the mode spectrum for a plastic crystal in the presence of pinning. We will keep the scale of pinning comparable to the wave-vector scale $\ell'\sim k$ throughout this discussion, although other scaling regimes could also be physically interesting and can be explored using our hydrodynamic theory. In the solid regime $\ell \ll k \ll L_T^{-1}$, we find that the already damped transverse sound mode is now pinned at the frequency $\omega_0$, i.e.
\begin{subequations}\label{eq:pinning-transverse-mode}
\begin{align}
    \omega
    &= \pm\sqrt{\omega_0^2 + v_\perp^2 k^2} 
    - \frac{i}{2} \Big( \Gamma_\perp k^2
    + \Omega_\perp(k) + \Gamma \Big)~.
\end{align}
The speed of sound and damping rate are modified to
\begin{align}
    v_\perp^2 &= \frac{\lambda^2 G}{\rho}, \nn\\
    \Omega_\perp(k)
    &= \frac{v_\perp^2 k^2}{\omega_0^2 + v_\perp^2 k^2} \Omega_G
    + \Omega_\phi~,
\end{align}
\end{subequations}
while the definitions of the attenuation rate $\Gamma_\perp$ remains the same as quoted in \eqref{eq:transverse-modes}. Note that we have allowed some $k$-dependence in the damping rate. We see that the damping of the sound mode is now dominated by the pinning-induced momentum relaxation $\Gamma$ and Goldstone phase relaxation $\Omega_\phi$. In this sense, pinning ``washes away'' the damping effects arising from the plasticity-induced relaxation $\Omega_G$ at small $k$. In the longitudinal sector, we instead find the pinned and damped sound mode and a damped crystal diffusion mode 
\begin{subequations}\label{eq:pinning-long-mode}
\begin{align}
    \omega 
    &= \pm \sqrt{\omega_0^2 + v_\|^2 k^2}
    - \frac{i}{2} \left(
    \Gamma_\|(k) k^2 
    + \Omega_\|(k)
    + \Gamma  \right), \nn\\
    \omega
    &= -iD_\| k^2 
    - i\Omega_D,
\end{align}
where the modified speed of sound, attenuation, and damping rates are given as
\begin{align}
    v_\|^2
    &= \frac{(n_m + \lambda_n)^2}{\rho\chi}
    + \frac{\lambda^2(B+G)}{\rho }, \nn\\
    \Gamma_\|(k)
    &= \frac{\zeta+2\frac{d-1}{d}\eta}{\rho} 
    + \frac{n_m^2\sigma k^2}{\rho  \chi ^2 
    (\omega_0^2+v_\|^2k^2)}
    + \frac{\rho\lb v_\|^2 - \frac{n n_m}{\rho\chi} \rb^2}
    {\sigma_\phi v_\|^2}
    \nn\\
    &\qquad 
    + \frac{2n_m\gamma_n}{\chi }
    \left(1-\frac{n n_m k^2}{\rho\chi
    (\omega_0^2+v_\|^2k^2)}\right), \nn\\
    \Omega_\|(k)
    &=
    \frac{v_\|^2 k^2}{\omega_0^2+v_\|^2k^2}
    \frac{\lb 1 + n_m\alpha_m/\chi \rb^2 B\Omega_B + 2\frac{d-1}{d}G\Omega_G}{\rho v_\|^2} \nn\\
    &\qquad 
    - \frac{n^2 n_m^2\Omega_\phi}{(\omega_0^2+v_\|^2k^2) 
    \chi^2\rho^2 v_\|^2}
    + \Omega_\phi,
\end{align}
where $n_m = n+\lambda B \alpha_m$.
The diffusion rate and the associated damping rate also modify to $k$-dependent expressions
\begin{align}
    D_\|(k)
    &= \frac{\sigma - 2n\gamma_n  + \frac{n^2}{\sigma_\phi}}{\chi}
    \lb 1 - \frac{n_m^2k^2}{\rho\chi(\omega_0^2+v_\|^2k^2)} \rb, \nn\\ 
    \Omega_D(k)
    &= 
    \frac{B\Omega_B \lb \rho\alpha_m\omega_0^2 - \lb n- 2\frac{d-1}{d} G\alpha_m\rb k^2 \rb^2}
    {\rho\chi\lb \omega_0^2 + \lb B+ 2\frac{d-1}{d} G \rb k^2\rb (\omega_0^2 + v_\|^2 k^2)} \nn\\
    &
    + \frac{2\frac{d-1}{d} G\Omega_G n_m^2 k^4}
    {\rho\chi\lb \omega_0^2 + \lb B+ 2\frac{d-1}{d} G \rb k^2\rb (\omega_0^2 + v_\|^2 k^2)}.
\end{align}
\end{subequations}
We see the same qualitative behaviour in the longitudinal sound mode as well: the damping is now dominated by the pinning-induced relaxation coefficients. However, the damping of the crystal diffusion mode, while gets affected by pinning, is still controlled by the plasticity-induced relaxation coefficients.

In addition, we get three non-hydrodynamic modes as opposed to just one found in the theory of plasticity without pinning; see \eqref{eq:non-hydro-mode}. This is because the lack of translational symmetry gives physical meaning to the $\phi^I$ fields independently of the reference metric $\bbh_{IJ}$, thereby breaking the $\Diff(\phi)$ symmetry mentioned in \eqref{eq:diffphi}. The effect this has is that we now have one non-hydrodynamic damped mode in the transverse sector 
\begin{subequations}
\begin{equation}
    \omega = - i\frac{\omega_0^2\Omega_G}{\omega_0^2 + v_\perp^2 k^2} + \ldots,
    \label{eq:pinning-transverse-nonhydro-mode}
\end{equation}
and two coupled modes in the longitudinal sector that can be disentangled assuming weak pinning as
\begin{align}
    \omega
    &= -i 
    \frac{B\Omega_G+ 2\frac{d-1}{d} G\Omega_B}{B+ 2\frac{d-1}{d} G}  \nn\\
    &\qquad 
    -i \frac{\rho B G (\Omega_B-\Omega_G)^2}{\lb B+ 2\frac{d-1}{d} G \rb^2 \lb B\Omega_G + 2\frac{d-1}{d} G \Omega_B\rb } 
    \frac{\omega_0^2}{k^2}
    + \ldots, \nn\\
    \omega 
    &= -i\frac{\rho\Omega_B\Omega_G}{B\Omega_G + 2\frac{d-1}{d} G \Omega_B} \frac{\omega_0^2}{k^2} + \ldots~.
    \label{eq:pinning-long-nonhydro-mode}
\end{align}
\end{subequations}
We can check that two of these modes go away when we switch off pinning, while the third reduces to \eqref{eq:non-hydro-mode}.

We can also obtain the mode spectrum in the liquid regime $k\ll \ell \ll L^{-1}_T$. The three transverse sector modes from \eqref{eq:pinning-transverse-mode} and \eqref{eq:pinning-transverse-nonhydro-mode} combine and give rise to a pair of pinned shear modes and a damped mode
\begin{align} \label{eq:pinning-transverse-fluid-modes}
    \omega &= \pm\omega_0 - \frac{i}{2} \lb \frac{\eta_f}{\rho} k^2 + \Gamma + \Omega_\phi \rb + \ldots, \nn\\
    \omega &= -i\Omega_G + \ldots~.
\end{align}
When pinning is absent, one of the branches of the shear modes goes away, while the other gives rise to the ordinary shear mode reported in \eqref{eq:modes-transverse-fluid}. The longitudinal sector is slightly more richer; the five modes from \eqref{eq:pinning-long-mode} and \eqref{eq:pinning-long-nonhydro-mode} combine to give rise to a pinned and damped liquid sound mode, an undamped and unpinned diffusion mode, and two damped modes 
\begin{align} \label{eq:pinning-long-fluid-modes}
    \omega 
    &= \pm \sqrt{\omega_0^2 + v_f^2 k^2}
    - \frac{i}{2} \left( \frac{\zeta_f + 2\frac{d-1}{d}\eta_f}{\rho} k^2
    + \frac{\sigma }{\chi_f} k^2 \right. \nn\\
    &\qquad 
    \left.
    - \frac{\omega_0^2 \left(\sigma - 2n\gamma_n  + \frac{n^2}{\sigma_\phi} \right)}{\chi_f (\omega_0^2 + v_f^2 k^2)} k^2
    + \Gamma + \Omega_\phi \right) + \ldots, \nn\\
    \omega 
    &= - \frac{\omega_0^2 \left(\sigma - 2n\gamma_n  + \frac{n^2}{\sigma_\phi} \right)}{\chi_f (\omega_0^2 + v_f^2 k^2)} k^2 + \ldots, \nn\\
    \omega 
    &= -i\Omega_B \lb 1 + \frac{B\alpha_m^2}{\chi} \rb + \ldots, \nn\\
    \omega 
    &= - i\Omega_G + \ldots~, 
\end{align}
where the speed of sound in this regime is modified to
\begin{align}
    v_f^2 = \frac{(n+\lambda_n)^2}{\rho(\chi + B\alpha_m^2)}~.
\end{align}
When pinning is absent, the additional diffusive mode drops out and the sound mode reverts to its original form seen in \eqref{eq:modes-long-fluid}.
The presence of an undamped diffusive mode is physically expected because pinning only causes the momentum to relax, not particle density. Therefore, after the proliferation of dislocations and impurities when both pinning and plasticity are strong, there will be only one long-lived hydrodynamic mode left associated with the diffusion of particles.

We can check that the modes presented here reduce to the pure plasticity results in \eqref{sec:plastic-modes} in the absence of pinning coefficients. On the other hand, they reduce to the results of \cite{Armas:2021vku} in the absence of plasticity effects, modulo the non-Galilean coefficients $\sigma$, $\gamma_n$, and $\lambda_n$.

\subsection{Correlation functions}
\label{sec:correlations-pinned}

\begin{figure*}[!t]
	\center
    {\includegraphics[width=0.4\linewidth]{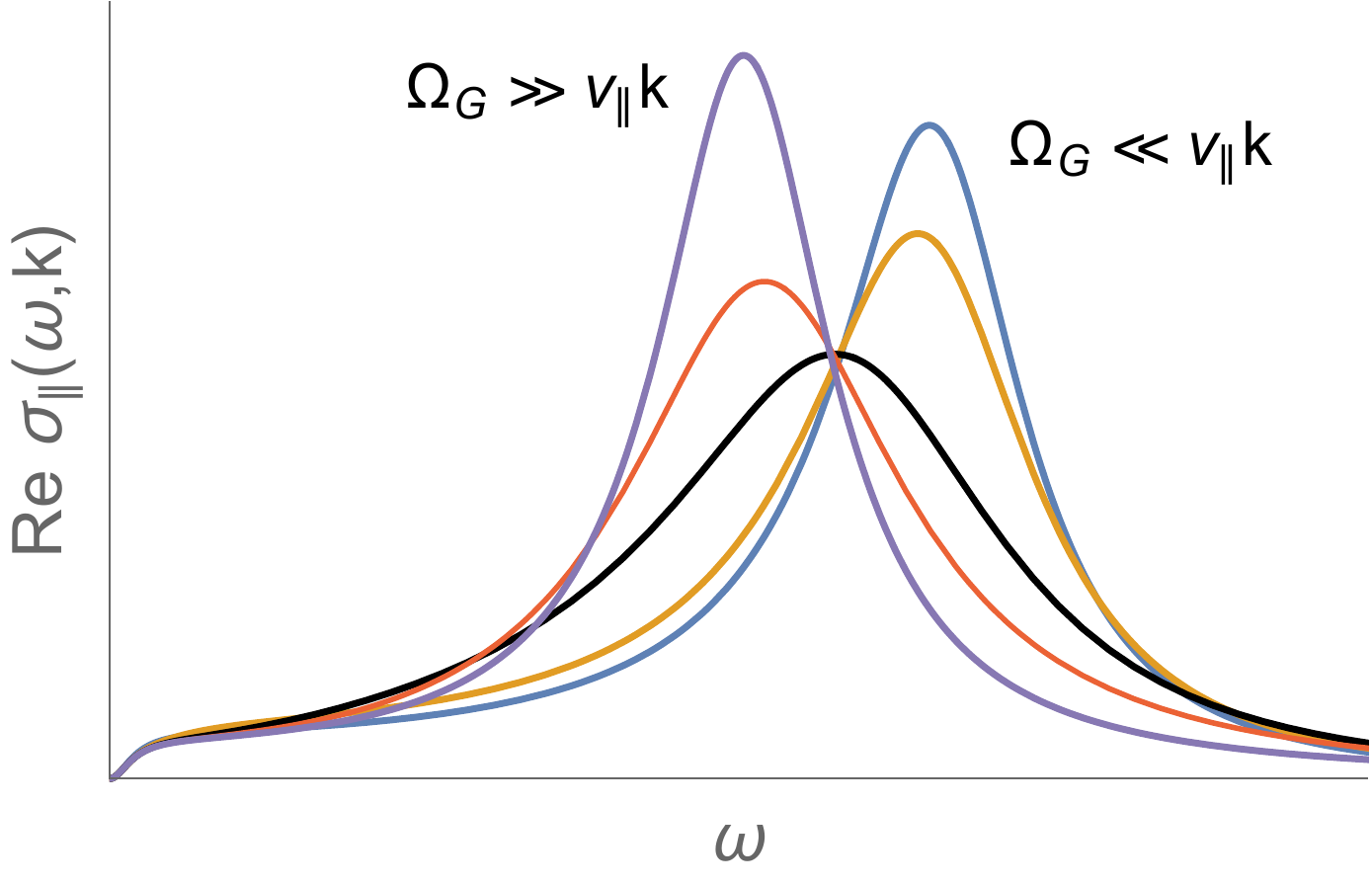}}
    \hspace{5em}
    {\includegraphics[width=0.4\linewidth]{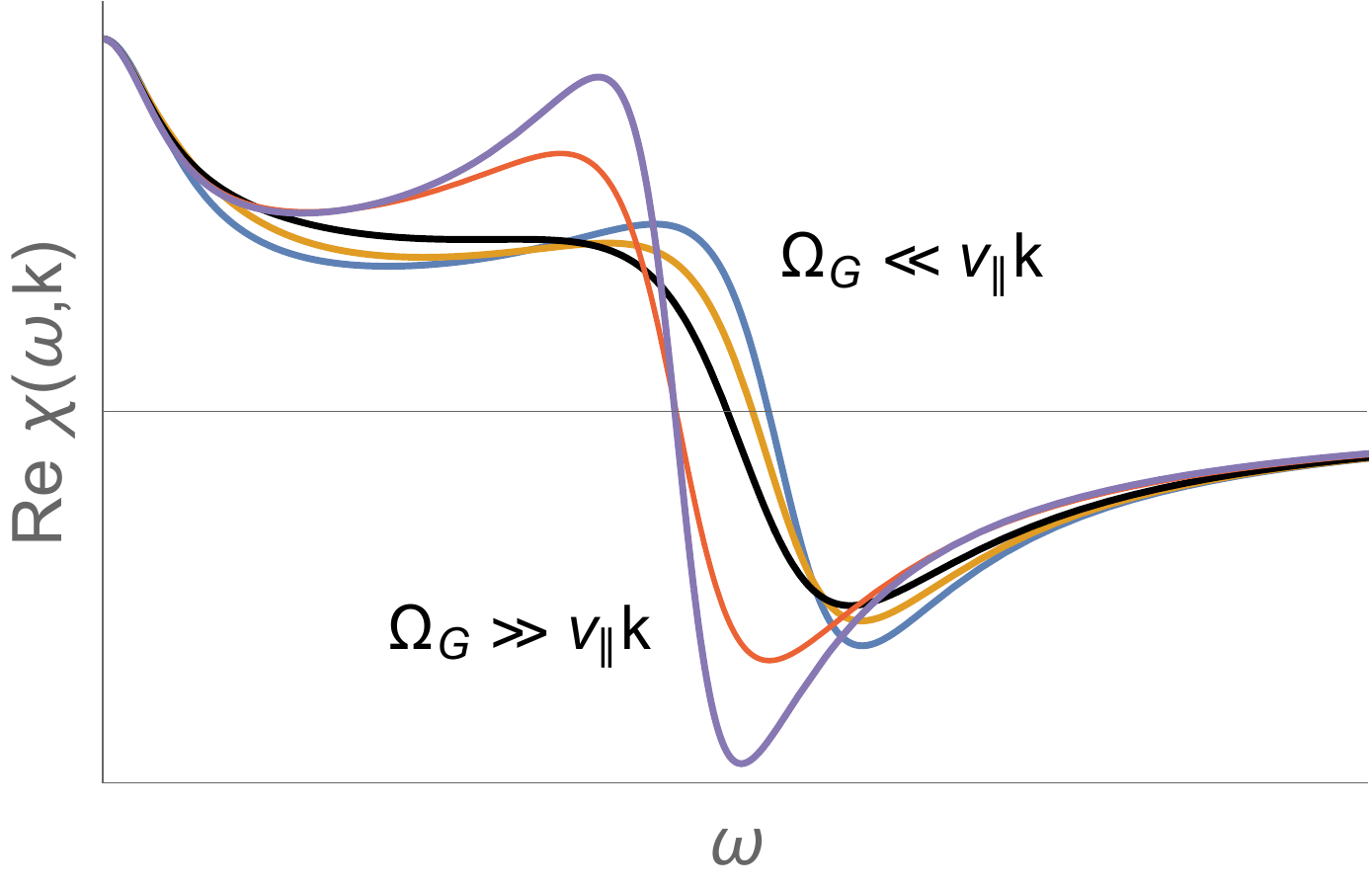}}
    
    \caption{Real parts of the longitudinal optical conductivity and charge/particle susceptibility at nonzero wavevector for increasing rate of plasticity-induced relaxation $\Omega_G$. The peaks in both the plots widen and shorten until the solid to liquid phase transition point, sharpening and rising back up again at a lower frequency after the transition. The black curves represent the phase transition point. The imaginary parts of the longitudinal optical conductivity and susceptibility are proportional to the imaginary parts of susceptibility and longitudinal optical conductivity respectively.
    \label{fig:peak-dance-long}}
\end{figure*}

We can couple the hydrodynamic equations to background sources, as given in appendix \ref{app:background}, and read off the respective correlation functions. Focusing on zero wave-vector, we find that the flux and crystal velocity correlators are sensitive to pinning effects and modify from their trivial form in \eqref{eq:current-correlators} to 
\begin{align} \label{eq:correlationfunctions-pinning}
    G^R_{j^ij^j}(\omega)
    &=
    \delta^{ij} \left(-i\omega\sigma 
    + \frac{\omega}{\rho}\frac{(n-\lambda_n)^2 (\omega +i\Omega_\phi)}
    {(\omega +i \Gamma ) (\omega +i\Omega_\phi) - \omega_0^2} \right. \nn\\
    &\qquad 
    \left. - \frac{\omega \gamma_{n} \omega_0^2}{\lambda^2}
    \frac{2i \lambda (n-\lambda_n)
    + \gamma_n \rho(\omega +i \Gamma )
    }
    {(\omega +i \Gamma ) (\omega +i\Omega_\phi) - \omega_0^2} \right), \nn\\[0.5em]
    G^R_{u^i_\phi u^j_\phi}(\omega)
    &= - \delta^{ij} \frac{\omega^2\lambda^2}{\rho\omega_0^2}
    \lb 1 - \frac{\omega (\omega +i \Gamma ) }
    {(\omega +i \Gamma ) (\omega +i\Omega_\phi) - \omega_0^2}  \rb, \nn\\[0.5em]
    G^R_{j^i u^j_\phi}(\omega)
    &= \omega^2
     \frac{\lambda (n-\lambda_n)/n
    -i\gamma_n (\omega + i\Gamma)
    }{(\omega +i \Gamma ) (\omega +i\Omega_\phi) - \omega_0^2}~.
\end{align}
These are a generalisation of our results in \cite{Armas:2021vku} to boost-agnostic crystals. Notably, as we discussed in the introduction, there are no signatures of plasticity in these correlators when $k=0$. In particular, optical conductivity at $k=0$ is insensitive to plasticity.

With sufficient computational power, we can also work out the correlation functions at nonzero wave-vector. In the transverse sector optical conductivity result reported in \eqref{eq:omega-k-conductivity}, we have ignored the non-Galilean coefficients $\sigma$, $\lambda_n$, $\gamma_n$ and set $\lambda\to 1$ for simplicity, along with the shear viscosity $\eta$. Further switching off the bulk viscosity $\zeta$ and the expansion coefficient $\alpha_m$, we find an analogous expression for the optical conductivity and charge/particle susceptibility in the longitudinal sector
\begin{widetext}
\begin{align}
    \sigma_\|(\omega,k)
    &= \frac{i}{\omega} G^R_{j_\|j_\|}(\omega,k) \nn\\
    &= - \frac{n^2}{\rho}
    \frac{(i\omega -\Omega_\phi) - \frac{i k^2 \omega}{\sigma_\phi}  \left(\frac{B}{i \omega -\Omega_B}
    + 2\frac{d-1}{d}\frac{G}{i \omega -\Omega_G}\right)}
    {\lb i \omega -\Gamma 
    + \frac{n^2}{\rho\chi}\frac{k^2}{i\omega} \rb
    (i\omega -\Omega_\phi )
    + \omega_0^2
    + i k^2 \omega \lb \frac{B}{i \omega -\Omega_B}
    + 2\frac{d-1}{d}\frac{G}{i \omega -\Omega_G} \rb 
    \lb \frac{1}{\rho }-\frac{1}{\sigma_\phi} \lb 
    i \omega - \Gamma
    + \frac{n^2}{\rho\chi}\frac{k^2}{i\omega} \rb\rb}~~, \\
    \chi(\omega,k)
    &= G^R_{nn}(\omega,k)
    = -\frac{k^2}{\omega^2} G^R_{j_\|j_\|}(\omega,k)
    = \frac{ik^2}{\omega}\sigma_\|(\omega,k) \nn\\
    &= \chi - \chi
    \frac{
    \lb i \omega -\Gamma  \rb
    (i\omega -\Omega_\phi)
    + \omega_0^2
    + i k^2 \omega \lb \frac{B}{i \omega -\Omega_B}
    + 2\frac{d-1}{d}\frac{G}{i \omega -\Omega_G} \rb 
    \lb \frac{1}{\rho }-\frac{1}{\sigma_\phi} \lb 
    i \omega - \Gamma \rb\rb
    }
    {\lb i \omega -\Gamma 
    + \frac{n^2}{\rho\chi}\frac{k^2}{i\omega} \rb
    (i\omega -\Omega_\phi)
    + \omega_0^2
    + i k^2 \omega \lb \frac{B}{i \omega -\Omega_B}
    + 2\frac{d-1}{d}\frac{G}{i \omega -\Omega_G} \rb 
    \lb \frac{1}{\rho }-\frac{1}{\sigma_\phi} \lb 
    i \omega - \Gamma
    + \frac{n^2}{\rho\chi}\frac{k^2}{i\omega} \rb\rb}~~.
\end{align}
\end{widetext}
Note that susceptibility is defined as the correlator of particle density and is proportional to the longitudinal optical conductivity due to Ward identities (conservation laws). We see a similar analytic form in these observables as the transverse optical conductivity in \eqref{eq:omega-k-conductivity} discussed in the introduction, except for the  $n^2/(\rho\chi)$ terms in the denominator that correspond to the fluid contribution to the speed of longitudinal sound mode.\footnote{This result reproduces the spatially resolved optical conductivity reported in~\cite{Delacretaz:2017zxd}, provided that we switch off the dislocation- or plasticity-induced relaxation coefficients $\Omega_B$, $\Omega_G$ and the crystal diffusion coefficient $1/\sigma_\phi$, while keeping the pinning-induced relaxation $\Omega_\phi$ and pinning frequency $\omega_0$ nonzero. In the years since, it has been understood that $\Omega_\phi$ and $\omega_0^2/\sigma_\phi$ are related by the ``damping-attenuation relation'' in \eqref{eq:damping-attentuation} and it is not possible to set one to zero while keeping the other nonzero~\cite{Armas:2021vku, Delacretaz:2021qqu}. 
} We also find a qualitatively similar peak interpolation behaviour across the melting phase transition as we increase the strength of plasticity-induced relaxation; see figure \ref{fig:peak-dance-long}.

    

On the other hand, we find that the stress and strain correlators at zero wave-vector do not get affected by pinning at all and are still given by their pure plasticity form in \eqref{eq:correlationfunctions}.

\subsection{Pinning vs plasticity}

Let us close this section with some comments on the distinction between the physical signatures of pinning and plasticity in a crystal. It is common knowledge that pinning and plasticity can both cause the strain tensor of a crystal to relax. In the former case, this relaxation is due to the presence of impurities or inhomogeneities that relax the crystal phase fields and thereby relax the strain tensor. Whereas, in the latter case, the relaxation is caused by the motion of dislocations present in a crystal and affects the strain tensor directly. Notably, plasticity does not relax the distortion strain of the crystal, while pinning does. The distinct origins of these phenomena are also highlighted by the recently discovered ``damping-attenuation relation'' \cite{Delacretaz:2021qqu, Armas:2021vku} given in \eqref{eq:damping-attentuation},
which posits that the pinning-induced relaxation rate of a crystal scales proportionately with its diffusion constant. We find that this relation continues to hold in the presence of plasticity as well, but no such relation applies for the plasticity-induced relaxation rate; see \eqref{eq:damping-attentuation}.

The hydrodynamic framework we developed allows us to probe these differences on a more quantitative level directly in the mode spectrum and correlation functions. When crystals are subjected to both weak plasticity and weak pinning, we find that the damping of the sound modes is dominated by the pinning-induced phase relaxation rate $\Omega_\phi$ and momentum relaxation rate $\Gamma$. On the other hand, the damping of the crystal diffusion mode is controlled by plasticity-induced strain relaxation rates $\Omega_B$, $\Omega_G$. When the strength of plasticity is increased in a pinned crystal, the crystal effectively melts giving rise to the mode spectrum of a ``pinned liquid phase''. Whereas, when the strength of pinning is increased, we instead are left with a single particle diffusion mode.

We find that the plasticity-induced relaxation has no effect on the optical conductivity at $k=0$, while pinning-induced relaxation does. One heuristic way to understand this for electronic crystals is to note that the optical conductivity at $k=0$ is the response of the material to a uniform time-varying electric field. Since this electric field is uniform in space, it only induces a uniform collective motion of the electrons making up the crystal and does not trigger any plasticity-related effects. On the other hand, this uniform motion is sensitive to the presence of a background lattice, impurities, or inhomogeneities, and does trigger pinning-related effects that shows up in the optical conductivity at $k=0$; see e.g.~\cite{Delacretaz:2016ivq, Delacretaz:2017zxd}. At $k\neq 0$, the electric (and magnetic fields) applied to the material are non-uniform and generically trigger both pinning and plasticity effects. As we noted in the introduction, the effects of plasticity on the optical conductivity at nonzero $k$ can be used to probe the solid-liquid phase transition.

We also computed the effects of pinning and plasticity on frequency-dependent viscosities. We found that pinning leaves no signatures on viscosities at $k=0$, while plasticity does. Returning to our heuristic picture, viscosities at $k=0$ parametrise the response of the electronic crystal to a uniform time-dependent shear/expansion of the background metric. Such deformations are sensitive to plasticity effects, but since the inhomogeneities and the crystalline matter experience the deformations simultaneously, the pinning effects are not triggered. The effect of plasticity on viscosities for an unpinned crystal was worked out in~\cite{Delacretaz:2017zxd} (see also \cite{1980PhRvB..22.2514Z}), although the effects of pinning were not explored.

Finally, we note that the rheology equations of an unpinned plastic crystal behave like a Jeffery material, prone to permanent distortions at long time scales. On the other hand, the rheology equations of a pinned elastic crystal, similar to that of an unpinned elastic crystal, behave like a Kelvin-Voigt material with no permanent distortions. However, upon combining these two effects, we find a new behaviour similar to a Zener material, in which the material admits ``permanent'' distortions at long time scales, but due to pinning eventually returns to its original state at very long time scales proportional to the product of pinning and plasticity time scales.


\section{Experimental observations}
\label{sec:expobs}

Experimentally observing the effects of plasticity on the dynamics of the collective excitations, i.e. fluctuations of the phasons $\phi^I$ such as charge density waves (CDW), requires a momentum resolved probe as pointed out above. The photon momentum involved in optical conductivity experiments is negligible compared to the typical electron momenta, and consequently the optical signatures of the collective dynamics remains largely unaffected. One possible exception may be the observation of creep dynamics in I(V) characteristics~\cite{2005PhRvB..71g5118O}. However, the phase slip dynamics reported there is beyond the scope of the current work and it remains to be seen if one can disentangle the different effects of pinning and plastic deformation from each other based on such experiments.

There are several other probes where the momentum dependence of the collective response can potentially be identified. In the frequency-momentum domain, inelastic x-ray scattering and neutron scattering experiments have been used to study the collective dynamics of CDW states. A particular difficulty is that one has to disentangle the phase and amplitude collective motion of the CDW order from the phonon modes. Given that the phonons are participants in the formation of the CDW phase, their response is intrinsically coupled. Nevertheless, collective modes have been observed with x-ray scattering methods in various CDW materials such as NbSe$_{3}$~\cite{requardt_JP_2002}, blue bronze~\cite{2004PhRvB..69k5113R}, and high T$_{c}$ superconductors~\cite{2016ARCMP...7..369C, 2017NatPh..13..952C, 2019SciA....5.3346M}. In particular, the improvements in energy resolution obtained in resonant inelastic x-ray scattering experiments has enabled the detection of collective modes. 

There are very few works that directly report the phason dispersion. In~\cite{2004PhRvB..69k5113R}, the momentum dependence of the phason mode around the ordering wavevector has been extracted at temperatures close to the CDW melting temperature. These authors also report the lifetime broadening of the collective mode. A later work on the same material reported the direct observation of dislocations using coherent x-ray diffraction imaging~\cite{2005PhRvL..95k6401L}. The situation is similar in cuprates~\cite{2017NatPh..13..952C, 2021JPSJ...90k1004L}, where signatures of the dispersive excitations have been observed. However, in all these experiments the collective mode is probed through the modulation of Bragg peaks associated with the CDW order. Our framework, on the other hand, describes the momentum dependence of the collective mode in an isotropic momentum space. 

An alternative approach to experimentally probe the effects of plasticity is through ultra-fast time resolved experiments. Using free electron laser sources, it is possible to probe energy scales down to 0.1 meV, which is in principle sufficient to probe the dynamics of the collective mode response. The authors in~\cite{2019SciA....5.3346M} report the exponential decay dynamics of collective excitations around the charge ordering wavevector in La$_{2-x}$Ba$_{x}$CuO$_{4}$. Interestingly, these authors report gapless excitations below the previously observed amplitude modes. Near-equilibrium,  \cite{2019SciA....5.3346M} reports an exponential decay of the collective mode with damping rate of the form $\Omega\sim\gamma_0+D k^2$, where $\gamma_0$ is a relaxation coefficient and $D$ a diffusion constant. Our results in \eqref{eq:pinning-transverse-mode}-\eqref{eq:pinning-long-fluid-modes} show the same qualitative behaviour of exponential decay and damping rate. Moreover, the very late-time dynamics reported \cite{2019SciA....5.3346M} shows a universal behaviour as a function of momentum with an approximate logarithmic time dependence. Such behaviour is characteristic of long-time tails in hydrodynamic correlation functions and could in principle be incorporated into our formalism by taking into account stochastic corrections as in \cite{Forster:1976zz, Chen-Lin:2018kfl, Jain:2020zhu}. We leave a precise comparison between our work and \cite{2019SciA....5.3346M} for the future.

These x-ray scattering methods also provide insight in the mode dynamics through the determination of the renormalised sound velocities. A very recent study of the collective mode dynamics in (TaSe$_{4}$)$_{2}$I reports a possible phase excitation that disperses approximately linearly with a mode velocity that is significantly different from the predicted sound velocity and momentum dependent damping rate~\cite{Nguyen:2022wlt}. Perhaps more detailed analysis of the experiments combined with further refinements of our theoretical framework will enable us to extract some of the parameters of the hydrodynamic theory.

To conclude, we mention three probes that could potentially be used to probe the dynamics of the CDW collective modes. The first probe is momentum resolved, electron energy loss spectroscopy (M-EELS). In this electron scattering experiment, one probes the density-density correlation function or charge susceptibility. In contrast to the transverse response probed with optical experiments, M-EELS probes the longitudinal response~\cite{2018arXiv180305439S}. As demonstrated in section \ref{sec:correlations-pinned}, this charge susceptibility is modified by plasticity and pinning and M-EELS could therefore be an interesting probe to detect the effects of plasticity and pinning. Several collective modes of electronic origin have been observed in CDW phases at finite momentum~\cite{rak_thesis}. However, the CDW collective modes have so far not been detected. The second probe we mention is ultra-fast electron diffraction where plasticity can be probed either directly through measuring the collective electron dynamics as well as through resolving real-space deformations~\cite{2020NatNa..15..761H}. Another experiment that can probe the finite momentum dynamics of ordered phases is near-field optical spectroscopy~\cite{2014PhRvB..90a4502S}. 

The caveat with each of these three probes is that one only probes the dynamics of the collective mode indirectly. Both in M-EELS and near-field optical spectroscopy, the collective modes will hybridze with polariton modes. Even though this prohibits the direct measurement of the momentum dependence introduced by plasticity, the framework presented in this work could be a useful starting point to describe the dynamics of such hybrid modes.

\section{Outlook}
\label{sec:outlook}

We have formulated a novel and systematic hydrodynamic framework for plastic deformations in electronic crystals, with and without pinning effects arising from translational disorder. We have used this framework to obtain the hydrodynamic mode spectrum and correlation functions in the presence of plasticity, interstitials/vacancies, and impurities or spatial inhomogeneities. As far as we are aware, this is the first time these results have appeared in full generality in various interpolating regimes of pinning and plasticity, parametrising the phase space of metals and electronic crystals. More broadly, this formalism is applicable to physical systems characterised by spontaneously (and explicitly) broken translation symmetry and residual rotation symmetry. Hence, it is useful for describing the near-equilibrium dynamics of isotropic electronic crystals, multi-component charge density waves, metals, as well as ordinary crystals.

\begin{figure}[!t]
	\center
    {\includegraphics[width=0.8\linewidth]{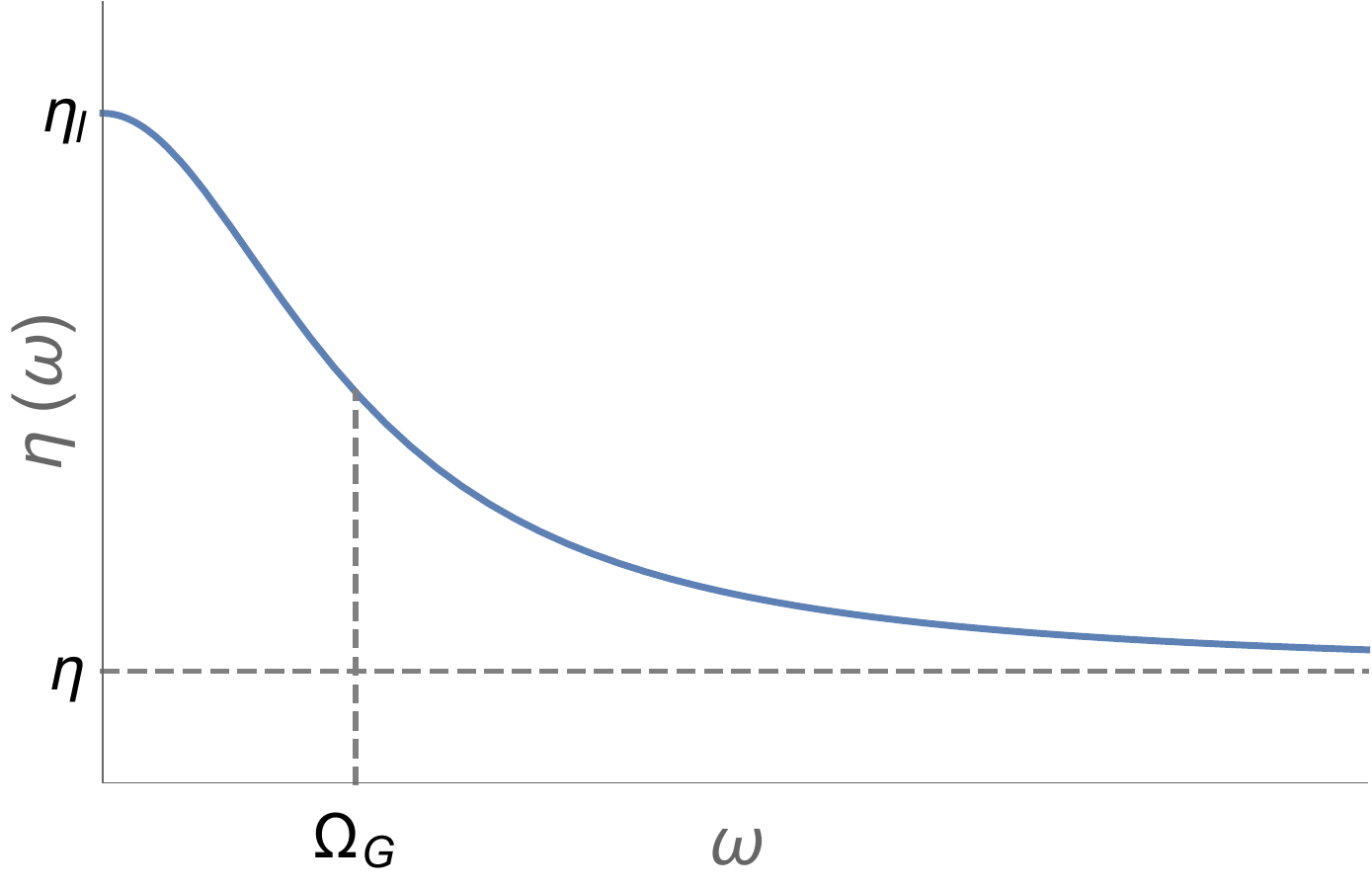}}
    
    \caption{Real part of frequency-dependent shear viscosity $\eta(\omega)$ for nonzero rate of plasticity-induced relaxation $\Omega_G$. The viscosity asymptotes to its ``solid value'' $\eta$ for $\omega\gg\Omega_G$ and to its ``liquid value'' $\eta_l$ for $\omega\ll\Omega_G$. The bulk viscosity $\zeta(\omega)$ also shows a similar qualitative behaviour with the bulk sector relaxation rate $\Omega_B$.
    \label{fig:viscosity}}
\end{figure}

Our primary motivation has been to identify the signatures of plasticity in electronic crystals, in particular isotropic Wigner crystals and multi-component charge density wave states. Starting with pure crystals without translational disorder or impurities, we found in section~\ref{sec:linear-viscoplasticity} that the introduction of weak plasticity causes the sound and diffusion modes of a crystal to become damped due to plasticity-induced strain relaxation rates $\Omega_B$, $\Omega_G$. In other words, the ``softening'' of the crystalline structure due to plasticity makes the perturbations die out faster. As the strength of plasticity is increased, or alternatively we probe the crystal at distance- and time-scales larger compared to the plasticity scale, the mode spectrum behaves like that of a liquid with increased effective viscosities; see figure \ref{fig:viscosity}. As we discussed in section~\ref{sec:dislocations}, the increase in plasticity can be understood as the proliferation of dislocations in a crystal, causing it to melt and hence the liquid-like behaviour. 

We also noted in section~\ref{sec:linear-viscoplasticity} that plasticity does not affect the optical conductivity of a crystal at zero wavevector. However, it does leave some interesting signatures in the optical conductivity peak at nonzero wavevector $k$ as the material undergoes the solid-liquid phase transition; see figure~\ref{fig:peak-dance}. For clear signatures of plasticity at $k=0$, one can look at the frequency-dependence of viscosities measured via the stress-stress correlator $G^R_{\tau\tau}(\omega)$ in \eqref{eq:correlationfunctions}; see figure \ref{fig:viscosity}. One can also look at the strain-strain correlator $G^R_{\kappa\kappa}(\omega)$ at $k=0$ given in \eqref{eq:correlationfunctions}, which is only nonzero in the presence of plasticity and can potentially be measured by applying an external stress to the crystal.


Combining plastic deformations with translational disorder or inhomogeneities in a crystal leads to an interesting interplay between plasticity-induced and pinning-induced effects. In particular, for weak plasticity and pinning, the sound modes are now pinned at the pinning frequency $\omega_0$ and their damping is controlled by the pinning-induced phase
relaxation coefficient $\Omega_\phi$, as opposed to the plasticity-induced strain relaxation $\Omega_G$, $\Omega_B$. On the other hand, the damping of the crystal diffusion mode receives corrections due to pinning, but is still controlled by $\Omega_B,\Omega_G$. When dislocations in a pinned crystal are proliferated and the strength of plasticity is strong, we find a spectrum analogous to that of a liquid, but with all the modes pinned at $\omega_0$. It will be interesting to further explore this ``pinned liquid'' phase as an effective hydrodynamic theory of its own, which might be relevant for metallic phases where electrons are pinned due to the ionic lattice but are not necessarily crystallised into an electronic crystal. Increasing the strength of pinning further removes all the hydrodynamic modes from the spectrum, leaving only the particle diffusion mode because particle conservation is not explicitly broken by pinning.

We found that at zero wavevector, conductivity (particle flux correlator) and viscosities (stress correlator) are exclusively sensitive to pinning-induced and plasticity-induced effects respectively. At nonzero wavevector, all correlators become contaminated with both plasticity- and pinning-induced effects. In particular,
we found that plasticity-induced relaxation ``de-pins'' the transverse optical conductivity at nonzero wavevector by lowering and broadening the optical conductivity peak as well as non-trivially affecting its position; see figure \ref{fig:peak-dance}. Furthermore, we have shown that the damping-attenuation relation for pseudo-Goldstones derived recently in \cite{Delacretaz:2021qqu, Armas:2021vku} still holds for the pinning-induced phase relaxation coefficient $\Omega_\phi$, relating it to the crystal diffusion coefficient $D_\phi^\perp$, while no such relation exists for the plasticity-induced strain relaxation coefficients $\Omega_B,\Omega_G$.

The fragility of electronic crystal phases in most experimental setups is often overcome by turning on background magnetic fields \cite{1990PhRvL..65.2189G}. Nonzero magnetic fields can lead to new physical effects, including the appearance of new vibrational magnetophonon modes \cite{1977PhRvB..15.1959B, 1978PhRvB..18.6245F} or modification of the pinning frequency \cite{2001PhRvB..65c5312C}. Weak and strong background magnetic fields can be accounted for in a hydrodynamic theory akin to the work of~\cite{Delacretaz:2019wzh, Amoretti:2021lll}. We have included the presence of weak electromagnetic fields in our hydrodynamic framework in appendix \ref{app:background}, but have not studied their physical effects. Accounting for strong magnetic fields would be especially interesting and requires a non-trivial extension of our formalism. We hope to return to these phenomena in the future.

It would be interesting to use the techniques developed here to study unidirectional charge density wave states, which can be thought of as the smectic phase of an electronic liquid crystal. While a lot of work has been done in this direction, we note that we are not aware of a complete formulation of the corresponding hydrodynamic theory along the lines of \cite{Armas:2019sbe, Armas:2020bmo, Armas:2021vku}. Such a theory, which we expect to be closely related to superfluids \cite{Bhattacharya:2011eea}, will likely reveal new transport properties. Topological defects, in this context, can be incorporated as vortices of a single phase field. A careful analysis of these systems, including the interplay with translational disorder and impurities, will be enlightening and shall be one of our main goals moving forward.

Our approach did not assume any particular type of boost symmetry and, as such, our results are valid for both Galilean and relativistic crystals, or crystals with no boost symmetry at all. The case of relativistic boost symmetry is interesting in its own right, given the large body of work on holographic models of pinned charge density waves (see e.g. \cite{Armas:2020bmo,Amoretti:2019kuf,Amoretti:2019cef,Amoretti:2018tzw,Amoretti:2017frz,Baggioli:2022pyb, Andrade:2022udb}), which could potentially be extended to accommodate the presence of topological defects and plasticity. We have given a manifestly relativistic formulation of pinned plastic crystals in appendix~\ref{app:relativistic}. The construction suggests that holographic models of dislocations and plasticity will require a bi-metric theory in the bulk, wherein the additional metric will give rise to a dynamical reference metric at the boundary. It would be interesting to pursue this line of thought further and construct holographic models for the solid-liquid phase transitions discussed here.

\acknowledgements 

We would like to thank Luca Delacretaz, Blaise Gouteraux, Maziyar Jalaal, Alexander Krikun, Edan Lerner, and Jan Zaanen for various helpful discussions and constructive comments on our manuscript. JA and AJ are partly supported by the Netherlands Organization for Scientific Research (NWO) and by the Dutch Institute for Emergent Phenomena (DIEP) cluster at the University of Amsterdam. AJ is funded by the European Union’s Horizon 2020 research and innovation programme under the Marie Sk{\l}odowska-Curie grant agreement NonEqbSK No. 101027527. RL was supported, in part, by the cluster of excellence ct.qmat (EXC 2147, project-id 390858490).

\newpage
 
\appendix

 \section{Viscoplastic hydrodynamics with background sources}
\label{app:background}

In this appendix, we revisit the framework of viscoplastic hydrodynamics while paying careful attention to the background sources. In addition to extending the applicability of the framework to plastic crystals on curved spacetime backgrounds and in the presence of background electromagnetic fields, this will also allow us to obtain the hydrodynamic correlation functions using background variational methods. We also include the effects of pinning in the following following the construction of \cite{Armas:2021vku}.

\subsection{Aristotelian geometry and conservation laws}

We wish to derive the conservation laws for viscoplastic hydrodynamics that served as the starting point of our discussion in section~\ref{sec:viscoplastic-hydro} and section \ref{sec:pinned-consti}. To this end, let us couple our crystal to an Aristotelian spacetime background, which is the appropriate geometric structure for systems without a boost symmetry~\cite{Armas:2020mpr, deBoer:2017ing, deBoer:2020xlc}; see also~\cite{Jain:2020vgc, Banerjee:2015hra, Jensen:2014ama, Jensen:2014aia, Son:2013rqa}. The spacetime background includes a \emph{clock form} $n_\mu$ (coupled to the energy density and flux) and a degenerate \emph{spatial metric} $h_{\mu\nu}$ (coupled to the momentum density and the stress tensor). The covariant indices $\mu,\nu,\ldots$ run over both the spatial indices $i,j,\ldots$ as well as the time coordinate $t$. 
The spatial metric admits a null eigenvector $v^\mu$, called the \emph{frame velocity}, satisfying $v^\mu h_{\mu\nu} = 0$ and $v^\mu n_\mu = 1$. Using this, it is also convenient to define a contravariant spatial metric $h^{\mu\nu}$ using $h^{\mu\nu}h_{\nu\rho} = \delta^\mu_\rho - v^\mu n_\rho$ and $h^{\mu\nu}n_\nu = 0$. The flat space (Cartesian) limit is given by
\begin{gather}
    n_\mu \df x^\mu = \df t, \qquad 
    v^\mu \dow_\mu = \dow_t, \nn\\
    h_{\mu\nu}\df x^\mu\df x^\nu = \delta_{ij}\df x^i\df x^j, \nn\\
    h^{\mu\nu} \dow_\mu \otimes \dow_\nu 
    = \delta^{ij} \dow_i \otimes \dow_j~.
    \label{eq:flat-sources}
\end{gather}
We also introduce a background gauge field $A_\mu$ to couple to the particle density and flux. 

We can define the covariant version of crystal frame fields as $e^I_\mu = \dow_\mu\phi^I$. We can also define the covariant crystal velocity $u^\mu_\phi$ such that $u_\phi^\mu e^I_\mu = 0$ and $u_\phi^\mu n_\mu = 1$.  The induced metric on the crystal space is defined naturally as $h^{IJ} = h^{\mu\nu}e_\mu^I e_\nu^J$, which can be used to define the strain tensor $\kappa_{IJ}$ same as \eqref{eq:strain-tensor}. The ``inverse frame fields'' $e^\mu_I$ can be defined via $e^\mu_I e^I_\nu = \delta^\mu_\nu - u^\mu_\phi n_\nu$ and $n_\mu e^\mu_I = 0$. It is worth noting that $e^\mu_I = h^{\mu\nu}e_\nu^J h_{IJ}$.

Consider a crystal described by some effective action $S$, expressed as a functional of the background fields as well as the dynamical crystal fields. An infinitesimal variation of $S$ can be parametrized as
\begin{align} \label{eq:actionVariation}
    \delta S   & =  \int\!\df t  \df^{d } x \sqrt{ \gamma } \bigg[  
    \lb \pi^{\mu} v^{\nu } +  \frac{1}{2} \tau^{\mu \nu } \rb \!\delta h_{\mu\nu} 
    - \epsilon^{\mu} \delta n_{\mu}
    + j^{\mu} \delta A_{\mu} 
    \nn\\  
    &\qquad\qquad
    + K_{I} \delta\phi^I 
     + \half U^{IJ} \lb \delta\psi_{IJ}
     - e_K^\mu \dow_\mu \psi_{IJ} \delta \phi^K \rb \nn\\
     &\qquad\qquad
     + \ell' L_I \delta\Phi^I
    \bigg],
\end{align}
with $\gamma$ denoting the determinant of $\gamma_{\mu \nu } = h_{\mu\nu} + n_{\mu} n_{\nu}$. This parametrisation defines the covariant momentum density $\pi^\mu$ (with $\pi^\mu n_\mu = 0$), stress tensor $\tau^{\mu\nu}$ (with $\tau^{\mu\nu}n_\nu = 0$), energy current $\epsilon^\mu$, and the particle number current $j^\mu$. See~\cite{Armas:2020mpr} for more details. We have parametrised the variations with respect to $\delta\phi^I$ and $\delta\bbh_{IJ} = \ell\delta\psi_{IJ}$ such that the dual the operators $K_I$ and $U^{IJ}$ transform covariantly under the $\Diff(\phi)$ symmetry in \eqref{eq:diffphi}. We have also allowed the action to depend on a set of background crystal fields $\Phi^I$, explicitly breaking the translation symmetry of the crystal and giving rise to the physics of pinning; see section~\ref{sec:pinning} or \cite{Armas:2021vku} for more discussion.

Taking the variations in \eqref{eq:actionVariation} to be physical spacetime diffeomorphisms and gauge transformations, we can obtain the respective conservation laws
\begin{subequations}    \label{feoiheoih22-}
\begin{align}  
    \lb\nabla_{\mu} + F^{n}_{\mu\lambda} v^{\lambda}\rb \epsilon^{\mu } \hspace{-5em}& \nn\\
    &= -  v^{\nu} \lb F_{\nu\mu} j^{\mu} - F^{n}_{\nu\mu} \epsilon^{\mu} \rb 
    - \tau^{\mu\lambda} h_{\lambda\nu} \nabla_{\mu} v^{\nu } 
    \nn\\  
    &\qquad
    - K_I v^{\mu} e^I_\mu 
    - \half U^{IJ } u^{\mu}_\phi \partial_\mu\psi_{IJ} \nn\\
    &\qquad 
    - \ell' L_I v^\mu \dow_\mu \Phi^I
    ~~,  \\ 
    \lb \nabla_{\mu} + F^{n}_{\mu\lambda} v^{\lambda}\rb \lb v^{\mu }  \pi^{\nu}  + \tau^{\mu \nu } \rb \hspace{-9.2em}& \nn\\
    &=   h^{\nu\lambda} \lb F_{\lambda\mu} j^{\mu} - F^{n}_{\lambda\mu} \epsilon^{\mu} \rb
    - \pi^{\mu} \nabla_{\mu} v^{\nu} \nn\\  
    &\qquad
    + K_I h^{\nu\mu} e^I_\mu
    + \ell' L_I h^{\nu\mu} \dow_\mu \Phi^I
    ~~,
 \\ 
   \lb\nabla_{\mu} + F^{n}_{\mu\lambda} v^{\lambda}\rb  j^\mu 
   = 0~~. \hspace{-7.2em} & 
    \end{align}
\end{subequations}
Here $F_{\mu\nu} = 2\dow_{[\mu}A_{\nu]}$ and $F^n_{\mu\nu} = 2\dow_{[\mu} n_{\nu]}$. The covariant derivative $\nabla_\mu$ is defined with respect to the torsional Aristotelian connection
\begin{equation}
    \Gamma^\lambda_{\mu\nu}
    = v^\lambda \dow_\mu n_\nu 
    + \half h^{\lambda\rho} 
    \lb \dow_\mu h_{\nu\rho} + \dow_\nu h_{\mu\rho} - \dow_\rho h_{\mu\nu} \rb.
\end{equation}
Restricting the background spacetime sources to their flat form in \eqref{eq:flat-sources},
the conservation laws reduce to their flat space versions given in \eqref{eq:conservation} and \eqref{eq:conservation-pinning}.

To derive the configuration equations, we need to consider the coupling of the action to background sources like \eqref{eq:freeEnergy}. However, in the dynamical case, we can also introduce an external momentum source $\Pi_\mu^\ext$ for the crystal velocity $u_\phi^\mu$ in addition to an external stress source $T^{\mu\nu}_\ext$ for the strain tensor $\kappa_{\mu\nu} = e_\mu^I e_\nu^J \kappa_{IJ}$. We take the source action to be
\begin{equation}
    S_{\text{source}} = \int \df t \df^d x \sqrt{\gamma} \bigg[
    T^{\mu\nu}_\ext \kappa_{\mu\nu} + \Pi_\mu^\ext u_\phi^\mu
    \bigg].
\end{equation}
These couplings are, of course, manifestly invariant under spacetime diffeomorphisms and gauge transformations. Variations of the source action can be obtained to be
\begin{align}\label{eq:extsources}
    \delta S_{\text{source}} 
    &= \int \df t \df^d x \sqrt{\gamma} \Bigg[
    \kappa_{\mu\nu} \delta T^{\mu\nu}_\ext + u_\phi^\mu \delta \Pi_\mu^\ext \nn\\
    &\qquad 
    + K_I^\ext \delta\phi^I 
    + \half U^{IJ}_\ext \lb \delta\psi_{IJ}
    - e_K^\mu \dow_\mu \psi_{IJ} \delta \phi^K \rb \nn\\
    &\hspace{-2em}
    - \lb 
    \bar T^{\mu\rho}_\ext h_{\rho\nu} u^\nu_\phi
    - T^{\rho\sigma}_\ext \kappa_{\rho\sigma} v^\mu
    + \Pi_\rho^\ext u_\phi^\rho (u^\mu_\phi - v^\mu) \rb 
     \delta n_\mu  \nn\\
    &\hspace{-2em}
    + \lb \bar T^{\mu\nu}_\ext + \lb T^{\rho\sigma}_\ext \kappa_{\rho\sigma}
    + \Pi_\rho^\ext u_\phi^\rho \rb h^{\mu\nu} \rb
    \half \delta h_{\mu\nu}
    \Bigg],
\end{align}
where $\bar T^{\mu\nu}_\ext = T^{\rho\sigma}_\ext 
(\delta_\rho^\mu - n_\rho u^\mu_\phi)(\delta_\sigma^\nu - n_\sigma u^\nu_\phi)$ is the external stress tensor projected against the crystal velocity and we have defined 
\begin{align}
    U^{IJ}_\ext 
    &= - \ell T^{\mu\nu}_\ext e_\mu^I e_\nu^J, \nn\\
    K_I^\ext 
    &= \lb \nabla_\mu + F^n_{\mu\lambda} v^\lambda \rb 
    \lb T^{\mu\nu}_\ext e_\nu^J \bbh_{JI}
    + \Pi_\nu^\ext e^\nu_{I} u^\mu_\phi 
    \rb \nn\\
    &\qquad 
    + \half U^{JK}_\ext e_I^\mu \dow_\mu \psi_{JK}~.
\end{align}
The terms in the last two lines of \eqref{eq:extsources} tell us how the definitions of the conserved currents obtained by varying the total action $S + S_{\text{source}}$ with respect to the background fields gets contributions from the crystal sources. These will play a role in the computation of correlation functions in \ref{app:correlation}. The terms in the second line tell us that the equations of motion for $\phi^I$ and $\psi_{IJ}$ are given by \eqref{eq:config-equations}. Finally, the terms in the first line are the coupling terms for the strain and crystal velocity operators.


\subsection{Covariant viscoplastic hydrodynamics}

We now redo the second law analysis for viscoplastic hydrodynamics from section \ref{sec:viscoplastic-hydro} and for pinned viscoplastic hydrodynamics from section \ref{sec:pinned-consti}, keeping track of all the background fields. We define the covariant fluid velocity $u^\mu$, normalised as $u^\mu n_\mu = 1$. We can isolate the spatial components of the fluid velocity with respect to the background frame velocity as $\vec u^\mu = u^\mu - v^\mu$, with $\vec u_\mu = h_{\mu\nu}\vec u^\nu$. The thermodynamic relations take the same form as \eqref{eq:thermodynamics} and \eqref{eq:pinning-pressure}, i.e.
\begin{align}
    \df\epsilon
    &= T\df s + \mu\df n + \frac{1}{2\rho} \df\vec\pi^2 
    - \half r_{IJ} \df h^{IJ}
    - \half \bbr^{IJ} \df \bbh_{IJ} \nn\\
    &\qquad
    + \ell'^2 m^2 h_{IJ} (\phi^J-\Phi^J)\, \df (\phi^I-\Phi^I)
    , \nn\\
    \df p
    &= s\df T + n\df\mu + \half \rho \df \vec u^2 
    + \half r_{IJ} \df h^{IJ}
    + \half \bbr^{IJ} \df \bbh_{IJ} \nn\\
    &\qquad 
    - \ell'^2 m^2 h_{IJ} (\phi^J-\Phi^J)\, \df (\phi^I-\Phi^I), \nn\\
    \epsilon
    &= -p + Ts + \mu n + u^\mu \pi_\mu.
\end{align}
The momentum density related to the fluid velocity as $\pi_\mu = \rho \vec u^\nu$. We have also used $\vec\pi^2 = \pi_\mu\pi_\nu h^{\mu\nu} = \rho^2\vec u^2$.

The statement of the second law of thermodynamics in the covariant language is that there must exist an entropy current $s^\mu$ such that 
\begin{equation}
    \lb\nabla_{\mu} + F^{n}_{\mu\lambda} v^{\lambda}\rb s^\mu = \Delta \geq 0.
\end{equation}
We start with an appropriately covariantised parametrisation of the constitutive relations \eqref{eq:ideal-consti} and \eqref{eq:consti-pinned}, i.e.
\begin{align}
    \epsilon^\mu 
    &= \epsilon u^\mu + p\,\vec u^\mu 
    + r_{IJ} e^{I\mu} v^\nu e^{J}_\nu
    + {\cal T}^{\mu\nu}\vec u_\nu + {\cal E}^\mu, \nn\\
    \tau^{\mu\nu}
    &= \rho\vec u^\mu \vec u^\nu + p\,h^{\mu\nu}
    - r_{IJ} e^{I\mu} e^{J\nu}
    + {\cal T}^{\mu\nu}, \nn\\
    j^\mu 
    &= n u^\mu + {\cal J}^\mu, \nn\\
    K_I 
    &= - \lb\nabla_{\mu} + F^{n}_{\mu\lambda} v^{\lambda}\rb\!\lb r_{IJ} e^{J\mu} \rb 
    + \frac{\ell}{2} \bbr^{JK} e_I^\mu \dow_\mu\psi_{JK} \nn\\
    &\qquad 
    - \ell'^2 m^2 h_{IJ}\lb\phi^J-\Phi^J\rb
    + {\cal K}_I, \nn\\
    U^{IJ}
    &= \ell \bbr^{IJ} + {\cal U}^{IJ}, \nn\\
    L_I &= \ell' m^2 h_{IJ}\lb\phi^J-\Phi^J\rb
    + {\cal L}_I~~,
\end{align}
where  and $e^{I\mu} = h^{\mu\nu}e^{I}_\nu$. The covariant parametrisation of the constitutive relations has been motivated from \cite{Armas:2020mpr}.
All the dissipative corrections are defined to be transverse to $n_\mu$. Let us pretend that the entropy current is given by the ideal form $s u^\mu$. By a straight-forward computation, we can find that
\begin{align}
    T\lb\nabla_{\mu} + F^{n}_{\mu\lambda} v^{\lambda}\rb (s u^\mu) \hspace{-8em}& \nn\\
    %
    &= 
    - {\cal E}^\mu \lb \frac1T \dow_\mu T - u^{\nu} F^{n}_{\nu\mu} \rb \nn\\
    &\qquad
    - {\cal T}^{\mu\nu} h_{\nu\rho} 
    \lb \nabla_\mu u^\rho
    - u^\rho u^{\lambda} F^{n}_{\lambda\mu} 
    \rb \nn\\
    &\qquad 
    - {\cal J}^\mu \lb T \dow_\mu \frac{\mu}{T} + u^\nu F_{\nu\mu} \rb \nn\\
    &\qquad
    - {\cal K}_I u^\nu e^I_\nu
    - \half {\cal U}^{IJ } u^{\mu}_\phi \partial_\mu\psi_{IJ}
    - \ell' {\cal L}_I u^\mu \dow_\mu \Phi^I
    \nn\\
    &\qquad 
    - T \lb\nabla_{\mu} + F^{n}_{\mu\lambda} v^{\lambda}\rb\! \lb \frac1T {\cal E}^\mu
    - \frac{\mu}{T} {\cal J}^\mu \rb.
\end{align}
From here, we can read out the corrected entropy current 
\begin{equation}
    s^\mu 
    = s u^\mu + \frac1T {\cal E}^\mu
    - \frac{\mu}{T} {\cal J}^\mu,
\end{equation}
while the dissipation rate is given as 
\begin{align}
    T\Delta
    &= - {\cal E}^\mu \lb \frac1T \dow_\mu T - u^{\nu} F^{n}_{\nu\mu} \rb \nn\\
    &\qquad
    - {\cal T}^{\mu\nu} h_{\nu\rho} 
    \lb \nabla_\mu u^\rho
    - u^\rho u^{\lambda} F^{n}_{\lambda\mu} 
    \rb \nn\\
    &\qquad 
    - {\cal J}^\mu \lb T \dow_\mu \frac{\mu}{T} + u^\nu F_{\nu\mu} \rb \nn\\
    &\qquad
    - {\cal K}_I u^\nu e^I_\nu
    - \half {\cal U}^{IJ } u^{\mu}_\phi \partial_\mu\psi_{IJ} \nn\\
    &\qquad 
    - \ell' {\cal L}_I u^\mu \dow_\mu \Phi^I,
\end{align}
which generalises \eqref{eq:dissipationrate} and \eqref{eq:entropy-production-pinning} to curved spacetime backgrounds. The discussion for one-derivative order constitutive relations in section \ref{sec:constitutive} and \ref{sec:pinned-consti} can be generalised to curved background by simply replacing
\begin{align}
    e^{Ii} \frac{1}{T} \partial_i T
    &\to e^{I\mu}\lb \frac1T \dow_\mu T - u^{\nu} F^{n}_{\nu\mu} \rb, \nn\\
    \dow_k u^k 
    &\to \lb \nabla_\mu 
    + F^{n}_{\mu\lambda} v^\lambda \rb u^\mu , \nn\\
    2 e^{\langle I}_i e^{J\rangle}_j \dow^i u^j 
    &\to 2 e^{\langle I\mu} e^{J\rangle\nu} h_{\nu\rho} 
    \lb \nabla_\mu u^\rho
    - u^\rho u^{\lambda} F^{n}_{\lambda\mu} 
    \rb, \nn\\
    e^{Ii} T\partial_i \frac{\mu}{T}
    &\to e^{I\mu}\lb T \dow_\mu \frac{\mu}{T} + u^\nu F_{\nu\mu} \rb, \nn\\
    (\dow_t + u^i\dow_i)\phi^I
    &\to u^\mu \dow_\mu \phi^I, \nn\\
    (\dow_t + u^i_\phi \dow_i) \psi_{IJ}
    &\to u^\mu_\phi \dow_\mu \psi_{IJ}, \nn\\
    (\dow_t + u^i\dow_i)\Phi^I
    &\to u^\mu \dow_\mu \Phi^I,
\end{align}
together with the parametrisation of the derivative corrections ${\cal E}^\mu = e^\mu_I {\cal E}^I$, ${\cal T}^{\mu\nu} = e^\mu_I e^\nu_J {\cal T}^{IJ}$, and ${\cal J}^\mu = e^\mu_I {\cal J}^I$.

\subsection{Without energy sources}

Let us record the version of the discussion above when the clock form $n_\mu$ is fixed to its flat spacetime form. This will allow us to compute correlations of all observables except the energy current, which we have ignored in the core of our paper as we neglected temperature fluctuations. Setting $n_\mu \df x^\mu = \df t$ means that
\begin{align}
    v^\mu \dow_\mu 
    &= \dow_t + v^i\dow_i, \nn\\ 
    h_{\mu\nu}\df x^\mu\df x^\nu 
    &= \vec v^2\df t^2 - 2v_i \df x^i\df t + g_{ij}\df x^i \df x^j, \nn\\
    h^{\mu\nu} \dow_\mu\otimes\dow_\nu 
    &= g^{ij} \dow_i\otimes \dow_j,
\end{align}
where $\vec v^2 = g_{ij}v^iv^j$. The spatial indices in the following are raised/lowered using the spatial metric $g_{ij}$ and its inverse $g^{ij}$. The coupling structure \eqref{eq:actionVariation} gives rise to
\begin{align}
    \delta S   & =  \int\!\df t  \df^{d } x \sqrt{g} \bigg[  
    - \pi_i\delta v^i
    + \frac{1}{2}  \tau^{ij} \delta h_{ij} 
    + n \delta A_{t} 
    + j^{i} \delta A_{i} 
    \nn\\  
    &\qquad\qquad
    + K_{I} \delta\phi^I 
     + \half U^{IJ} \lb \delta\psi_{IJ} - e_K^k \dow_k \psi_{IJ} \delta \phi^K \rb \nn\\
     &\qquad\qquad
     + \ell'L_{I} \delta\Phi^I 
    \bigg].
\end{align}
Since only the spatial components of the strain tensor $\kappa_{ij}$ and crystal velocity $u^i_\phi$ are independent, we can switch off the time-components of their respective sources $T^{tt}_\ext$, $T^{ti}_\ext$, $\Pi_t^\ext$ for simplicity. The variation of the source action in \eqref{eq:extsources} then becomes
\begin{align}
    \delta S_{\text{source}} 
    &= \int \df t \df^d x \sqrt{\gamma} \Bigg[
    \kappa_{ij} \delta T^{ij}_\ext + u_\phi^i \delta \Pi_i^\ext \nn\\
    &\qquad 
    + K_I^\ext \delta\phi^I 
    + \half U^{IJ}_\ext \lb \delta\psi_{IJ}
    - e_K^k \dow_k \psi_{IJ} \delta \phi^K \rb \nn\\
    &\hspace{-2em}
    + \lb T^{ij}_\ext + \lb T^{kl}_\ext \kappa_{kl}
    + \Pi_k^\ext u_\phi^k \rb g^{ij} \rb
    \half \delta g_{ij}
    \Bigg],
\end{align}
where the definitions of $U^{IJ}_\ext $ and $K_I^\ext$ become 
\begin{align}
    U^{IJ}_\ext 
    &= - \ell T^{ij}_\ext e_i^I e_j^J, \nn\\
    K_I^\ext 
    &= \frac{1}{\sqrt g}\dow_t
    \lb \sqrt{g}\,\Pi_i^\ext e^i_{I} \rb 
    + \nabla_i\!\lb T^{ij}_\ext e_j^J \bbh_{JI}
    + \Pi_k^\ext e^k_{I} u^i_\phi 
    \rb \nn\\
    &\qquad 
    + \half U^{JK}_\ext e_I^k \dow_k \psi_{JK}~.
\end{align}



The conservation equations, on the other hand, reduce to the non-covariant form
\begin{subequations}
\begin{align}  
    \frac{1}{\sqrt{g}}\dow_t (\sqrt{g}\,\epsilon)
    + \nabla_i\epsilon^i \hspace{-6em}& \nn\\
    &= E_i \lb j^{i} - n v^i \rb
    - \tau^{ij} \lb \nabla_{i} v_j + {\textstyle\half}\dow_t g_{ij} \rb \nn\\  
    &\qquad
    - K_I \lb e^I_t + v^ie^I_i \rb 
    - \half U^{IJ } (\partial_t+u^i_\phi\dow_i)\psi_{IJ} \nn\\
    &\qquad 
    - \ell' L_I \lb \dow_t\Phi^I + v^i \dow_i\Phi^I \rb~~,  \\ 
    \frac{1}{\sqrt{g}}\dow_t(\sqrt{g}\, \pi_i)
    + \nabla_j \lb v^{j} \pi_i  + \tau^{j}_{~i} \rb 
    \hspace{-11em}& \nn\\
    &= E_i n
    +  F_{ij} (j^{j} - n v^j)
    - \pi_k \nabla_i v^k
    \nn\\  
    &\qquad
    + K_I e^{I}_i
    + \ell' L_I \dow_i\Phi^I
    ~~,
 \\ 
   \frac{1}{\sqrt{g}}\dow_t (\sqrt{g}\,n)
    + \nabla_i j^i
    = 0~~. \hspace{-9em}&
    \end{align}
\end{subequations}
Here $E_i = F_{it} + F_{ij}v^j$ is the electric field with respect to the background frame velocity $v^i$.
Also, $\nabla_i$ is the spatial covariant derivative with respect to $g_{ij}$. We have also identified the densities $\epsilon = \epsilon^\mu n_\mu$ and $n = j^\mu n_\mu$. We can see that the energy fluctuations decouple from the momentum and density fluctuations and thus can be ignored in the isothermal limit.

The constitutive relations are given now given in their non-covariant form
\begin{align}
    \epsilon^i
    &= \epsilon\, u^i + p\,\vec u^i
    + r_{IJ} e^{Ii} \lb e^{J}_t + v^k e^J_k \rb
    + {\cal T}^{ij}\vec u_j + {\cal E}^i, \nn\\
    \tau^{ij}
    &= \rho\vec u^i \vec u^j + p\,h^{ij}
    - r_{IJ} e^{Ii} e^{Jj}
    + {\cal T}^{ij}, \nn\\
    j^i
    &= n u^i + {\cal J}^i, \nn\\
    K_I 
    &= - \nabla_i\!\lb r_{IJ} e^{Ji} \rb 
    + \frac{\ell}{2} \bbr^{JK} e_I^k \dow_k\psi_{JK} \nn\\
    &\qquad 
    - \ell'^2 m^2 h_{IJ}\lb\phi^J-\Phi^J\rb
    + {\cal K}_I, \nn\\
    U^{IJ}
    &= \ell \bbr^{IJ} + {\cal U}^{IJ}, \nn\\
    L_I &= \ell' m^2 h_{IJ}\lb\phi^J-\Phi^J\rb
    + {\cal L}_I~~.
\end{align}
Note that $u^i$ and $\vec u^i$ are generically different when the frame velocity $v^i\neq 0$, i.e. $\vec u^i = u^i - v^i$. Both of these are appropriate notions of fluid velocity. On the one hand, $u^i$ is aligned parallel to the (ideal order) particle flux, while $\vec u^i$ is aligned parallel to the flow of momentum $\pi_i = \rho \vec u_i$. The constitutive relations in section \ref{sec:constitutive} and \ref{sec:pinned-consti} can be generalised to this particular background by replacing 
\begin{align}
    \dow_k u^k 
    &\to \nabla_k u^k + \half g^{ij}\dow_t g_{ij}, \nn\\
    2 e^{\langle I}_i e^{J\rangle}_j \dow^i u^j 
    &\to 2 e^{\langle I i} e^{J\rangle j} \lb \nabla_i u_j + \half \dow_t g_{ij} \rb, \nn\\
    e^{Ii} T\partial_i \frac{\mu}{T}
    &\to e^{Ii}\lb T \dow_i \frac{\mu}{T} - E_i - F_{ij} \vec u^j \rb.
\end{align}

\subsection{Correlation functions via background variation}
\label{app:correlation}

We can compute the retarded correlation functions of hydrodynamic observables by inspecting how they respond to changes in the associated sources. We will ignore the energy density and flux for clarity, although the procedure works exactly the same as follows. First, we need the one point functions of various operators in the presence of background sources. These can be obtained by taking variations of the total onshell action $S_{\text{tot}} = (S + S_{\text{source}})_{\text{onshell}}$ with respect to the appropriate background fields. We find that
\begin{align}
    \langle \pi_i\rangle 
    &= - \frac{\delta S_{\text{onshell}}}{\delta v^i}
    = \sqrt{g}\,\pi_i, \nn\\
    \langle \tau^{ij} \rangle
    &= 2\frac{\delta S_{\text{onshell}}}{\delta g_{ij}}
    = \sqrt{g}\Big( \tau^{ij} + T^{ij}_\ext \nn\\
    &\hspace{5em}
    + T^{kl}_\ext \kappa_{kl} g^{ij}
    + \Pi_k^\ext u^k_\phi g^{ij} \Big), \nn\\
    \langle n \rangle
    &= \frac{\delta S_{\text{onshell}}}{\delta A_t}
    = \sqrt{g}\, n, \nn\\
    \langle j^i \rangle
    &= \frac{\delta S_{\text{onshell}}}{\delta A_i}
    = \sqrt{g}\,j^i, \nn\\
    \langle \kappa_{ij} \rangle
    &= \frac{\delta S_{\text{onshell}}}{\delta T^{ij}_\ext}
    = \sqrt{g}\,\kappa_{ij}, \nn\\
    \langle u^i_\phi \rangle
    &= \frac{\delta S_{\text{onshell}}}{\delta \Pi_{i}^\ext}
    = \sqrt{g}\, u^i_\phi.
\end{align}
All the objects on the right should be understood as evaluated on the solutions of the equations of motion, so they only depend on the background sources. In the absence of sources, they reduce to the equilibrium values of the respective operators. Note that for a Galilean system, even when the crystal sources are absent, the particle flux $j^i$ is the same as the momentum density $\pi_i$ only up to the frame velocity $v^i$ (in addition to the metric factor). This difference manifests itself as slight differences in their correlation functions.

Using the short-hand notation for the operators and respective sources
\begin{align}
    {\cal O} 
    &= \Big(\pi_i,\tau^{ij}, n, j^i, \kappa_{ij}, u^i_\phi \Big)   , \nn\\
    J 
    &= \Big( {-v^i},{\textstyle\half} g_{ij}, A_t, A_i, T^{ij}_\ext, \Pi_i^\ext\Big),
\end{align}
the retarded correlation functions can be read off as 
\begin{equation}
    G^R_{{\cal O}_1{\cal O}_2}
    = - \frac{\delta \langle {\cal O}_1\rangle }{\delta J_2} \bigg|_{\text{flat}}.
\end{equation}
This procedure has been used to compute the correlation functions reported in sections \ref{sec:correlations} and \ref{sec:correlations-pinned}.

If the microscopic description underlying the crystal under consideration is invariant under time-reversal symmetry, the correlation functions in momentum space satisfy the so-called Onsager reciprocal relations
\begin{equation}
    G^R_{{\cal O}_1{\cal O}_2}(\omega,k)
    = \eta^T_1\eta_2^T G^R_{{\cal O}_2{\cal O}_1}(\omega,-k),
    \label{eq:onsager-formula}
\end{equation}
where $\eta^T=(-,+,+,-,+,-)$ are the time-reversal eigenvalues of the respective operators. These lead to the relations between the primed and un-primed off-diagonal dissipative coefficients in \eqref{eq:onsager} and \eqref{eq:onsager-pinning}. These constraints can be adapted to alternative discrete symmetries that the crystal might possess, like CPT, by using the appropriate eigenvalues of operators in \eqref{eq:onsager-formula}.

\section{Material diagrams for total matter displacement}
\label{app:diagramsprocess}

In this appendix, we continue the discussion initiated in section \ref{sec:rheology} regarding the type of materials described by our framework. 
As we noted towards the end of that section, the form of the rheology equations depend on the observable being probed, e.g. displacements of just the crystalline matter or displacements of the total matter including interstitials. In view of the fact that the distinction between the crystalline and interstitial matter in a plastic crystal is a bit artificial because the two are not individually conserved, let us look at the rheology equations from the perspective of the total matter evolution, crystal plus interstitials. To this end, we define the shear tensor with respect to the particle flux
\begin{align}
    K_{ij} 
    &= \frac{1}{n} \dow_{(i} j_{j)} 
    = \dot\varepsilon_{ij} + \frac{1}{n} \dow_{(i} j^\Delta_{j)} \nn\\
    &= \dot\varepsilon_{ij} 
    - D_\Delta^\| \dow_i\dow_j \kappa^k_{~k} \nn\\
    &\qquad 
    - 2 D_\Delta^\perp\!
    \lb \dow_{(i}\dow^k \kappa_{j)k} {\,-\,} \dow_i\dow_j \kappa^k_{~k}  \rb.
\end{align}
For a Galilean crystal, this object is the same as the fluid shear tensor $\dow_{(i}u_{j)}$ and serves as the equivalent of $\dot\varepsilon_{ij}$ for total matter evolution. Note that these expressions are only exact linearly and when both $\mu$ and $T$ are fixed to their equilibrium values. We can now express the rheology equations \eqref{eq:rheocrystal} in terms of $K_{ij}$ instead of $\dot\varepsilon_{ij}$. We find
\begin{subequations}
    \begin{align}
    \tau_{ij} 
    &= - \delta_{ij} \lb B \kappa^k_{~k} 
    + \zeta K^k_{~k}  \rb
    \nn\\
    &\quad 
    - 2G \kappa_{\langle ij\rangle}
    - 2\eta K_{\langle ij\rangle}, 
    \label{eq:rheo-1-j} \\
    \dot\kappa_{ij}
    &= \frac{1}{d} \delta_ {ij}\lb K^k_{~k}
    - \Omega_B \kappa^k_{~k} \rb \nn\\
    &\quad 
    + K_{\langle ij\rangle} 
    - \Omega_G \kappa_{\langle ij\rangle}~~ \nn\\
    &\quad 
    + D_\Delta^\| \dow_{i}\dow_{j} \kappa^k_{~k}
    + 2D_\Delta^\perp\!
    \lb \dow_{i}\dow^k \kappa_{jk} {\,-\,}
    \dow_{i}\dow_{j} \kappa^k_{~k}  \rb,
    \label{eq:rheo-2-j}
\end{align}
\label{eq:rheo-21}%
\end{subequations}
where we have ignored subleading terms in the derivative expansion as well as the thermodynamic pressure, which becomes a constant when $T$ and $\mu$ are frozen.

Due to the presence of diffusive corrections, the rheology equations \eqref{eq:rheo-21} do not admit a neat split into shear and bulk sectors. Instead, to be able to draw the respective material diagrams, we need to perform a wavevector decomposition of $K_{ij}$ and $\tau_{ij}$ as follows
\begin{align} \label{feuehiu200929009}
    K_{ij} 
    &= \frac{k_ik_j}{k^2} K^\| + \frac{k_{(i}}{k} K_{j)}^\perp, \nn\\
    \tau_{ij} 
    &= \frac{k_ik_j}{k^2} \tau^\| + 2\frac{k_{(i}}{k} \tau_{j)}^\perp
    + \lb \delta_{ij} - \frac{k_ik_j}{k^2} \rb \tau^{\sfT}
    + \tau^{\sfTT}_{ij}.
\end{align}
The transverse vector components $K_{i}^\perp$, $\tau_{i}^\perp$ are transverse to $k^i$, while $\tau^\sfTT_{ij}$ is transverse and traceless. Note that $K_{ij}$ is a pure gradient and does not get purely transverse contributions. In addition $K^\|$ and $\tau^\|$ are the longitudinal parts of $K_{ij}$ and $\tau_{ij}$ respectively, while $\tau^{\sfT}$ is the transverse trace of $\tau_{ij}$. Note that the crystal strain tensor $\kappa_{ij}$ was already decomposed in this format in~\eqref{eq:kappa-k-decomposition}.

Given this wavevector decomposition, we find that the transverse traceless sector decouples trivially with 
\begin{align} \label{eq:tensorequations}
    \tau_{ij}^\sfTT
    &= - 2G \kappa_{ij}^\sfTT, \nn\\
    \dot\kappa_{ij}^\sfTT
    &= - \Omega_G \kappa^\sfTT_{ij}, \nn\\
    \implies
    \dot \tau_{ij}^\sfTT
    &+ \Omega_G \tau^\sfTT_{ij} 
    = 0~,
\end{align}
and has no contribution to the matter distortion strain.  

\begin{figure*}[t]
	\centering
    \subfigure[Circuit representation of the vector sector \eqref{eq:vectordecomposition}. The dashpot component in the lower arm measures $K_{i}^{\perp}$. The dashpot component in the top elastic arm has $k$-dependence. \label{fig:vector}]
    {\includegraphics[width=0.49\linewidth]{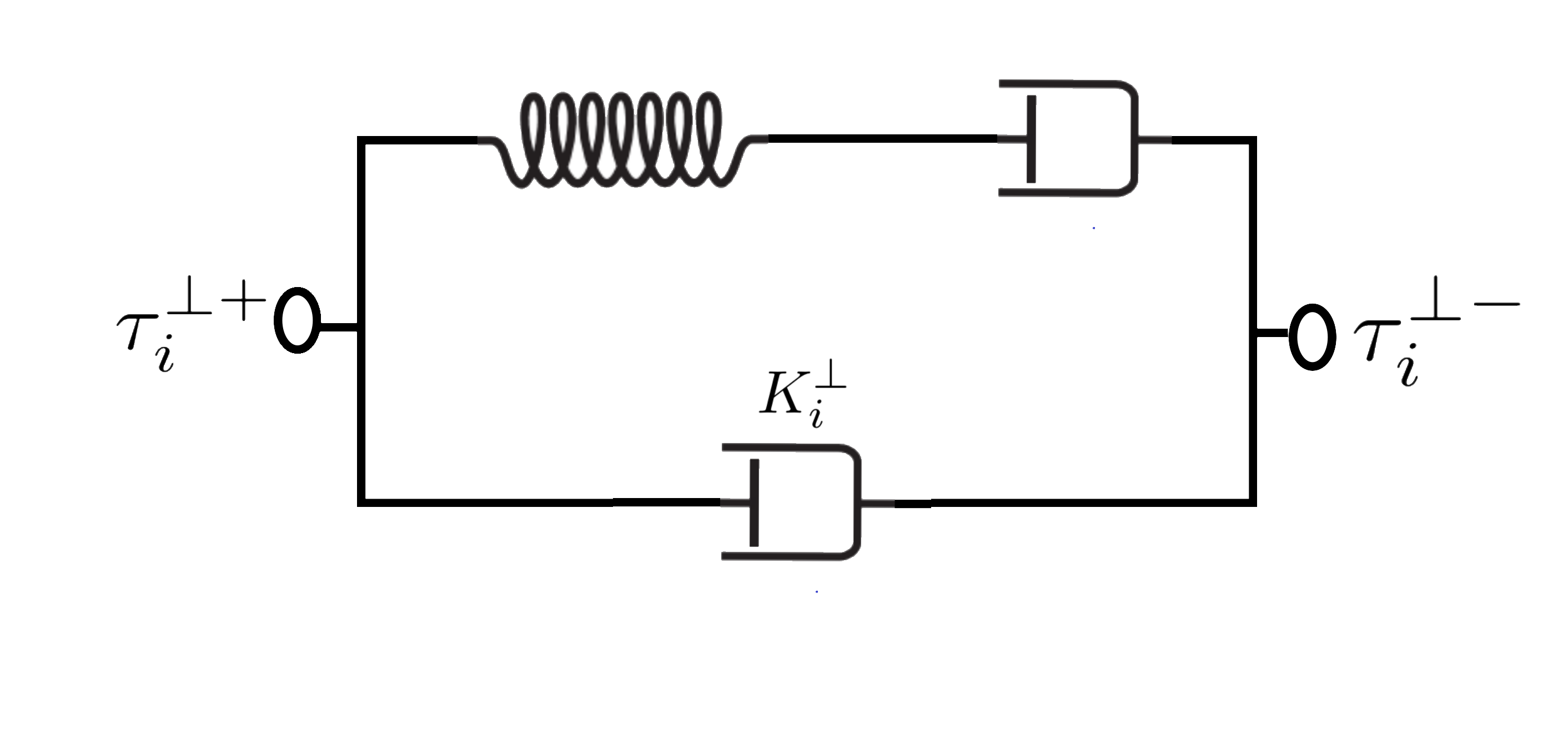}}
     \hfill
    \subfigure[Circuit representation of the scalar sector \eqref{eq:scalarsector}. The shared dashpot component measures $K^{\parallel}$ and connects two non-parallel circuits. All the other components have $k$-dependence. \label{fig:scalar}]
    {\includegraphics[width=0.49\linewidth]{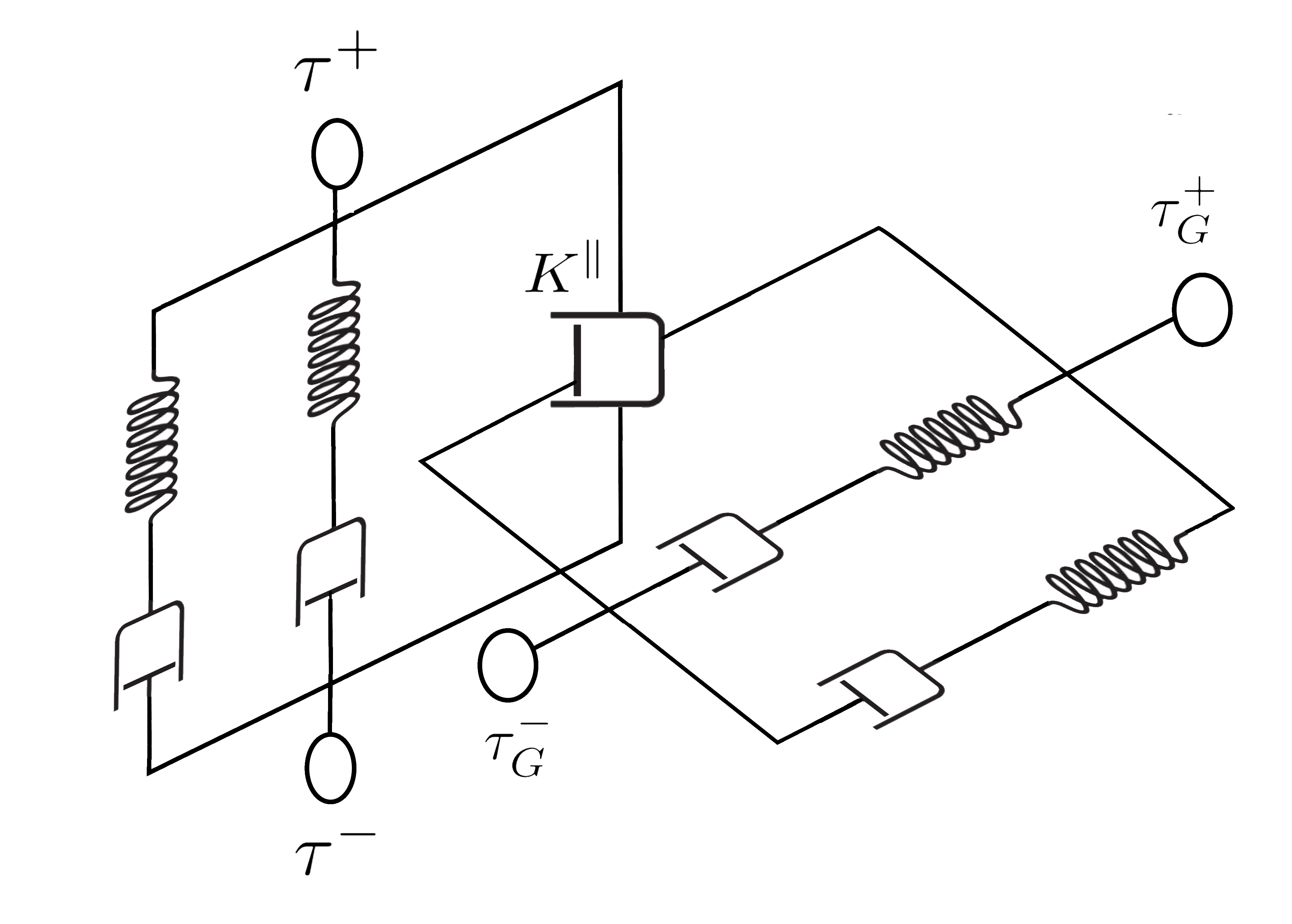}}
	\caption{Circuit representation of the rheology equations from the perspective of total matter displacement.}
	\label{fig_stability228}
\end{figure*}

Next, we have the vector sector with
\begin{align} \label{eq:vectordecomposition}
    \tau^\perp_i
    &= - G \kappa^\perp_i
    - \eta K^\perp_i \nn\\
    \dot\kappa^\perp_i
    &= K^\perp_{i}
    - \lb \Omega_G
      + k^2 D_\phi^\perp \rb \kappa^\perp_i, \nn\\
  \implies 
  \dot\tau^\perp_i 
      &+ \lb \Omega_G
      + k^2 D_\Delta^\perp \rb \tau^\perp_i
    =  \nn\\
    &\hspace{-1em}
    - \lb \lambda_G^2 G + \eta \lb \Omega_G
    + k^2 D_\Delta^\perp \rb \rb \dot K^\perp_i
    - \eta \ddot K^\perp_i,
\end{align}
which behaves like a ``$k$-dependent'' Jeffrey material, with the relaxation rate $\Omega_G$ replaced with $\Omega_G  + k^2 D_\Delta^\perp$. We have given the corresponding material diagram in figure~\ref{fig:vector}. In detail, we can label the spring and dashpot components in the top elastic arm of the diagram by $E_1$ and $\upsilon_1$ respectively, while the dashpot in the bottom arm by $\upsilon_2$. Using circuit rules, the diagram gives rise to the stress-strain relations
\begin{equation}
    \dot\tau^\perp_i + \frac{E_1}{\upsilon_1} \tau^\perp_i
    = - E^\perp_1\lb 1 + \frac{\upsilon_2}{\upsilon_1} \rb K^\perp_i
    - \upsilon_2\dot K^\perp_i.
\end{equation}
We can compare this to \eqref{eq:vectordecomposition} to read off 
\begin{equation}
    E_1 = G, \qquad 
    \upsilon_1 = \frac{G}{\Omega_G + k^2 D^\perp_\Delta}, \qquad 
    \upsilon_2 = \eta.
\end{equation}
This is the usual Jeffrey's model with $k$-dependence in one of the dashpots.
One learns that the effect of Goldstone (interstitial) diffusion in the rheology equation is that the dashpot component in the elastic arm of the circuit representation gets a wavenumber dependence.

Finally, the diffusive corrections non-trivially couple the two scalar sectors. It is neater to work in a basis with isotropic trace stress $\tau =   \frac{1}{d} \tau^\| +  \frac{d-1}{d}\tau^\sfT$ and scalar shear stress $\tau_G = \frac{1}{2} \tau_\| - \frac{1}{2} \tau^\sfT$, and similarly for the strain tensor $\kappa = \kappa^\| + (d-1) \kappa^\sfT$ and $\kappa_G = \kappa_\| - \kappa^\sfT$. In the trace sector, we find
\begin{subequations}
\begin{align} \label{eq:tauwithoutGequations}
    \tau
    &=  - B \kappa  -  \zeta K^\|, \nn\\
    \dot\kappa 
    &= K^\| 
     - \lb \Omega_B
     +  D_\Delta^\| k^2
     - {\textstyle 2\frac{d-1}{d}} D_\Delta^\perp k^2
     \rb \kappa \nn\\
     &\qquad
    - {\textstyle 2\frac{d-1}{d}} D_\Delta^\perp k^2 \kappa_G.
\end{align}
On the other hand, in the scalar shear sector we have
\begin{align} \label{eq:tauwitheGquations}
    \tau_G 
    &= 
    - G \kappa_G
    - \eta K^\|,  \nn\\
    \dot\kappa_G
    &= K^\| 
    - \lb \textstyle \Omega_G
    + 2\frac{d-1}{d} D_\Delta^\perp k^2 \rb \kappa_G
    \nn\\
    &\qquad 
  - k^2
    \lb \textstyle D_\Delta^\|
    - 2\frac{d-1}{d} D_\Delta^\perp \rb \kappa~.
\end{align}
\label{eq:scalarsector}%
\end{subequations}
If we ignore the $k$-dependence in these equations for the moment, both these sectors behave as $k$-independent Jeffrey materials, but with the shared $K_\|$. This sharing is merely a manifestation of the fact that $K_{ij}$, like $\dot\varepsilon_{ij}$, is a pure gradient and has only one scalar component which couples to both the scalar components of the stress tensor $\tau$ and $\tau_G$. Restoring the $k$-dependence leads to additional non-trivial couplings between the two sectors due to the diffusive terms. After eliminating $\kappa$ and $\kappa_G$, we arrive at ``generalised Maxwell materials''~\cite{christensen1982xi} in both the sectors, still sharing the same $K_\|$, to wit
\begin{align} \label{eq:scalarsetofequations}
   \ddot{\tau}  + \Omega_1  \dot{\tau} + \Omega_2 \tau  
   &= 
    - \alpha_1  K^\| 
    - \alpha_2 \dot{K}^\|  
   - \zeta \ddot{K}^\|  ,   \nn \\ 
  \ddot{\tau}_G  + \Omega_1  \dot{\tau}_G + \Omega_2 \tau_G 
  &= 
    - \alpha_1^G  K^\| 
  - \alpha_2^G \dot{K}^\|  
  - \eta \ddot{K}^\|. 
\end{align}
The explicit expressions for the coefficients appearing here are given by
\begin{align}
   \Omega_1 & = \tilde \Omega_B +\tilde \Omega_G, \nn\\
   \Omega_2 &= \tilde \Omega_B  \tilde \Omega_G - X_BX_G ,  \nn\\   
   \alpha_1 & = \zeta  (\tilde \Omega_B  \tilde \Omega_G 
   - X_BX_G) 
   + B (\tilde\Omega_G-X_G) \nn\\
   \alpha_2 &= \zeta  (\tilde \Omega_B +\tilde \Omega_G ) + B  ,  \nn\\  
   \alpha_1^G &= \eta (  \tilde \Omega_B  \tilde \Omega_G 
   - X_B X_G)+ G (\tilde\Omega_B - X_B) \nn\\
    \alpha_2^G  &= \eta  (\tilde \Omega_B +\tilde \Omega_G )
        + G, 
\end{align}
where we defined
\begin{align}
   \tilde \Omega_B   
    &=  \Omega_B + 
    \left(D_\Delta^\|-2\frac{d-1}{d} D_\Delta^\perp\right)k^2   ~~ , \nn\\ 
    \tilde \Omega_G   
    &= \Omega_G
    + 2\frac{d-1}{d} D_\Delta^\perp k^2 ~~ , \nn\\ 
    X_B &= \left(D_\Delta^\|-2\frac{d-1}{d}D_\Delta^\perp\right)  k^2~~, \nn\\
    X_G &= 2\frac{d-1}{d} D_\Delta^\perp k^2  ~~. 
\end{align}
Note that all these coefficients have $k$-dependence.
The rheology equations \eqref{eq:scalarsetofequations} can be represented by the material diagram given in figure~\ref{fig:scalar}, involving two circuits for $\tau$ and $\tau_G$ that share a dashpot measuring $K^\|$. Focusing on the $\tau_G$ circuit, we can label the springs and dashpots in the two elastic arms by $E_1$, $E_2$ and $\upsilon_1$, $\upsilon_2$ respectively, while the dashpot shared with the $\tau$ circuit with $\upsilon_3$. We find the stress-strain relations 
\begin{align}
    \ddot\tau 
    &+ \lb \frac{E_1}{\upsilon_1} + \frac{E_2}{\upsilon_2} \rb \dot\tau 
    + \frac{E_1E_2}{\upsilon_1\upsilon_2} \tau  \nn\\
    &= - E_1E_2 \lb \frac{1}{\upsilon_1} + \frac{1}{\upsilon_2}
    + \frac{\upsilon_3}{\upsilon_1\upsilon_2}
    \rb K^\| \nn\\
    &\quad 
    - \lb E_1 + E_2
    + \upsilon_3\lb \frac{E_1}{\upsilon_1} + \frac{E_2}{\upsilon_2} \rb \rb \dot K^\| 
    - \upsilon_3 \ddot K^\|.
\end{align}
Comparing this expression to \eqref{eq:scalarsetofequations}, we can read off 
\begin{equation}
    \upsilon_3 = \eta, 
\end{equation}
while the explicit solution for the remaining coefficients $E_1$, $E_2$, $\upsilon_1$, $\upsilon_2$ can be also be obtained. The discussion for the $\tau$ sector proceeds similarly.

\section{Comparison to Zippelius et. al}
\label{app:Zpp}

In this appendix, we compare our results to the work of~\cite{1980PhRvB..22.2514Z}. Instead of the grand canonical ensemble, the authors in~\cite{1980PhRvB..22.2514Z} work in the canonical ensemble controlled by the temperature $T$, interstitial density $n_\Delta$, and momentum $\pi_i$. This is characterised by the free energy density
\begin{equation}
    {\cal F} = - p + (\mu-\mu_0)n + u^i\pi_i,
    \label{eq:zpp-free-energy}
\end{equation}
where we perform the canonical transformation with respect to $\mu-\mu_0$ instead of just $\mu$ to avoid generating a linear term in strain in the free energy.
Using the definition of interstitial density from \eqref{eq:interstitials}, 
we can check that
\begin{align}
    \df{\cal F} 
    &= - s \df T + (\mu-\mu_0) \df n_\Delta
    + u^i \df \pi_i \nn\\
    &\qquad 
    - \frac{1}{2} \lb r_{IJ} - (\mu-\mu_0)m_0 v h_{IJ} \rb \df h^{IJ}  \nn\\
    &\qquad 
    - \frac{1}{2} \lb {\mathbb r}^{IJ} - (\mu-\mu_0) m_0 v \bbh^{IJ} \rb \df \bbh_{IJ},
\end{align}
where $\bbh^{IJ}$ is the inverse of $\bbh_{IJ}$.
Plugging in the thermodynamic objects from \eqref{eq:linearised-thermo}, and specialising to the isothermal limit, we can linearly expand the differential of ${\cal F}$ to get
\begin{align}
    \df{\cal F} 
    &= \frac{1}{\chi}\delta n_\Delta \df n_\Delta  + \frac{1}{\rho} \pi^i \df \pi_i \nn\\
    &\qquad 
    - \frac{n+B \alpha_m}{\chi} \lb \kappa^k_{~k} \df n_\Delta
    + \delta n_\Delta \df \kappa^k_{~k} \rb \nn\\
    &~~
    + \lb B + \frac{(n+B \alpha_m)^2}{\chi} \rb \kappa^k_{~k} \df \kappa^k_{~k} \nn\\
    &~~
    + 2G \kappa^{\langle ij\rangle} \df \kappa_{\langle ij \rangle}.
\end{align}
We have used that, linearly, $\delta n = \chi\delta\mu + B\alpha_m\kappa^k_{~k}$, and $\delta n_\Delta = \delta n + n\kappa^k_{~k}$. We have also take the constant $m_0$  in the definition of interstitial density in \eqref{eq:interstitials} to be the equilibrium number density $n_0$. Comparing this to (2.4) of~\cite{1980PhRvB..22.2514Z}, we arrive at the identification of thermodynamic coefficients
\begin{gather}
    \hat n_0 = \rho, \qquad  
    \hat\chi_0 = \frac{\chi}{n^2}, \qquad 
    \hat\gamma_0 = -\frac{n(n+B \alpha_m)}{\chi}, \nn\\
    \hat\mu_0 = G, \qquad 
    \hat\lambda_0 + \hat\mu_0 = B + \frac{(n+B \alpha_m)^2}{\chi}.
\end{gather}
We have denoted the coefficients in~\cite{1980PhRvB..22.2514Z} with a ``hat''.
The first relation stems from the fact that~\cite{1980PhRvB..22.2514Z} works with a Galilean system, wherein the particle density is the same as the momentum susceptibility; we have taken the ``mass per unit particle'' proportionality factor $m$ in~\cite{1980PhRvB..22.2514Z} to be 1. 

To compare the dissipative coefficients, we need to perform a redefinition of the linearised crystal fields
\begin{equation}
    \delta\phi'^i = \delta\phi^i - a u^i,
\end{equation}
which results in a transformation of the strain tensor and crystal velocity as
\begin{align}
    \kappa'_{ij} &= \kappa_{ij} - a\dow_{(i} u_{j)}, \nn\\
    u'^i_\phi &= u^i_\phi - a\dow_t u^i \nn\\
    &= u^i_\phi + a\frac{n+B\alpha_m}{\rho} \dow^i\mu 
    - a \frac{B}{\rho} \dow^i\kappa^k_{~k} \nn\\
    &\qquad 
    - a \frac{2G}{\rho} \dow_j\kappa^{\langle ij\rangle}
    + {\cal O}(\dow^2).
\end{align}
The interstitial density and flux with respect to the transformed crystal fields are given as
\begin{align}
    n'_\Delta
    &= n_\Delta - an \dow_k u^k, \nn\\
    j'^i_\Delta 
    &= j^i_\Delta + an \dow_t u^i \nn\\
    &= j^i_\Delta - an\frac{n+B\alpha_m}{\rho} \dow^i\mu 
    + an \frac{B}{\rho} \dow^i\kappa^k_{~k} \nn\\
    &\qquad 
    + an \frac{2G}{\rho} \dow_j\kappa^{\langle ij\rangle}
    + {\cal O}(\dow^2).
\end{align}
To keep the thermodynamic relation $\delta n = \chi\delta\mu + B\alpha_m\kappa^k_{~k}$ intact, we also need to transform the chemical potential
\begin{equation}
    \delta\mu'
    = \delta\mu + \frac{aB\alpha_m}{\chi}\dow_i u^i.
\end{equation}

The authors in \cite{1980PhRvB..22.2514Z} work with Galilean-invariant crystals, for which we need to set $\rho = n$ and remove the dissipative transport coefficients $\sigma$ and $\gamma_n$ in order to make $j^i=\pi^i$. Having done this, we can compare the interstitial flux $j'^i_\Delta$ to the one obtained in \cite{1980PhRvB..22.2514Z}. The interstitial flux in \cite{1980PhRvB..22.2514Z} does not have any contributions from the derivatives of the strain tensor. This can be achieved by choosing
\begin{equation}
    a = \frac{\rho}{\sigma_\phi}~,
\end{equation}
which leads to
\begin{align} 
    j'^i_\Delta 
    &= - \frac{n^2}{\sigma_\phi} \dow^i\mu~.
    \label{eq:zzp-comparison-1}
\end{align}
The evolution equation for the transformed strain tensor can be read off using \eqref{eq:strain-evolution-linear}, leading to
\begin{align}
    \dot\kappa'_{ij}
    &= \dow_{(i}u_{j)}
    + \frac{n}{\sigma_\phi} \dow_i\dow_j\delta\mu'  \nn\\
    &\qquad
    - \frac{1}{d} \delta_{ij} \Omega_B\lb 
    \kappa'^k_{~k} - \alpha_m \delta\mu' \rb
    - \Omega_G\kappa'_{\langle ij\rangle}~.
    \label{eq:zzp-comparison-2}
\end{align}
We have set $\lambda_B=\lambda_G=1$ following our discussion in section \ref{eq:strain-redef}.
Finally, we can obtain the stress tensor in the transformed basis as
\begin{align}
    \tau^{ij} 
    &= \lb p + B\alpha_m \delta\mu' \rb \delta^{ij} 
    - B \delta^{ij} \kappa'^k_{~k}
    - 2G\kappa'^{\langle ij\rangle}
    \nn\\
    &\qquad 
    - \lb \zeta 
    + \frac{n B}{\sigma_\phi} \lb 1
    + \frac{\alpha_m}{\chi} (n+B\alpha_m) \rb \rb \delta^{ij} \dow_k u^k \nn\\
    &\qquad 
    - 2 \lb \eta + \frac{nG}{\sigma_\phi} \rb \dow^{\langle i} u^{j\rangle}.
    \label{eq:zzp-comparison-3}
\end{align}

Before we read out the translation between the respective transport coefficients, we need to consider that the discussion in~\cite{1980PhRvB..22.2514Z} also includes a dynamical angular part of the strain tensor that does not contribute to plasticity. See our discussion following \eqref{eq:plasticity-dislocation-map}. To switch these effects off we need to take the following limit of the coefficients in their work
\begin{gather}
    \hat n_f \hat a_0^2\to\infty, \qquad 
    \hat D_\perp, \hat D_\| \to 0, \nn\\
    \text{keeping}~~
    \hat D_\perp \hat n_f \hat a_0^2,~ \hat D_\| \hat n_f \hat a_0^2 ~~\text{finite}.
    \label{eq:mode-killer}
\end{gather}
With these in mind, we can directly compare \eqref{eq:zzp-comparison-1}, \eqref{eq:zzp-comparison-2}, \eqref{eq:zzp-comparison-3} to section IV\,D and V\,A of~\cite{1980PhRvB..22.2514Z}. We find the mapping
\begin{gather}
    \hat\eta  = \eta + \frac{nG}{\sigma_\phi}, \qquad 
    \hat\zeta = \zeta 
    + \frac{n B}{\sigma_\phi} \lb 1
    + \frac{\alpha_m}{\chi} (n+B\alpha_m) \rb, \nn\\
    \hat\Gamma =
    \frac{1}{\sigma_\phi}, \nn\\
    \frac{\hat n_f \hat a_0^2}{T} \lb\hat D_\| + \hat D_\perp\rb 
    = 2\hat\nu_0 = \frac{\Omega_G}{G}, \nn\\
    \frac{\hat n_f \hat a_0^2}{T} 2\hat D_\perp = \frac{\Omega_B}{B}.
\end{gather}

Using the mappings above, we can directly compare the linearised mode spectrum reported in~\cite{1980PhRvB..22.2514Z} to the one found in section \ref{sec:plastic-modes}. The matching works perfectly, except that~\cite{1980PhRvB..22.2514Z} has an extra mode in the transverse sector associated with the dynamics of the angular degree of freedom. This mode is removed from the spectrum upon using the limit prescribed in \eqref{eq:mode-killer}.

\section{Relativistic plastic hydrodynamics}
\label{app:relativistic}

In this appendix, we specialise the results from the bulk of the paper to plastic crystals with relativistic boost symmetry. In principle, our boost-agnostic construction covers relativistic systems as a special case (see e.g.~\cite{Armas:2020mpr}). However, it is neater to consider this case separately manifesting the full spacetime covariance at each step. This will allow us to directly compare with the work of~\cite{Fukuma:2011pr} and is also more relevant for holographic models of crystals such as~\cite{Amoretti:2019kuf,Amoretti:2019cef,Amoretti:2018tzw,Amoretti:2017frz,Baggioli:2022pyb}.

\subsection{Relativistic conservation laws}
\label{app:rel-conservation}

Throughout this appendix, we will keep the relativistic background metric $g_{\mu\nu}$ turned on, as well as the background gauge field $A_\mu$. The covariant indices $\mu,\nu,\ldots$ run over both space and time.
We will use $g_{\mu\nu}$ to raise/lower the covariant indices, while the covariant derivative $\nabla_\mu$ shall also be defined with respect to $g_{\mu\nu}$; these conventions differ from the boost-agnostic discussion in appendix \ref{app:background}. The fluid velocity $u^\mu$ (normalised as $u^\mu u_\mu = -1$), temperature $T$, and chemical potential $\mu$ denote the respective relativistic notions, related to the boost-agnostic definitions used in the rest of the paper via Lorentz factors; see~\cite{Armas:2020mpr}.

In a relativistic crystal, we can combine the space and time derivatives of the crystal fields $\phi^I$ into the covariant crystal frame fields $e^I_\mu = \dow_\mu \phi^I$; see~\cite{Armas:2019sbe,Armas:2020bmo}. The crystal velocity $u^\mu_\phi$ can be defined as the unique time-like vector transverse to the frame fields, i.e. $u^\mu_\phi e^I_\mu = 0$. We normalise it so that $u^\mu_\phi u^\nu_\phi g_{\mu\nu}=-1$. In the relativistic context, it is also more useful to define the induced inverse crystal metric $h^{IJ}$ covariantly using the full relativistic metric $g_{\mu\nu}$, i.e. we take $h^{IJ} = g^{\mu\nu}e^I_\mu e^I_\nu$. The induced metric $h_{IJ}$, the reference metric $\bbh_{IJ}$, and the strain tensor $\kappa_{IJ}$ are defined in the same way as our discussion in section \ref{sec:elastic-plastic}. We can also use this to define an inverse frame field $e_I^\mu = h_{IJ}e^{I\mu}$; note that this is \emph{not} the inverse of $e^I_\mu$, which is not a square anymore. In particular, we have $e^I_\mu e_J^\mu = \delta^I_J$ but $e^I_\mu e^\nu_I = \delta^\mu_\nu + u^\mu_\phi u^\phi_\nu$.

In the relativistic case, the coupling of background sources to currents takes a familiar form 
\begin{align} \label{fhiuehieu}
    \delta S 
    &= \int \df^{d+1} x \sqrt{-g} \bigg[\frac{1}{2} T^{\mu \nu } \delta g_{\mu \nu } + J^{\mu} \delta  A_{\mu} \nn\\ 
    &\qquad\qquad 
    +  K_I \delta \phi^I 
    + \frac{1}{2} U^{IJ} \lb \delta \psi_{IJ} - e^\mu_K \dow_\mu\psi_{IJ} \delta\phi^K \rb \nn\\
    &\qquad\qquad
    + \ell' L_I \delta\Phi^I
    \bigg]. 
\end{align}
From these, we can obtain the relativistic version of the energy-momentum and particle conservation laws
\begin{subequations} \label{eoiehu1110998du}
\begin{align}
    \nabla_{\mu} T^{\mu \nu }  
    &= K_{I}e^{I\nu} 
    - \frac{1}{2} U^{IJ} u^\mu_\phi u^\nu_\phi \partial_\nu \psi_{IJ}
    + F^{\nu\rho} J_{\rho} \nn\\
    &\qquad 
    + \ell' L_I \nabla^\nu \Phi^I~~,    
    \label{eq:emcons-rel}  \\ 
    \nabla_{\mu} J^{\mu}   &  =0    ~~ .
    \end{align}
\end{subequations}

On the other hand, the crystal sources can be defined via the source action
\begin{equation}
    S_{\text{source}}
    = \int \df^{d+1}x\sqrt{-g} \, \bigg[
    T^{\mu\nu}_\ext \kappa_{\mu\nu}
    + \Pi_\mu^\ext u^\mu_\phi
    \bigg].
\end{equation}
Performing a variation of the source action, we find
\begin{align}
    \delta S_{\text{source}}
    &= \int \df^{d+1}x\sqrt{-g} \,\Bigg[ 
    \kappa_{\mu\nu} \delta T^{\mu\nu}_\ext
    + u^\mu_\phi \delta \Pi_\mu^\ext
    \nn\\ 
    &\hspace{-3em}
    + K^\ext_I\delta \phi^I 
    + \half U^{IJ}_\ext \lb \delta \psi_{IJ} 
    - e^\mu_K \dow_\mu\psi_{IJ} \delta\phi^K \rb
    \nn\\ 
    &\hspace{-3em}
    + \frac12 
    \lb \bar T^{\mu\nu}_\ext 
    + T^{\rho\sigma}_\ext \kappa_{\rho\sigma} g^{\mu\nu} 
    + \Pi_\rho^\ext u^\rho_\phi P^{\mu\nu}_\phi 
    \rb
    \delta g_{\mu\nu}
    \Bigg],
\end{align}
where $\bar T^{\mu\nu}_\ext = P_\phi^{\mu\rho}P_\phi^{\nu\sigma} T_{\rho\sigma}^\ext$ is defined to be transverse to $u^\mu_\phi$, with the projector operator $P^{\mu\nu}_\phi = g^{\mu\nu} + u^\mu_\phi u^\nu_\phi$. We have also defined
\begin{align}
    U^{IJ}_\ext 
    &= - \ell e^I_\mu e^J_\nu T^{\mu\nu}_\ext, \nn\\
    K^\ext_I
    &= \nabla_\mu \lb 
    T^{\mu\nu}_\ext e^J_\nu \bbh_{IJ}
    + T^{\rho\nu}_\ext u^\phi_\rho 
    e_{I\nu} u^\mu_\phi
    + \Pi^\ext_\nu e^\nu_I u^\mu_\phi
    \rb \nn\\
    &\qquad 
    + \half U^{JK}_\ext e^\mu_I \dow_\mu\bbh_{JK},
\end{align}
which result in the configuration equations for $\phi^I$ and $\psi_{IJ}$ as in \eqref{eq:josehpson-together}.

\subsection{Relativistic viscoplastic hydrodynamics}

The equation of state for the thermodynamic pressure $p$ can be expressed in terms of the relativistic temperature $T$ and chemical potential $\mu$. The thermodynamics takes the appropriate form
\begin{align}
    \df \epsilon &= T\df s + \mu \df n - \half r_{IJ} \df h^{IJ} 
    - \half \bbr^{IJ}\df \bbh_{IJ} \nn\\
    &\qquad 
    + \ell'^2 m^2 h_{IJ} (\phi^J-\Phi^J)\, \df (\phi^I-\Phi^I), \nn\\
    \df p &= s\df T + n\df\mu + \half r_{IJ} \df h^{IJ} 
    + \half \bbr^{IJ}\df \bbh_{IJ} \nn\\
    &\qquad 
    - \ell'^2 m^2 h_{IJ} (\phi^J-\Phi^J)\, \df (\phi^I-\Phi^I), \nn\\
    \epsilon &= Ts + \mu n - p.
\end{align}
Note that there is no explicit momentum term in the thermodynamic relations due to Lorentz-invariance. Momentum appears implicitly through the Lorentz factors contained within the ``proper'' relativistic densities.

We now consider the hydrodynamic constitutive equations for $T^{\mu\nu}$, $J^\mu$, $K_I$, and $U^{IJ}$. With hindsight from our boost-agnostic calculation, we choose the parametrisation of the constitutive relations as
\begin{align}
    T^{\mu \nu }  &= (\epsilon  + p ) u^{\mu } u^{\nu}  + p g^{\mu \nu }    
    -  r_{IJ} e^{I\mu}e^{J\nu}    + \mathcal{T}^{\mu \nu }   , \nn \\
    J^{\mu}  &= n u^{\mu} + \mathcal{J}^{\mu}    ~~ ,    \nn\\    
    K_{I}  &  =  - \nabla_{\mu}\!\lb r_{I J }   e^{J \mu } \rb 
    + \frac{\ell}{2}\bbr^{JK} e^{\mu}_I\dow_\mu \psi_{JK} \nn\\
    &\qquad 
    - \ell'^2 m^2 h_{IJ}\lb\phi^J-\Phi^J\rb
    + \mathcal{K}_I    ~~ ,  \nn\\
    U^{IJ} &  = \ell \bbr^{IJ }
    - \frac{1}{u_\mu u^\mu_\phi}\mathcal{U }^{IJ}~~, \nn\\
    L_I &= \ell' m^2 h_{IJ}\lb\phi^J-\Phi^J\rb
    + {\cal L}_I~~,
\end{align}
where ${\cal T}^{\mu\nu}$, ${\cal J}^\mu$, ${\cal K}_I$, ${\cal U}^{IJ}$, and ${\cal L}_I$ denote the respective dissipative corrections. We shall work in the Landau frame \cite{landau1959fluid, Kovtun:2012rj}, which is to say that we choose the direction of the fluid velocity so that there is no dissipation in the energy density, energy flux/momentum density, or particle density in the comoving frame of the fluid. This amounts to choosing 
\begin{equation}
    {\cal T}^{\mu\nu} u_\nu = {\cal J}^\mu u_\mu = 0.
\end{equation}
These constitutive relations must satisfy the second law of thermodynamics, i.e. there must exist an entropy current $S^\mu$ such that 
\begin{equation}
    \nabla_\mu S^\mu = \Delta \geq 0,
\end{equation}
on solutions of the equations of motion.
We choose the entropy current to take the canonical form
\begin{align}
    S^{\mu} & = s u^{\mu}  - \frac{\mu}{ T} \mathcal{J}^{\mu}   ~~ .
\end{align}
Going through with the entropy divergence calculation as in appendix \ref{app:background}, we find that the relativistic dissipation rate is given by
\begin{align}
    T\Delta
    &= - {\cal T}^{\mu\nu} \nabla_\mu u_\nu 
    - {\cal J}^\mu \lb T \dow_\mu \frac{\mu}{T} + u^\nu F_{\nu\mu} \rb  \nn\\
    &\qquad 
    - {\cal K}_I u^\mu e^I_\mu 
    - \half {\cal U}^{IJ} u^\mu_\phi \dow_\mu \psi_{IJ} \nn\\
    &\qquad 
    - \ell' {\cal L}_I u^\mu \dow_\mu \Phi^I.
\end{align}

Up to first order in derivatives, we can work out the hydrodynamic constitutive relations as in \ref{sec:constitutive}. First, in the vector sector we have~\cite{Armas:2019sbe,Armas:2020bmo}
\begin{subequations}
\begin{equation}
    \begin{pmatrix}  
        \mathcal{J}^{I}    \\ 
        {\cal K}^I \\
        {\cal L}^I
    \end{pmatrix}  = -  
    \begin{pmatrix}
        \sigma_{n} & \gamma_{n  \phi } & \gamma_{n\Phi}   \\
        \gamma'_{n\phi}   &    \sigma_{\phi} & \sigma_\times  \\
        \gamma'_{n\Phi} & \sigma'_\times & \sigma_\Phi 
    \end{pmatrix}   
    \begin{pmatrix}  
        P^{I\mu}(T \dow_\mu \frac{\mu}{T} + u^\nu F_{\nu\mu})  \\ 
        u^\mu e^I_\mu \\
        \ell' u^\mu \dow_\mu \Phi^I
    \end{pmatrix},
\end{equation}
where we have chosen the parametrisation of dissipative corrections ${\cal J}^\mu = P^\mu_I{\cal J}^I$. We have defined the projected frame fields as $P^I_\mu = e^I_\mu + u_\mu u^\nu e^I_\nu$. As before, we get plasticity effects in the scalar and symmetric-traceless tensor sector
\begin{align}
    \begin{pmatrix}
        {\cal T}^{IJ} \nn\\
        {\cal U}^{IJ}
    \end{pmatrix}
    &= 
    - h^{IJ}\begin{pmatrix}
        \zeta_\tau & \zeta_{\tau\bbh} \\
        \zeta'_{\tau\bbh} & \zeta_\bbh
    \end{pmatrix}
    \begin{pmatrix}
        P_I^\mu P^I_\nu \nabla_\mu u^\nu \\
        \half h^{KL} u^\mu_\phi \dow_\mu \psi_{KL}
    \end{pmatrix} \nn\\
    &
    - \begin{pmatrix}
        \eta_\tau & \eta_{\tau\bbh} \\
        \eta'_{\tau\bbh} & \eta_\bbh
    \end{pmatrix}
    \begin{pmatrix}
        2 P^{\langle I}_\mu P^{J\rangle}_\nu \nabla^\mu u^\nu \\
        h^{K\langle I}h^{J\rangle L} u^\mu_\phi \dow_i \psi_{KL}
    \end{pmatrix},
    \label{eq:consti-scalar-tensor-rel}
\end{align}
\end{subequations}
where we have again defined ${\cal T}^{\mu\nu} = P^\mu_I P^\nu_J {\cal T}^{IJ}$.

We get the same constraints as sections \ref{sec:constitutive} and \ref{sec:correlations-pinned} for Onsager's relations, i.e.
\begin{subequations}
\begin{gather}
    \gamma'_{n\phi} = - \gamma_{n\phi}, \qquad 
    \eta'_{\tau\bbh} = \eta_{\tau\bbh}, \qquad 
    \zeta'_{\tau\bbh} = \zeta_{\tau\bbh}, \nn\\
    \gamma'_{n\Phi} = - \gamma_{n\Phi}, \qquad 
    \sigma'_\times = \sigma_\times.
\end{gather}
On the other hand, demanding the entropy production to be sign-definite, we get the positivity constraints 
\begin{gather}
    \eta_\tau \geq 0, \qquad 
    \zeta_\tau \geq 0, \qquad 
    \sigma_\phi \geq 0, \nn\\
    \sigma_n \geq 0, \qquad 
    \eta_\tau\eta_\bbh \geq \eta_{\tau\bbh}^2, \qquad 
    \zeta_\tau\zeta_\bbh \geq \zeta_{\tau\bbh}^2, \nn\\
    \sigma_\Phi \sigma_\phi \geq \sigma_\times^2.
\end{gather}
\end{subequations}

\subsection{Modes and correlators}

Let us now specialise to the isothermal regime like the bulk of the paper. In the relativistic context, this decouples the $u_\mu$ projection of the energy-momentum conservation equation \eqref{eq:emcons-rel}. Following through with the computation, we find the exact same modes as discussed in sections \ref{sec:plastic-modes} and \ref{sec:pinned-modes}, only the momentum susceptibility $\rho$ should be replaced with its relativistic version $\epsilon+p$ (where the speed of light $c$ has been set to 1). The same holds for all the correlation functions discussed in sections \ref{sec:correlations} and \ref{sec:correlations-pinned}.

\newpage

%
\end{document}